\documentclass[fleqn,usenatbib]{mnras}

\usepackage{newtxtext,newtxmath}
\usepackage[T1]{fontenc}

\DeclareRobustCommand{\VAN}[3]{#2}
\let\VANthebibliography\thebibliography
\def\thebibliography{\DeclareRobustCommand{\VAN}[3]{##3}\VANthebibliography}

\usepackage{graphicx}	% Including figure files
\usepackage{amsmath}	% Advanced maths commands

\usepackage{mathtools}
\usepackage{multirow}
\usepackage{siunitx}
\usepackage{textgreek}
\usepackage[dvipsnames]{xcolor}
\DeclareSIUnit \h {\ensuremath{\mathit{h}}}
\DeclareSIUnit \parsec {pc}
\DeclareSIUnit \solarmass {\ensuremath{\mathit{M_\odot}}}
\DeclareSIUnit \dex {dex}

\newcommand{\hMsun}{h^{-1}\mathrm{M_\odot}}
\newcommand{\hsqMsun}{h^{-2}\mathrm{M_\odot}}
\newcommand{\hkpc}{h^{-1}\mathrm{kpc}}
\newcommand{\hMpc}{h^{-1}\mathrm{Mpc}}
\newcommand{\hGpc}{h^{-1}\mathrm{Gpc}}

\newcommand{\kms}{\mathrm{km\,s^{-1}}}

\newcommand{\sqdeg}{\mathrm{deg}^2}

\newcommand{\magr}{{}^{0.1}M_r^h}
\newcommand{\kcorrr}{{}^{0.1}k_r}

\title[Uchuu-SDSS galaxies]{The Uchuu-SDSS galaxy lightcones: a clustering, RSD and BAO study}

\author[C.~A.~Dong-P\'aez et al.]{
C. A.~Dong-P\'aez$^{1,2,3}$\thanks{E-mail: chiandongpaez@gmail.com}, A.~Smith$^{3,4,5}$, A. O.~Szewciw$^{6}$, J.~Ereza$^{1}$, 
M. H.~Abdullah$^{7,8}$, C.~Hern\'andez-Aguayo$^{9,10}$,
\newauthor
S.~Trusov$^{11}$,
F.~Prada$^{1}$, A.~Klypin,$^{12,13}$, T.~Ishiyama$^{7}$,
A.~Berlind$^{6}$, P.~Zarrouk$^{11}$, J.~L\'opez Cacheiro$^{14}$, J.~Ruedas$^{1}$
\vspace*{4pt} \\ 
\scriptsize $^{1}$Instituto de Astrof\'isica de Andaluc\'ia (CSIC), Glorieta de la Astronom\'ia, E-18080 Granada, Spain \vspace*{-2pt}\\
\scriptsize $^{2}$Institut d’Astrophysique de Paris, Sorbonne Universit\'es, CNRS, UMR 7095, 98 bis bd Arago, 75014 Paris, France \vspace*{-2pt}\\
\scriptsize $^{3}$Institute for Computational Cosmology, Department of Physics, Durham University, South Road, Durham DH1 3LE, UK \vspace*{-2pt}\\
\scriptsize $^{4}$IRFU, CEA, Universit\'e Paris-Saclay, F-91191 Gif-sur-Yvette, France \vspace*{-2pt}\\
\scriptsize $^{5}$Institute for Astronomy, University of Edinburgh, Royal Observatory, Blackford Hill, Edinburgh EH9 3HJ, UK \vspace*{-2pt}\\
\scriptsize $^{6}$Vanderbilt University, 2201 West End Ave, Nashville, TN, 37235 USA \vspace*{-2pt}\\
\scriptsize $^{7}$Institute of Management and Information Technologies, Chiba University, 1-33, Yayoi-cho, Inage-ku, Chiba 263-8522, Japan \vspace*{-2pt}\\
\scriptsize $^{8}$Department of Astronomy, National Research Institute of Astronomy and Geophysics, Cairo, 11421, Egypt \vspace*{-2pt}\\
\scriptsize $^{9}$Max-Planck-Institut f\"ur Astrophysik, Karl-Schwarzschild-Str 1, D-85748 Garching, Germany \vspace*{-2pt}\\
\scriptsize $^{10}$Excellence Cluster ORIGINS, Boltzmannstrasse 2, D-85748 Garching, Germany \vspace*{-2pt}\\
\scriptsize $^{11}$ Sorbonne Universit\'e, Universit\'e Paris Diderot, Sorbonne Paris Cit\'e, CNRS,
Laboratoire de Physique Nucléaire et de Hautes Energies (LPNHE), 4 place Jussieu, F-75252, Paris
Cedex 5, France \vspace*{-2pt}\\
\scriptsize $^{12}$Astronomy Department, New Mexico State University, Las Cruces, NM, USA \vspace*{-2pt}\\ 
\scriptsize $^{13}$Department of Astronomy, University of Virginia, Charlottesville, VA, USA \vspace*{-2pt}\\
\scriptsize $^{14}$Centro de Supercomputaci\'on de Galicia (CESGA), Avenida de Vigo, s/n Campus Sur, E-15705 Santiago de Compostela, Spain \vspace*{-2pt}\\
}

\date{Accepted XXX. Received YYY; in original form ZZZ}

\pubyear{2022}

\begin{document}
\label{firstpage}
\pagerange{\pageref{firstpage}--\pageref{lastpage}}
\maketitle

\begin{abstract}
%%%%%% SHORTER VERSION, CUT DOWN TO 250 WORDS %%%%%%%
We present the data release of the Uchuu-SDSS galaxies: a set of 32 high-fidelity galaxy lightcones constructed from the large Uchuu 2.1 trillion particle $N$-body simulation using Planck cosmology. We adopt subhalo abundance matching to populate the Uchuu-box halo catalogues with SDSS galaxy luminosities. These cubic box galaxy catalogues generated at several redshifts are combined to create the set of lightcones with redshift-evolving galaxy properties. The Uchuu-SDSS galaxy lightcones are built to reproduce the footprint and statistical properties of the SDSS main galaxy survey, along with stellar masses and star formation rates. This facilitates direct comparison of the observed SDSS and simulated Uchuu-SDSS data. Our lightcones reproduce a large number of observational results, such as the distribution of galaxy properties, the galaxy clustering, the stellar mass functions, and the halo occupation distributions. Using the simulated and real data we select  samples of bright red galaxies at $z_\mathrm{eff}=0.15$ to explore Redshift Space Distortions and Baryon Acoustic Oscillations (BAO) utilizing a full-shape analytical model of the two-point correlation function. 
We create a set of 5100 galaxy lightcones using GLAM N-body simulations to compute covariance errors. 
We report a $\sim 30\%$ precision increase on $f\sigma_8$, due to our better estimate of the covariance matrix. From our BAO-inferred $\alpha_{\parallel}$ and $\alpha_{\perp}$ parameters, we obtain the first SDSS measurements of the Hubble and angular diameter distances
$D_\mathrm{H}(z=0.15) / r_d = 27.9^{+3.1}_{-2.7}$, $D_\mathrm{M}(z=0.15) / r_d = 5.1^{+0.4}_{-0.4}$. Overall, we conclude that the Planck \textLambda CDM cosmology nicely explains the observed large-scale structure statistics of SDSS. All data sets are made publicly available.

\end{abstract}


%%%%%%%%%%%%%%%%%%%%%%%%%%%%%%%%%%%%%%%%%%%%%%%%%%

%%%%%%%%%%%%%%%%% BODY OF PAPER %%%%%%%%%%%%%%%%%%

\section{Introduction}
\label{sec:introduction}

Galaxy redshift surveys, measuring the spatial distribution of galaxies throughout cosmic time, have proven to be key observational probes for constraining cosmological models and astrophysical phenomena. For example, the distribution of galaxies on large scales can be used to infer cosmological parameters \citep[e.g.][]{Alam21}, as well as to constrain the properties of dark energy that drives the accelerated expansion of the Universe \citep{Riess98,Perlmutter99}. On smaller scales, the galaxy clustering signal encodes information on how galaxies populate dark matter haloes, on star formation, feedback and other baryonic processes that shape galaxy formation \citep[see][for a review]{Wechsler18}.

In order to connect galaxy redshift surveys to theoretical predictions, it is essential to generate high-fidelity galaxy catalogues from cosmological simulations that capture the expected properties and clustering of the observed galaxy sample \citep[e.g.][]{delatorre13,White14,RodriguezTorres16,Lin20}. 
%The mock data must resemble the galaxy survey at hand, and provide the same galaxy properties.
%Mock catalogues are usually created from cosmological simulations, which model the growth of structure across cosmic time by solving numerically the relevant physical equations.
Hence, they can be used to increase the amount of information that can be extracted from galaxy survey data. Firstly, the cosmological parameters of the simulation are known exactly. 
%The effectiveness of the analysis tools used to extract cosmological parameters from real survey data can therefore be assessed by applying them to the simulated galaxy survey, and thus, contrasting these estimates against the known cosmological parameters. %The effectiveness of the tools used to extract cosmological parameters from real survey data can be assessed by applying these tools in the mock data. Since the cosmological parameters corresponding to the mock data are known, they can be contrasted against the estimates obtained from mock data. 
This allows us to assess the systematic and statistical errors of cosmological measurements from galaxy surveys, to compute the covariance errors, to test the performance of statistical analyses, and to aid the theoretical interpretation of survey results. Secondly, high-fidelity catalogues based on simulations allow us to assess the systematics arising from such observational effects as selection function and fibre collisions. These effects are a source for incompleteness in the survey sample. They must to be understood in order to minimise their impact in the measured clustering signal \citep[see][and references therein]{Smith19}.

In the last two decades, vast observational efforts such as the 2dF Galaxy Redshift Survey (2dFGRS) \citep{Colless01} and the Sloan Digital Sky Survey (SDSS) saga \citep{Alam21},
%the Galaxy And Mass Assembly (GAMA) survey \citep{Baldry10} and the Sloan Digital Sky Survey (SDSS) \citep{York00,Abazajian2009,Eisenstein11} 
have driven most of the major discoveries about the Large-Scale Structure (LSS) of our Universe. The relevance of the SDSS surveys is not purely historical -- to date, SDSS continues to be the largest reference galaxy database, offering sky positions, redshifts, spectra and images for millions of galaxies. New technological advances continue to push the depth and volume of galaxy surveys -- the new generation of galaxy surveys, which include the Dark Energy Spectroscopic Instrument (DESI) survey \citep{DESI16}, the Large Synoptic Survey Telescope (LSST) \citep{Ivecic19}, the Subaru Prime Focus Spectrograph (PFS) \citep{Takada14} and the Euclid survey \citep{Laureijs11}, aim to produce unprecedentedly large data sets in an effort to map the Universe to even higher precision. 

In order to generate high-fidelity simulated galaxy lightcones for these large surveys, cosmological simulations with high-resolution in a large volume are needed. Although it would be desirable to create mocks from hydrodynamical simulations, in which the formation of galaxies is modelled self-consistently by solving the coupled evolution of both baryons and dark matter (DM), such simulations are complex and computationally expensive. The largest to date do not exceed a few hundred Mpc 
in box size \citep[e.g.][]{Dubois14,Schaye15,Pillepich18,Springel18}. Hydrodynamical simulations are also strongly affected by the modelling uncertainties of complex baryonic processes. Thus, one usually resorts to DM-only simulations. Since galaxy formation physics is not included, dark matter halos must be populated with galaxies in order to produce the desired simulated galaxy catalogue. This can be done using a variety of methods, each comprising different assumptions about the galaxy--halo connection \citep{Wechsler18}.%Although it would be desirable to create mock catalogues from hydrodynamical simulations, in which the formation of galaxies in dark matter (DM) haloes is modelled self-consistently by solving the coupled evolution of both baryons and DM, such simulations are complex and very computationally expensive. The largest hydrodynamical simulations to date do not exceed a few hundred $\si{\per\h\mega\parsec}$ in box size \citep[e.g.][]{Dubois14,Schaye15,Pillepich18}. Additionally, a large amount of baryonic processes relevant to galaxy formation occur on scales which are much smaller than the resolution of the simulation, and consequently have to be implemented in a subgrid fashion. Some properties of the simulated galaxies are strongly model-dependent, and struggle to reproduce the observed population. Thus, one usually resorts to DM-only simulations, which are significantly cheaper than their hydrodynamic counterparts. Nonetheless, since baryons are not included, in order to produce a mock galaxy catalogue, galaxies must be "painted" onto the DM distribution. This can be done using a variety of methods, each comprising a different set of assumptions about the galaxy--halo connection \citep{Wechsler18}.

A common option to generate galaxies from DM halos is to use empirically based methods, such as subhalo abundance matching \citep[SHAM; e.g.][]{Marinoni02,Vale04,Kravtsov04,Conroy06,RodriguezTorres16,Safonova21}. 
%SHAM is a technique that straightforwardly connects the galaxies in large surveys to the evolution of their dark matter haloes%, allowing the resulting galaxy population to readily reproduce a large number of observational results
In its simplest form, the main underlying assumption is that every halo and subhalo contains a galaxy, and that there are correlations between halo and galaxy properties. For instance, the most massive and luminous galaxies are generally assumed to reside in the most massive haloes. At face value, one can assign galaxies to haloes by generating a rank-ordered relation between observed galaxy luminosities and simulated halo masses. However, such a one-to-one relation between galaxies and halos is incompatible with observations~\citep[e.g.][]{Trujillo-Gomez11,Shu12} -- an intrinsic scatter must be incorporated in the method to produce realistic catalogues. Furthermore, the clustering of haloes is observed to depend on properties other than halo mass, including the halo formation time, concentration and spin~\citep[e.g.][]{Wechsler01, Gao05, Wechsler06}. 
%In order to account for this phenomenon, known as assembly bias, a suitable proxy for the halo mass incorporating some of the required additional information may be needed.
%A few possible choices for this halo parameter are discussed later in Section~\ref{subsubsec:luminosity_SHAM}. 
Empirical models such as SHAM or the popular HOD method \citep[e.g.][]{Zehavi11} are computationally fast, and are able to workaround the uncertainties in the physics of galaxy formation by constraining the model parameters directly with data. Applications of the SHAM method are able to reproduce observed properties of galaxies in large surveys as the luminosity and stellar mass functions or the luminosity and colour-dependent clustering to high accuracy \citep[e.g.][]{Trujillo-Gomez11,RodriguezTorres16}. 
%\ca{I have removed mentions to HOD, which might be useful for section \ref{subsec:GLAM_construction}? Let me know if you want me to restore it}

%---------------------------------------
% RSD introduction (draft)
%---------------------------------------

From real and simulated survey data, the nature of dark energy can be probed for instance by measuring the growth rate of structure, defined in linear theory as $f(a) = d \mathrm{ln}D(a)/d \mathrm{ln}a$, where $a$ is the scale factor and $D(a)$ is the linear growth function. This parameter can be regarded as a measure of the energy content of the Universe, allowing us to constrain different theories of gravity and dark energy \citep[see e.g.][for a review]{Peebles1980_book,Guzzo2008}. %This parameter allows us to probe the energy content of the Universe through the relation $f(z)\approx\Omega_\mathrm{m}(z)^{\gamma}$, where $\gamma$ is related to the equation of state of dark energy and is predicted in general relativity to be $\gamma \approx 0.55$. This quantity has therefore the power to constrain different theories of gravity and dark energy \citep[see e.g.][for a review]{Peebles1980_book,Guzzo2008}.
%When it comes to analysing the galaxy catalogues, one immediately notices that the g
Galaxy peculiar velocities introduce anisotropies in their observed redshifts. This effect, first described in \citet{Kaiser1987}, is known as Redshift Space Distortions (RSD). Measurements of RSD in galaxy surveys yield $f\sigma_8$, where $\sigma_8(z)$ is the normalization of the power spectrum at redshift $z$ on a scale of $8~\hMpc$. Another source of anisotropy comes from the choice of the fiducial cosmology adopted in the clustering analysis, used for converting redshifts and angles to comoving coordinates. If the fiducial cosmology differs from the observed one, galaxy clusters will appear flattened or elongated%, depending on the choice of the fiducial cosmology \citep{Alcock1979}
. The study of this so-called Alcock-Paczynski effect can provide an additional source of cosmological information from the data, as well as a means to validate a cosmological model from a simulated galaxy survey.
%Thus, RSD measurements allow the cosmological parameters of interest to be measured independently of the fiducial cosmology, as the difference between it and the real one will be accounted for.

%-------------------------------------

%---------------------------------------
% BAO introduction (draft)
%---------------------------------------
In addition to RSD, one can measure the Baryon Acoustic Oscillation (BAO) feature from the two-point clustering statistics in galaxy surveys% in the survey
. %galaxy surveys provide enough data to estimate the scale of Baryon Acoustic Oscillations (BAO) from the measured galaxy clustering (two-point correlation function or power spectrum), which 
This allows us to determine the expansion rate of the Universe \citep[e.g.][]{Eisenstein:2005su,Cole05}. Recently, \citet{Alam21} reported cosmological implications from two decades of SDSS spectroscopic surveys based on clustering measurements from galaxies and quasars in the redshift range $0.1 < z < 3$. %\citep{eboss1,eboss2,eboss3,eboss4}. 
This includes the SDSS DR7 main galaxy sample (MGS), at low redshift $z\sim0.1$, which clustering statistics is compared in this work with the high-fidelity simulated galaxy light-cones built from on our 2.1 trillion $N$-body Uchuu simulation  \citep{UchuuDR1}.

%-------------------------------------

%The aim of this paper is to create high-fidelity mocks tailored for the analysis and interpretation of SDSS data. By applying a mixture of SHAM and statistical methods on the large, high-resolution, dark-matter-only Uchuu simulation \citep{UchuuDR1}, we construct a set of 32 lightcone catalogues from a set of simulation snapshots at different redshifts, accounting naturally for the evolution of galaxy properties with redshift. 

Our Uchuu-SDSS light-cones are built using the SHAM method to reproduce the basic properties of the SDSS galaxy population, match its sky footprint and selection function, and include the effect of fibre collisions, which facilitates their straightforward comparison with the SDSS MGS data. Furthermore we study and measure RSD and BAO in the real and simulated data to test the Planck base $\Lambda$CDM cosmology model \citep{Planck18}.
%In order to test the quality of our mocks, we compare their properties against a galaxy sample from the seventh data release of SDSS \citep{Abazajian2009}. 
The Uchuu-SDSS galaxy catalogues include sky positions, redshifts, $ugriz$ apparent and absolute magnitudes  
stellar masses and star formation rates (SFR), as well as several halo properties. The catalogues are made publicly available at the Skies \& Universes website\footnote{\url{http://www.skiesanduniverses.org/Simulations/Uchuu/}}. 
%Our Uchuu-SDSS galaxy catalogues, including sky positions, redshifts, apparent and absolute magnitudes in several bands, observed and rest-frame $g-r$ colours, stellar masses and star formation rate, as well as many other halo properties, are made publicly available at \url{http://www.skiesanduniverses.org}.

The structure of this paper is as follows. In Section~\ref{sec:sdss}, we introduce the SDSS MGS data and define the volume-limited galaxy samples used to validate our simulated galaxy catalogues. In Section~\ref{sec:mock_construction}, we describe the Uchuu simulation and the methodology behind the creation of galaxies from the dark matter halo properties. This includes the construction of light-cones from cubic boxes, the implementation of fibre collisions, and the assignment of additional SDSS galaxy properties.
%In Section~\ref{sec:mock_construction}, we describe the Uchuu simulation, the methodology behind the construction of the mock box catalogues from the simulation, and the construction of mock SDSS lightcone catalogues, as well as the assignment of additional galaxy properties and our implementation of fibre collisions in our lightcone catalogues. 
Section~\ref{sec:properties} presents the basic properties of the Uchuu-SDSS light-cones, compared to the SDSS data. In particular, we explore the statistics of several galaxy properties, the galaxy clustering dependence on luminosity, color, and stellar mass, and the galaxy HOD. In Section~\ref{sec:RSD_BAO}, we present our RSD and BAO measurements both in the real and simulated data, and compare them with the fiducial Planck cosmology. Finally, in Section~\ref{sec:conclusions}, we present a summary of our results.

\section{SDSS Galaxy Samples}
\label{sec:sdss}

\subsection{Parent and volume-limited samples}
\label{subsec:sdss_samples}

\begin{table}
\centering
    \begin{tabular}{cccccc}
        \hline
        $^{0.1}M_r^\mathrm{max}$ & $z_\mathrm{max}$ & $N$ & $n_g$ & $V_\mathrm{eff}$ & $f_\mathrm{NN}$ \\
        \hline
        -18.0 & 0.041 & 35359 & 31.95 & 1.11 & 0.043 \\
        -18.5 & 0.053 & 49272 & 20.41 & 2.54 & 0.046 \\
        -19.0 & 0.064 & 62534 & 14.47 & 4.55  & 0.050 \\
        -19.5 & 0.085 & 112652 & 11.09 & 10.68 & 0.058 \\
        -20.0 & 0.106 & 119734 & 6.13 & 20.53 & 0.062 \\
        -20.5 & 0.132 & 112496 & 3.03 & 39.03 & 0.068 \\
        -21.0 & 0.159 & 71795 & 1.13 & 66.95 & 0.079 \\
        -21.5 & 0.198 & 33505 & 0.28 & 125.63 & 0.099 \\
        -22.0 & 0.245 & 9820 & 0.045 & 218.32 & 0.149 \\
        \hline
    \end{tabular}
\caption{For each SDSS volume-limited sample, the columns list (from left to right): the absolute magnitude threshold, the maximum redshift, the number of galaxies, the galaxy number density (in $10^{-3}~h^{3}\mathrm{Mpc}^{-3}$), the effective volume of each sample (in $10^{6}~h^{-3}\mathrm{Mpc}^{3}$), and the fraction of galaxies requiring the nearest neighbour correction. The lower redshift cut for all samples is $z_\mathrm{min}=0.02$. Magnitudes of all galaxies are $k$–corrected and passively evolved to the survey's median redshift of $z = 0.1$ and computed assuming $h=1$.}
\label{tab:sdss}
\end{table}

In this work, we build and compare Uchuu simulated galaxies to observational data from the seventh data release \citep[DR7;][]{Abazajian2009} of the SDSS \citep[][]{York00}. More specifically, we make use of the large scale structure catalog of the SDSS MGS from the NYU Value Added Galaxy Catalogue \citep[NYU-VAGC;][]{Blanton2005}.
We restrict our sample to consist only of galaxies in the contiguous northern footprint in regions which have a completeness of $> 90\%$. 
This parent sample covers an effective area of $\sim6511\ \mathrm{deg}^2$ and contains $\sim497\,000$ galaxies with Petrosian $r$-band magnitudes in the range $14.5<r_\mathrm{pet}<17.6$.
In this sample, $\sim6\%$ of targeted galaxies lack a spectroscopically measured redshift due to fibre collisions.
We apply a nearest neighbour correction to these galaxies, assigning to them the redshift of the galaxy with which they collided.

Apparent magnitudes can be linked to absolute magnitudes via the distance modulus equation,
\begin{equation}\label{eq:appmag}
r = \magr + 5\log_{10}D_L(z)+ 25 + \kcorrr(z),
\end{equation}
where $D_L(z)$ is the luminosity distance (in units of $\hMpc$), $\kcorrr(z)$ is the $k$-correction,
which takes into account the shift in the bandpass with redshift. 

Throughout this paper, we denote $k$-corrected unevolved absolute magnitudes in the $r$-band as $\magr \equiv {}^{0.1}M_r - 5\log_{10} h$, where the superscript 0.1 indicates that the rest-frame magnitude has been $k$-corrected to a reference redshift of $z_\mathrm{ref}=0.1$ \citep{Blanton2003b}. 
A similar notation is used for magnitudes in other bands. 
We also denote the rest-frame $g-r$ colours as ${}^{0.1}(g-r)$, $k$-corrected to the same reference redshift.

%The absolute magnitudes of these galaxies are computed assuming $h=1$ and have been $k$-corrected to rest-frame magnitudes at redshift $z=0.1$.
Additionally, the absolute magnitudes are corrected for passive evolution using the model of \citet{Blanton2006}.
Specifically, the `evolved' magnitude ${}^{0.1}M_r^{h,e}$ is given by
\begin{equation}
    {}^{0.1}M_r^{h,e}=\magr + E(z),
    \label{eq:e-correction}
\end{equation}
where the evolution correction $E(z)$ is given by
\begin{equation}
   E(z) = q_0(1+q_1(z-q_{z0}))(z-q_{z0}).
\end{equation}
This model contains three parameters which we set to values of $q_0=2$, $q_1=-1$, and $q_{z0}=0.1$.

From this parent sample of galaxies, we construct nine different volume-limited samples, each of which is complete down to a specified $r$-band magnitude, $^{0.1}M_r^\mathrm{max}$, which is $k$-corrected and corrected for evolution. 
For our samples, we use the same magnitude thresholds and maximum redshifts as \citet{Guo15}.
In all samples, to minimize the impact of peculiar velocities and spureous objects, we keep only galaxies with redshifts $z>0.02$.
In Table~\ref{tab:sdss}, for each sample we provide the magnitude threshold, maximum redshift, number of galaxies, number density of galaxies, and effective volume.
These quantities can be compared directly to those in Table 1 of \citet{Guo15}.
We also show the fraction of galaxies in each sample which require the nearest neighbour correction due to fibre collisions, see Section~\ref{subsec:fibre} for the details.

Finally, in order to study in greater detail the BAO signal in the SDSS MGS, we define a BAO sample in which this signal is enhanced, analogously to \citet{Ross15}. This sample, named SDSSbao, is defined by the cuts, %$0.07<z<0.2$, $\magr<21.2$ and ${}^{0.1}(g-r)>0.8$.
\begin{equation}  
        \begin{array}{l}
        0.07 < z < 0.2 \\
        \magr<-21.2\\
        {}^{0.1}(g-r)>0.8\\
        14.5 < r < 17.6\\
        c>0.9,
        \end{array}
    \label{eq:hod_cuts}
\end{equation}
where $c$ is the completeness.

% \Adam{If there is a desire to also show clustering comparisons between Guo and Uchuu without fibre collisions, this would be a place to mention that. Currently the Guo comparisons are what's in the paper. I think there's value in doing both comparisons: 1. Uchuu w/o fibre collisions vs. Guo; 2. Uchuu w/ fibre collisions vs. the sample described in this section.}

\subsection{Stellar masses and star formation rates}

\label{sec:stellar_mass_sfr}

Uchuu-SDSS galaxy catalogues list stellar masses $M_\ast$ and star formation rates SFRs using two independent sources, which are the MPA/JHU catalogue\footnote{\url{ http://www.mpa-garching.mpg.de/SDSS/DR7/}} (the Max Planck Institute for Astrophysics and the Johns Hopkins University) and the Granada Group catalogue\footnote{\url{https://www.sdss.org/dr17/spectro/galaxy_granada/}. We use the data from stellarMassFSPSGranEarlyDust table.}

For the MPA/JHU catalogue, $M_\ast$ was calculated following the methodologies presented in  \citet{Kauffmann03} and \citet{Salim07}. 
%Stellar masses are calculated using a Bayesian approach and theoretical stellar populations models by \citep{Kauffmann03}, and assuming a Kroupa initial mass function \citep{Kroupa01}. 
%The galaxy spectra are measured through a $3\arcsec$ aperture, which is in general not representative of the whole extent of a galaxy. Therefore, the MPA/JHU model is built on the $ugriz$ galaxy photometry only. The photometry is corrected for the small contribution from nebular emission using the galaxy spectra. The stellar mass within the SDSS spectroscopic fibre aperture of $3\arcsec$ is estimated using SDSS-fibre magnitudes while for the total stellar mass SDSS-model magnitudes are used. %They provided the stellar mass corresponding to the median and the 2.5\%, 16\%, 84\% and 97.5\% percentiles of the probability distribution function.
%Galaxy stellar masses calculated by the MPA/JHU team differ slightly from the masses obtained in \citet{Kauffmann03}, which uses spectral indices. 
%of $\mathrm{D}_{4000}$ and $\mathrm{H}_\delta$, and a total of five indices in the case of \citet{Gallazzi05}. 
The stellar masses in the MPA/JHU catalogue, have been found to be consistent with other estimates \citep[e.g.][]{Taylor11,Chang15,Duarte15,Leslie16}. 
%Throughout the paper we use the median as $M_\ast$ of a galaxy \Paco{What does it mean?}. 
%The largest differences is due to aperture corrections and changing M/L with radius in galaxies.

For galaxies classified as star forming (SF) in the MPA/JHU catalogue, SFRs are calculated using the nebular emission lines within the spectroscopic fibre aperture of $3\arcsec$ as described in \citet{Brinchmann04}. SFRs are calculated using the empirical calibration of $\mathrm{H}_\alpha$~emission lines \citep{Kennicutt98} and corrected from the dust extinction with the Balmer decrement $\mathrm{H}_\alpha/\mathrm{H}_\beta$ \citep{Charlot00}, assuming a Kroupa initial mass function \citep{Kroupa01}. The SFR contribution outside of the fibre is estimated from the galaxy photometry following \citet{Salim07}. 
%For galaxies with weak emission lines classified as non-SF or passive, SFRs were estimated from their photometry. The MPA/JHU team estimates their SFRs from an empirical relation between the SFR and the spectral index $\mathrm{D}_{4000}$  (\citealp{Bruzual83,Balogh99,Brinchmann04}). 
Finally, the specific star formation rate (sSFR, defined as SFR/$M_\ast$) were calculated by combining the SFR and stellar mass likelihood distributions as outlined in Appendix A of \citet{Brinchmann04}. %MPA/JHU team reported both the fibre and the total sSFR at the median and the 2.5\%, 16\%, 84\% and 97.5\% percentiles of the probability distribution function.
Throughout the paper we use the median values of the resulting probability distribution functions as the sSFR of a galaxy.

For the Granada Group catalogue, $M_\ast$ and SFR are calculated using the SDSS spectroscopic redshift and $ugriz$ magnitudes by means of broad-band spectral energy distribution (SED) fitting via Flexible stellar population synthesis technique (FSPS, \citealp{Conroy09}). 
%The flexibility of the model templates allows for a variety of formation time scenarios% including early star formation time
%, and a wide range of star formation times. %The technique also included two scenarios with dust and no dust configurations \citep{Charlot00}. Calculations carried out for both \citet{Salpeter55} and \citet{Kroupa01} initial mass functions. Note that the Granada galaxy stellar masses for all configurations are usually larger than those estimated by MPA/JHU team.
Our Uchuu-SDSS galaxy catalogues include $M_\ast$ and sSFR obtained from the early formation time and dust attenuation model \citep{Charlot00} with \citet{Kroupa01} initial mass functions. %This model configuration returns the smallest galaxy stellar masses, but are nonetheless generally higher than the MPA/JHU estimates. However, in the following we focus primarily on the $M_\ast$ and sSFR MPA/JHU values (see Section~\ref{sec:SMF}).
%%%%%%%%%%%%%%%%%%%%%%%%%%%%%%%%%%%%%%%%%%%%%%%%%%%%%%%%%%%%%%%%%%%%%%%%%%%%%%%%%%%%%%%%%%%55

\section{Constructing the Uchuu-SDSS catalogues}
\label{sec:mock_construction}

\subsection{The Uchuu Simulation}

The Uchuu simulation is a large high-resolution $N$-body cosmological simulation, the largest simulation in the Uchuu suite~\citep{UchuuDR1}. It follows the evolution of $2.1$ trillion ($12\,800^3$) dark matter particles with particle mass resolution of \SI{3.27e8}{\per\h\solarmass} in a (\SI{2.0}{\per\h\giga\parsec})$^3$ comoving periodic box. The simulation adopts the Planck \textLambda CDM cosmological parameters: $\Omega_\mathrm{m} = 0.3089$, $\Omega_\mathrm{b}=0.0486$, $\Omega_\Lambda=0.6911$, $h=0.6774$, $n_\mathrm{s}=0.9667$, and $\sigma_8=0.8159$~\citep{Planck16}.
% For a summary of the cosmological parameters and properies of the Uchuu simulation, see Table ref.
Starting at $z=127$, the subsequent gravitational evolution was solved down to $z=0$ using the TreePM code \textsc{GreeM}~\citep{Ishiyama09,Ishiyama12}, with a gravitational softening length of \SI{4.27}{\per\h\kilo\parsec}.

A set of 50 particle snapshots ranging from $z=14$ to $z=0$ were saved, from which bound structures were identified by running the \textsc{RockStar} phase-space halo/subhalo finder~\citep{Behroozi13}.
Then merger trees were constructed using the \textsc{Consistent Trees} code~\citep{Behroozi2013b}. All Uchuu data products are publicly available.

The characteristics of Uchuu make it ideal for the creation of simulated galaxy catalogues. Its high resolution allows to resolve dark matter haloes down to small halo masses on a large volume. This renders the simulation suitable for the application of SHAM, which requires that subhaloes are identified. The very large volume of Uchuu allows for detailed statistics, as well as the study of large-scale clustering features such as the BAO.

% \subsection{Populating dark matter halos/subhalos}
\subsection{Box galaxy catalogues}
\label{subsubsec:luminosity_SHAM}

In order to construct  Uchuu galaxy lightcones, we first start with applying SHAM algorithm to assign luminosities to all halos and subhaloes in simulation boxes.
%to add galaxy luminosities% and colours to the halo catalogues.
Following the basic principle of SHAM, we  assign galaxy luminosities to DM haloes by matching the galaxy luminosity function to the cumulative distribution function of a
 halo property that serves as a proxy of galaxy stellar mass. A possible choice for this halo property is the maximum circular velocity, $V_\mathrm{max}$ \citep[e.g.][]{Conroy06,Trujillo-Gomez11}, defined as the maximum value of the halo circular velocity at the redshift of interest,
\begin{equation}
    V_\mathrm{max}(z)=\max\left(\sqrt{\frac{GM(<r,z)}{r}}\right).
\end{equation}
Using $V_\mathrm{max}$ generally yields better results compared to the halo mass. Since the maximum velocity is generally reached at smaller scales compared to the halo radius, it characterises both the halo concentration \citep{Campbell18} and the depth of the potential at the typical galactic scales \citep{Chaves-Montero16}. It is also less affected by the tidal stripping suffered by subhaloes upon
being accreted by larger haloes \citep{Hayashi03}. 
Another typical choice, which we adopt in this work, is to use the peak circular velocity, $V_\mathrm{peak}$, defined as the peak value of $V_\mathrm{max}$ over the history of the halo.
%{\bf Using Vpeak has no relation to assembly history. It is about tidal stripping that can affect even central regions of subhalos. We do not even need to explain why we use Vpeak} By considering the historical peak of the maximum circular velocity, we incorporate information about the assembly history of the halo. {\bf Several previous works have found => 
This makes estimates that reproduce even more closely the properties of observed  data  \citep[e.g.][]{Reddick13,Chaves-Montero16,Safonova21}. 

The subhaloes in Uchuu are $90\%$ complete down to $V_\mathrm{peak} \gtrsim \SI{70}{\kilo\meter\per\second}$, while distinct haloes are $90\%$ complete down to $V_\mathrm{peak} \gtrsim \SI{50}{\kilo\meter\per\second}$. This allows us to reach low galaxy luminosities in our Uchuu galaxy catalogues.

We match $V_\mathrm{peak}$ with an observationally motivated target luminosity function, $\phi_\mathrm{target}$. We adopt a recipe closely following \citet{Smith17}, which interpolates between the measured luminosity function from SDSS, $\phi_\mathrm{SDSS}$~\citep{Blanton03}, at low redshifts and the luminosity function from the GAMA survey, $\phi_\mathrm{GAMA}$~\citep{Loveday12}, at higher redshifts, as follows
\begin{equation}
\begin{split}
    \phi_\mathrm{target}(\magr, z) &= (1-w(z)) \phi_\mathrm{SDSS}(\magr, z) \\ &+w(z)\phi_\mathrm{GAMA}(\magr, z),
    \label{eq:target_LF}
\end{split}
\end{equation}
where $w(z)$ is a sigmoid function describing this smooth transition at z=0.15, which is given by
\begin{equation}
    w(z) = \left(1+e^{-100(z-0.15)}\right)^{-1}.
\end{equation}
Both luminosity functions are modelled with an evolving Schechter fit
\begin{equation}
    \phi(M) = 0.4 \ln{10} \, \phi^\ast \left(10^{0.4(M^\ast-M)}\right)^{1+\alpha}\exp{\left(-10^{0.4(M^\ast-M)}\right)},
\end{equation}
where the redshift evolution of the Schechter parameters is modelled as
\begin{equation}  
        \begin{array}{l}
        \alpha(z) = \alpha(z_0) \\
        M^\ast(z) = M^\ast(z_0) - Q(z-z_0) \\
        \phi^\ast = \phi^\ast(0)10^{0.4Pz},
        \end{array}
\end{equation}
where $z_0 = 0.1$~\citep{Blanton03,Loveday12}. The corresponding parameters for the SDSS and GAMA function are shown in Table~\ref{tab:Schechter_params}. Note that this is different to \citet{Smith17}, which used the tabulated measurement from SDSS.
\begin{table}
\centering
    \begin{tabular}{c|cc}
        \hline
        Parameter & SDSS & GAMA\\
        \hline
        $\alpha$ & $-1.05$ & -1.23  \\
        $\phi^\ast / h^3\mathrm{Mpc}^{-3}$ & $1.49 \times 10^{-2}$ & $0.94 \times 10^{-2}$ \\
        $M^\ast(z_0)$ & $-20.44$ & $-20.70$ \\
        $P$ & $0.18$ & $1.8$ \\
        $Q$ & $1.62$ & $0.7$ \\
        \hline
    \end{tabular}
\caption{Schechter parameters obtained from a fit to SDSS \citep{Blanton03} and  GAMA \citep{Loveday12} data. These two luminosity functions are interpolated in order to obtain our target luminosity function (eq.~\ref{eq:target_LF}).}
\label{tab:Schechter_params}
\end{table}
In our galaxy assignment algorithm, we use the cumulative LF, which gives the number density of galaxies brighter than magnitude threshold.%The cumulative luminosity function $n_\mathrm{g}(<\magr)$, which will be used directly in the assignment algorithm, can be calculated by integrating $\phi_\mathrm{target}$ on $\magr$ for the simulation volume.

Starting from the halo $V_\mathrm{peak}$ values, we use a SHAM algorithm to assign galaxy luminosities by matching the $V_\mathrm{peak}$ cumulative number density function to our target galaxy luminosity function, with some added intrinsic scatter. We detail below our algorithm, which is based on the method introduced in \citet{McCullagh17,Safonova21},
\begin{enumerate}
    \item Sort the haloes by $V_\mathrm{peak}$ in descending order, to compute the $V_\mathrm{peak}$ cumulative number density function, $n_\mathrm{h}(>V_\mathrm{peak})$%, which describes the number of haloes above a given value of $V_\mathrm{peak}$
    .
    \item Assign an `unscattered' galaxy $r$-band magnitude, $\magr$, to each halo by matching the above $V_\mathrm{peak}$ cumulative halo number density function to the target galaxy luminosity function, preserving the ranking such that large $V_\mathrm{peak}$ haloes host high-luminosity galaxies.
    \begin{equation}
        n_\mathrm{g}(<\magr) = n_\mathrm{h}(>V_\mathrm{peak})
    \end{equation}
    \item Define a new `scattered' value of the magnitude, ${}^{0.1}M_r^{h,\mathrm{scat}}$, 
    \begin{equation}
        {}^{0.1}M_r^{h,\mathrm{scat}} = \mathcal{N}(0, \sigma^2) + \magr,
    \end{equation}
    where $\mathcal{N}(0, \sigma^2)$ is a number drawn from a normal distribution with mean $0$ and variance $\sigma^2$.
    \item Sort the haloes by ${}^{0.1}M_r^{h,\mathrm{scat}}$, and compute the ${}^{0.1}M_r^{h,\mathrm{scat}}$ cumulative distribution, $n_\mathrm{h}(<{}^{0.1}M_r^{h,\mathrm{scat}})$.
    \item Assign the final $r$-band magnitude by matching the cumulative distribution of ${}^{0.1}M_r^{h,\mathrm{scat}}$ to the target cumulative distribution function,
    \begin{equation}
        n_\mathrm{g}(<\magr) = n_\mathrm{h}(<{}^{0.1}M_r^{h,\mathrm{scat}}).
    \end{equation}
    In other words, the brightest final luminosities are assigned to the haloes with brightest ${}^{0.1}M_r^{h,\mathrm{scat}}$. %In other words, the brightest luminosities that were computed in step (ii) are assigned to the galaxies with brightest ${}^{0.1}M_r^{h,\mathrm{scat}}$.

\end{enumerate}

%The modifications with respect to the original method of~\citet{RodriguezTorres16} -- which, instead of a log-normal distribution, used a normal distribution with constant variance -- allow us to use a $V_\mathrm{peak}$-dependent form of $\sigma^2$ while avoiding unphysicalities in the final $(V_\mathrm{peak}, \magr)$ distribution. 

\begin{figure}
    \centering
    \includegraphics[width=\columnwidth]{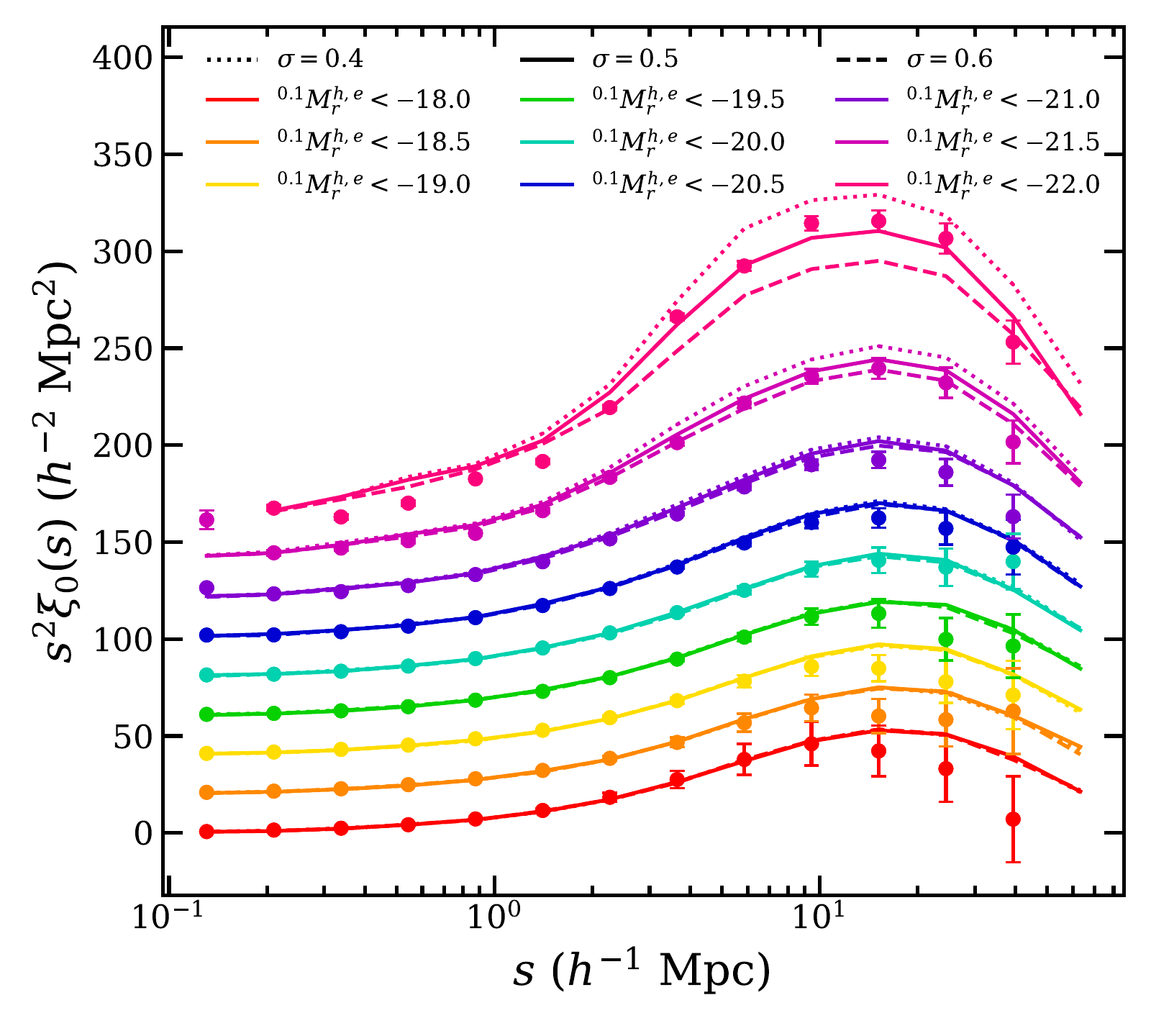}
    \caption{The monopole of the two-point correlation functions from the Uchuu box galaxy catalogue at $z=0.093$ (solid lines) and the SDSS data (filled symbols) for several volume-limited samples corresponding to luminosity cuts as listed in Table~\ref{tab:sdss}. 
    %The solid lines represent the correlation function of the Uchuu catalogue. Data points represent clustering measurements data from our SDSS sample. %For clarity, lines and points have been offset from their original values by successive intervals of $\SI{0.15}{\dex}$ from the sample corresponding to $\magr<-20.5$. 
    For clarity, lines corresponding to different luminosity cuts have been offset by successive intervals $\SI{20}{\per\h\squared\mega\parsec\squared}$, starting from the lowest luminosity sample. The clustering is shown for different values of the scatter parameter $\sigma = \{0.4, 0.5, 0.6\}$, by the dotted, sold and dashed curves, respectively.}
    \label{fig:tpcf_box_SDSS}
\end{figure}

All the degrees of freedom in the procedure above are contained in the choice of the scatter parameter $\sigma$, which we can regulate. In our case, we reduce the number of tunable parameters of our model and neglect any possible redshift or luminosity dependence of $\sigma$ by fixing it to a constant value of $0.5 \, \mathrm{mag}$. This choice, despite its simplicity, is able to reproduce the observed galaxy clustering while avoiding unphysicalities in the $(V_\mathrm{peak},\magr)$ galaxy distribution. This is shown in Fig.~\ref{fig:tpcf_box_SDSS}, where we show the monopole of the two-point correlation function of our Uchuu galaxy catalogue at $z=0.093$, the catalogue closest to the median redshift of SDSS, compared to the results from our SDSS samples. 

Our model is able to recover the SDSS results to good accuracy for a large range of scales and volume-limited samples.
The value of $\sigma$ is calibrated to reproduce the observed SDSS galaxy clustering. For more details about the calibration of $\sigma$ and its effect on galaxy clustering, we direct the reader to Appendix~\ref{App:scatter}. We also recover the target LF by construction.

\begin{figure}
    \centering
    \includegraphics[width=\columnwidth]{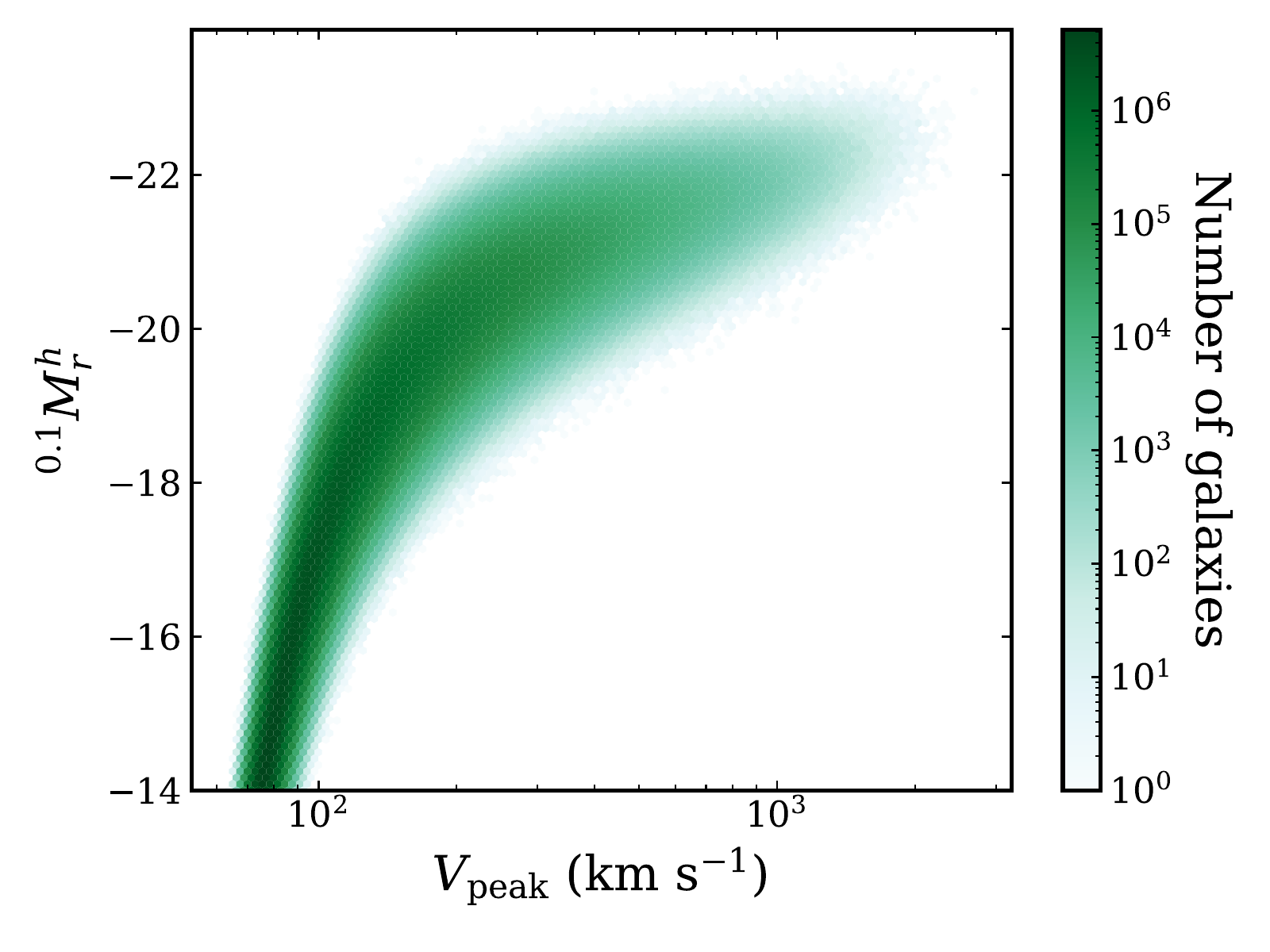}
    \caption{The $(V_\mathrm{peak}, \magr)$ distribution in the Uchuu $z=0.093$ box, assuming a constant scatter parameter $\sigma=0.5$. Colour indicates the number of galaxies in each hexagonal bin, in logarithmic scale.}
    \label{fig:Vpeak_vs_mag}
\end{figure}

The resulting relation between $V_\mathrm{peak}$ and $\magr$  is shown in Fig.~\ref{fig:Vpeak_vs_mag} for galaxies with $^{0.1}M_r <-14.0$ in the $z=0.093$ box. The large volume of Uchuu allows to reach a huge dynamical range in galaxy luminosity  $-23 < \magr < -14$.

\subsection{Uchuu-SDSS galaxy lightcones}
\label{subsec:light_mocks}
 We use a total of 6 snapshots between $z=0$ and $z=0.5$, which are 
separated in redshift by approximately 0.1 ($z_\mathrm{snap}=0, 0.093, 0.19, 0.3, 0.43, 0.49$).
Lightcones are created from the snapshots by joining them together 
in spherical shells. 
% An observer is first placed in the box, and the Cartesian galaxy
% coordinates in each snapshot are converted to equatorial coordinates, and the redshift
% is calculated, taking into account the line-of-sight velocity. To extend
% the redshift range of the mock beyond $z=0.36$, periodic replications of the cubic
% box must be applied \Julia{I dont get this. You say you need to do replications to reach z=0.36, but you are already using a 0.43 and 0.49 box...}\Alex{Galaxies in the lightcone at z=0.43 would come from the z=0.43 snapshot, but the distance to the observer is > 1000 Mpc/h, so the cubic box needs to be replicated}. In each simulation snapshot, galaxies in the redshift range
% $(z_\mathrm{snap-1} + z_\mathrm{snap})/2 < z < (z_\mathrm{snap} + z_\mathrm{snap+1})/2$ 
% are selected, which are then joined together to build the lightcone.
% \Julia{Maybe something like this?: 
% An observer is first placed in the box, and the Cartesian galaxy
% coordinates in each snapshot are converted to equatorial coordinates, 
% and the redshift is calculated, taking into account the line-of-sight velocity. 
% In each simulation snapshot, galaxies in the redshift range $(z_\mathrm{snap-1} + z_\mathrm{snap})/2 < z < (z_\mathrm{snap} + z_\mathrm{snap+1})/2$ are selected, which are then joined together to build the lightcone. \textbf{Periodic replications must be applied in those cubic boxes not big enough to reach the required redshift range.}}
An observer is first placed in the box, and the Cartesian galaxy
coordinates in each snapshot are converted to equatorial coordinates, 
and the redshift is calculated, taking into account the line-of-sight velocity. 
In each simulation snapshot, galaxies in the redshift range $(z_\mathrm{snap-1} + z_\mathrm{snap})/2 < z < (z_\mathrm{snap} + z_\mathrm{snap+1})/2$ are selected, which are then joined together to build the lightcone. If the redshift shell is too big to fit inside a single cubic box, periodic replications are applied. In the final lightcone there are no periodic replications below $z=0.36$.

The full-sky lightcone is then cut to the northern contiguous region of the 
SDSS survey footprint, using a healpix map \citep{Gorski1999, Blanton2005, Swanson2008}.
The original healpix map has $N_\mathrm{side}=512$, but we increase the size of the pixels, using $N_\mathrm{side}=128$, and keeping pixels where the completeness in the data is greater than 0.9. This results in a footprint with area $7261~\sqdeg$. 
The SDSS footprint can be replicated across the full sky to
create 4 independent SDSS simulated catalogues. By generating lightcones from
eight observer positions (with coordinates at either 0 or $1~\hGpc$, along each of the three dimensions), we are therefore
able to create a total of 32 Uchu-SDSS lightcones.
The first 8 mocks (constructed with observer at the origin, and at the centre of the box) are independent below $z=0.3$. There is some overlap in the volume between the lightcones at higher redshifts than this, but the fraction of galaxies with $z>0.3$ is very small ($\sim 0.1\%$). We refer to these lightcones as the 8 independent lightcones. The full set of 32 lightcones are independent below $z=0.175$, and we use all 32 to improve the statistics of our RSD and BAO measurements (see Section~\ref{sec:RSD_BAO}). The BAO galaxy sample (eq.~\ref{eq:hod_cuts}) has a maximum redshift of $z=0.2$, so there is some overlap between mocks, and $\sim 25\%$ of galaxies in the sample have $z>0.175$.

Combining multiple snapshots in this way to construct a lightcone %is not perfect, and 
has the issue that there are discontinuities at the boundaries between snapshots.
It is possible for the same halo to appear twice at either side of the boundary,
or to not appear at all, and the duplicated haloes artificially boost the pair counts on very small scales.
% Increasing the total number of snapshots used will reduce the probability of this
% happening for an individual halo (and reducing the separation of the duplicated haloes), 
% but this comes at the expense of increasing the total number of discontinuities in the lightcone, and potentially
% increasing the total number of duplicated haloes.
We investigate this in \citet{Smith2022a}, and find that there is a boost in the 
real-space clustering on small scales due to this effect. 
In redshift space, when velocities are included, the effect is greatly reduced. 
Using snapshots separated by $\sim 0.1$ in redshift is a good compromise which adds evolution
to the lightcone, without excessively boosting the clustering below $1~\hMpc$.
The clustering measurements on the scales used in a typical RSD analysis ($\gtrsim 20~\hMpc$) are insensitive to the number of snapshots used.

\subsubsection{Magnitude evolution}
\label{subsubsec:mag_colour_evolution}

Each simulation snapshot was constructed to reproduce the target luminosity function
%and ${}^{0.1}(g-r)$ colour distributions 
at the redshift of the snapshot, $z_\mathrm{snap}$. In the lightcone, 
this leads to discontinuities in these properties at the boundaries between snapshots.

In order to create a simulated lightcone with a smoothly evolving luminosity function, 
and smooth $n(z)$, we rescale the galaxy absolute magnitudes as a function of redshift. 
The original magnitude assigned to each galaxy is first mapped to the corresponding number density, 
using the luminosity function at $z_\mathrm{snap}$. This number density can then be mapped back to 
an absolute magnitude at the redshift of the galaxy in the lightcone, $z$, using the target luminosity 
function at the same redshift, $z$. By construction, this reproduces the smooth evolution of the target luminosity 
function. 

% To achieve smoothly evolving ${}^{0.1}(g-r)$ colour distributions, we re-assign a colour
% to each galaxy independently of its original colour in the box, using the redshift $z$ of the galaxy 
% and the re-scaled value of absolute magnitude. This is done using the colour assignment algorithm of \citet{Smith17}, which randomly decides whether a galaxy is red or blue based on its magnitude, redshift and whether it is a central or satellite. A random colour is then drawn from the red or blue Gaussian. This algorithm is different to the cubic boxes (Section~\ref{subsubsec:colour_assignment}), and was used since it is straightforward to apply to an evolving lightcone.

\subsubsection{Colour assignment}
\label{subsubsec:colour_assignment}

After galaxy luminosities have been assigned, we add ${}^{0.1}(g-r)$ colours to our simulated galaxy sample.
We use an evolving empirical model for the ${}^{0.1}(g-r)$ colours, based on \citep{Skibba2009, Smith17}, but improved to better
reproduce the colour-magnitude diagram from the GAMA survey, at a range of redshifts \citep{Smith2022b}.
The bi-modality of the colour probability distribution functions is modelled as a sum of two Gaussian distributions. Defining $p({}^{0.1}(g-r); \magr, z) \equiv dN/N/d{}^{0.1}(g-r)$, where $N$ is the number of galaxies,
\begin{equation}
\begin{split}
    p({}^{0.1}(g-r); \magr, z) &= f_\mathrm{blue} \, \mathcal{N}_\mathrm{blue}(\mu_\mathrm{blue},\sigma_\mathrm{blue}) \\ 
    &+ (1-f_\mathrm{blue}) \, \mathcal{N}_\mathrm{red}(\mu_\mathrm{red},\sigma_\mathrm{red}),
\end{split}
\end{equation}
where $\mathcal{N}_\mathrm{blue}$ and $\mathcal{N}_\mathrm{red}$ are Gaussian probability distribution functions corresponding to blue and red galaxies, respectively, and $f_\mathrm{blue}$ is the fraction of blue galaxies. The parameters describing this double-Gaussian distribution all depend on $\magr$ and $z$.

At a fixed redshift, the mean and rms of each Gaussian, $\mu_\mathrm{red}(\magr)$, $\sigma_\mathrm{red}(\magr)$,
$\mu_\mathrm{blue}(\magr)$ and $\sigma_\mathrm{blue}(\magr)$, are modelled as broken linear functions. These
functions were fit to the colour-magnitude diagram measured in GAMA, in several redshift bins, which we then interpolate
between. Basing the colour distributions on the data from GAMA allows us to create simulated galaxy catalogues to very faint magnitudes,
which will be useful e.g. for the DESI Bright Galaxy Survey. 

However, we find good agreement between these colours and the measurements from
the SDSS data. The fraction of blue galaxies at given luminosity, $f_\mathrm{blue}(\magr)$, is modelled differently for central and satellite galaxies. Colours are then drawn randomly from the ${}^{0.1}(g-r)$ colour distributions described above. This method is able to reproduce by construction our target colour distributions.

\subsubsection{Apparent magnitudes}

The apparent $r$-band magnitude is computed from the absolute magnitude using eq.~\ref{eq:appmag}. 
In the $r$-band, we use a set of
colour-dependent $k$-corrections from the GAMA data \citep[see figure 13 of][]{Smith17}. 
These are a set of polynomial $k$-corrections, measured in several bins of ${}^{0.1}(g-r)$ colour.
These $k$-corrections allow us to calculate apparent magnitudes using the information that was
originally available in the simulated catalogue ($r$-band magnitude and $g-r$ colour).
In the other bands, we use the $k$-corrections of \citet{Blanton2003b}. 

We have compared the GAMA and SDSS $r$-band $k$-corrections, and find that the median $k$-corrections at each redshift are in good agreement, differing by no more than $\sim 0.01$.
The $1\sigma$ scatter is at a level $< 0.04$.
At $z=0.1$, both $k$-corrections agree exactly, since
${}^{0.1}k(z=0.1)=-2.5 \mathrm{log}_{10}(1.1) \approx -0.103$.
% The GAMA and SDSS $k$-corrections in the $r$-band are compared in Fig.~\ref{fig:k_correction_comparison},
% as a function of redshift. While there is some scatter, the median $k$-corrections are in good agreement,
% differing by no more than $\sim 0.01$. At $z=0.1$, both $k$-corrections agree, since
% ${}^{0.1}k(z=0.1)=-2.5 \mathrm{log}_{10}(1.1) \approx -0.103$.

% \begin{figure}
%     \centering
%     \includegraphics[width=\columnwidth]{figures/k_corrections.pdf}
%     \caption{The difference between the SDSS and GAMA $r$-band $k$-corrections for each SDSS galaxy,
%     $\Delta \kcorrr(z) = \kcorrr^\mathrm{SDSS}(z) - \kcorrr^\mathrm{GAMA}(z)$. The red solid line
%     shows the median, while the dotted red lines are the 16th and 84th percentiles.{\bf AK: what are the conclusions? Add from the text something like :the median $k$-corrections are in good agreement,
% differing by no more than $\sim 0.01$. At $z=0.1$, both $k$-corrections agree. }  \ca{this figure is very nice, although in order to save space it might be good to remove it and summarise its conclusions in the text}}
%     \label{fig:k_correction_comparison}
% \end{figure}

\subsubsection{Assigning galaxy properties using SDSS data}
\label{sec:assign_gal_properties_to_mock}

The method we have used to create the lightcone assigns each galaxy a rest-frame $r$-band absolute magnitude,
$\magr$, and ${}^{0.1}(g-r)$ colour. In order to assign more properties to the mock, we match galaxies
to the SDSS data. We use a k-d tree to find the closest SDSS galaxy, based on $z$, $\magr$ and ${}^{0.1}(g-r)$.
Each mock galaxy is then assigned the absolute magnitude in the $u$, $g$, $i$ and $z$-band of the 
closest-matching galaxy, in addition to its stellar mass and specific star formation rate.

% There is freedom in how the redshifts, magnitudes and colours are weighted when finding the 
% distances in this 3-dimensional space. We multiply both the redshifts and colours by a factor of
% 5 to increase their dynamic range, but we find that the exact choice of weighting does not have
% a strong effect on how well the mock and SDSS galaxies match. \ca{i removed this in order to save space, feel free to add it back}

The $u$, $g$, $i$ and $z$-band apparent magnitudes are calculated at the redshift
of the mock galaxy from the absolute magnitudes, using eq.~\ref{eq:appmag} and the SDSS $k$-corrections. %\ca{rewrote a bit more concisely}
% We do not assign the mock galaxies the same apparent magnitudes as the SDSS galaxies, since there
% is some scatter in the redshift. We instead recalculate the apparent magnitudes from the absolute magnitudes, at the redshift
% of the mock galaxy.

\subsubsection{Modeling fibre collisions}
\label{subsec:fibre}

\begin{figure}
    \centering
    \includegraphics[width=\columnwidth]{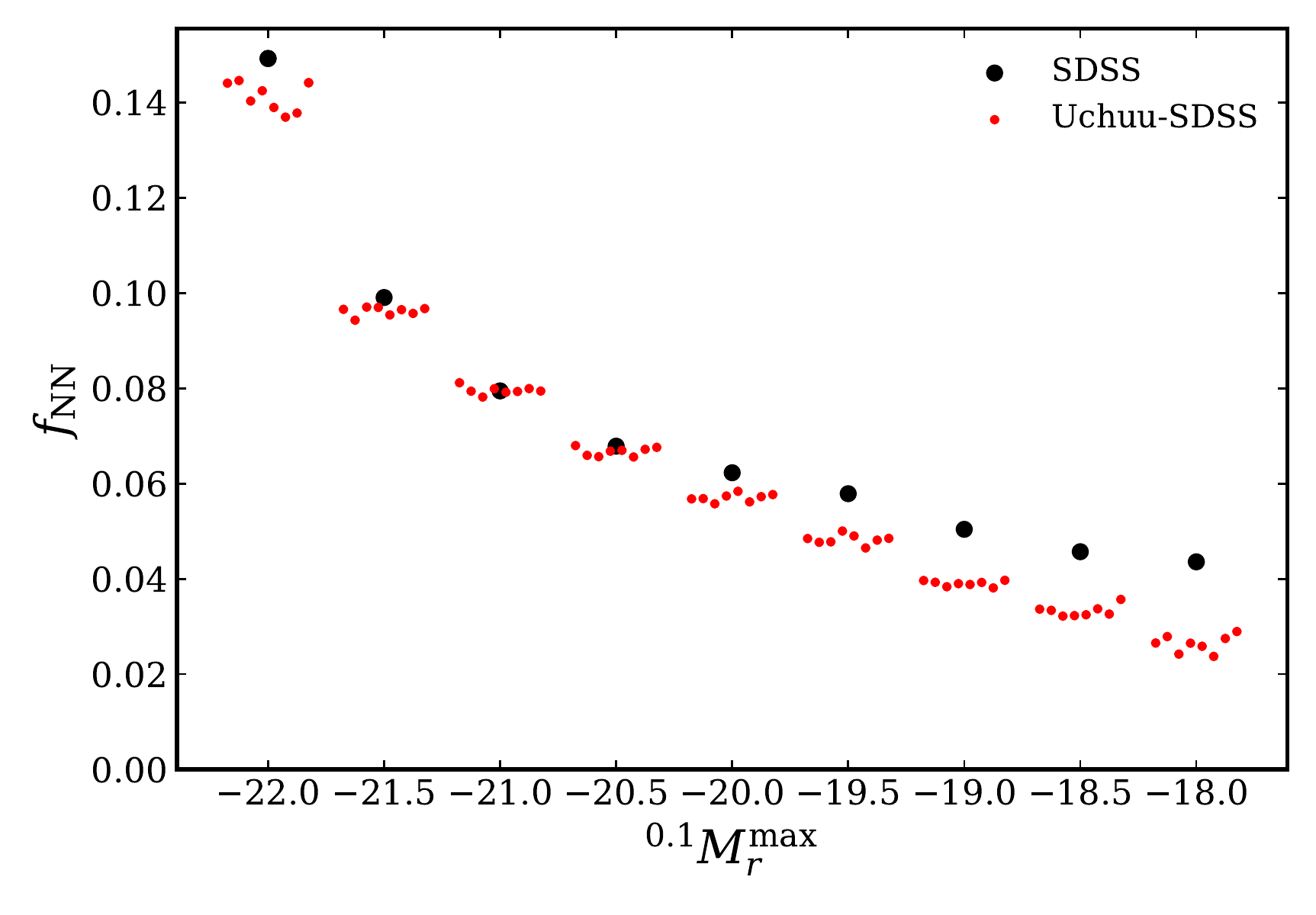}
    \caption{The fraction of galaxies affected by fibre collisions in Uchuu-SDSS and SDSS lightcones. For each of our SDSS volume-limited sample, the black points show the fraction of galaxies receiving the nearest neighbour correction, $f_\mathrm{NN}$. The red points show $f_\mathrm{NN}$ when we construct volume-limited samples from each of our eight parent Uchuu-SDSS catalogues.%{\bf AK: mention and comment on apparent disagreement between SDSS and Uchuu. Are they important? You cannot just leave it as is.}
    }
    \label{fig:fnn}
\end{figure}

In SDSS, spectroscopic fibres on a single plate cannot be placed closer to each other than the diameter of the fibre plugs.
As a result, if two galaxies are in close proximity ($<55''$), a spectrum can only be obtained for one of them.
The tiling of SDSS plates slightly alleviates this problem, with $\sim30\%$ of the SDSS footprint covered by multiple plates.
These plate `overlap regions' allow for spectroscopic redshifts to be obtained for several galaxies that are within $55''$ of one another.
Still, $\sim6\%$ of targeted galaxies in our SDSS sample lack a spectroscopically measured redshift due to their proximity to a neighbouring galaxy.

To mimic the fibre collision effect of SDSS, when constructing our mock galaxy catalogues, we employ a procedure adapted from \citet{Szewciw2022}.
First, we link together galaxies into friends-of-friends `collision groups'.
A galaxy is part of a collision group if its angular distance to any galaxy in that group is $<55''$.
Next, we decide whether each galaxy in a group will be `observed' (i.e. receive its spectroscopic redshift) or `unobserved'.
When making this choice, we first attempt to maximize the number of galaxies in a collision group that could receive spectroscopic fibres from a single SDSS plate.
% In many cases, there will be more than one set of galaxies that maximizes the number of observed galaxies.
% For example, if the collision group only consists of two galaxies, then we can observe either one.
If there does exist more than one set that maximizes the number of observed galaxies, then we randomly choose one of these sets to be observed.
Next, to simulate the effect of SDSS plate overlap regions, we randomly select $\sim30\%$ of the unobserved galaxies to receive their original observed spectroscopic redshifts.
% These galaxies are selected randomly from among the unobserved galaxies.
The remaining unobserved galaxies are then assigned the redshift of their angular nearest neighbour.
Finally, given the new redshifts of these galaxies, we recompute their $k$-corrections \citep[using the colour-dependent $k$-corrections described in][]{Smith17} and absolute magnitudes.
% Thus, the galaxies are assigned the absolute magnitudes we would calculate if they had been spectroscopically unobserved and assigned the redshift of their nearest neighbour.

It is important to note that this procedure does not fully mimic the role that plate overlap regions play in recovering the redshifts of collided galaxies.
In our procedure the collided galaxies whose redshifts are recovered are chosen at random from the full sky.
In SDSS, by contrast, recovered galaxies lie in plate overlap regions and thus are spatially correlated. %, as determined by the SDSS tiling algorithm.
Furthermore, in SDSS, each overlap region of the sky is covered by a different number of intersecting plates.
The number of overlapping plates dictates the number of galaxies whose redshifts can be spectroscopically measured in a given collision group.
Our procedure, however, is agnostic with respect to the number of plates required to recover redshifts for any randomly chosen set of collided galaxies.

Despite these differences, we apply the procedure described above to our simulated galaxy catalogues.
% From these parent mocks, we construct volume-limited samples using the same magnitude thresholds applied to SDSS (Table~\ref{tab:sdss}).
% These mocks can be directly compared to the SDSS samples.
With this relatively simple procedure described above, the global fraction of galaxies affected by fibre collisions in the simulated lightcone ($f_\mathrm{NN} \sim 5.2 \%$) is quite similar to that of SDSS ($f_\mathrm{NN} \sim 5.9 \%$).
% We also examine this fraction for each of our volume-limited samples.
In Fig.~\ref{fig:fnn}, we show a comparison of $f_\mathrm{NN}$ between SDSS (black points) and each of our eight independent Uchuu-SDSS lightcones (red points) for the different volume-limited samples defined in Table~\ref{tab:sdss}.
In both Uchuu and SDSS, we see the same qualitative trend -- $f_\mathrm{NN}$ increases as we move to more luminous samples. This is as expected, since more luminous galaxies tend to be more strongly clustered and thus are more likely to be affected by fibre collisions.
There is good agreement for bright volume-limited samples, although for  $^{0.1}M_r^\mathrm{max} \leq -19.5$ Uchuu underestimates $f_\mathrm{NN}$ by up to a $2\%$ offset. %However, the effect of fibre collisions in the galaxy clustering is smaller for these dimmer samples.
%The values of $f_\mathrm{NN}$ are in reasonable agreement for most of the samples, although Uchuu tends to underpredict $f_\mathrm{NN}$ systematically by $\sim 2\%$.\Alex{[comment] in the plot, the fNN values are systematically smaller for Uchuu compared to SDSS. Is there a reason for this?}
%\Adam{The plot I had uploaded before showed a different trend with luminosity (between Uchuu and SDSS) but showed little to no systematic offset. The current plot was not made by me.}
% Even so, these small discrepancies are important to consider when comparing clustering statistics between Uchuu and SDSS.

After running the fibre assignment algorithm on the 32 Uchuu-SDSS lightcones, we evaluate the fibre assignment completeness in healpix pixels with $N_\mathrm{side}=512$. The lightcones are then cut to pixels where the average completeness $> 0.9$ (averaged over the 32 lightcones). This results in a final area of $6642~\sqdeg$, which is comparable to the footprint of the SDSS data.

\vspace{1em}

The 32 Uchuu-SDSS lightcone catalogues described in this section, containing \textit{ugriz} magnitudes, stellar masses, star formation rates and fibre assignment information, are made publicly available at Skies \& Universes. A subset of the cubic box catalogues used in the construction of the lightcones, and the companion SDSS data set are also made available. Note that our released box catalogues also include ${}^{0.1}(g-r)$ computed using a similar method as described in Section~\ref{subsubsec:colour_assignment}. % A list and brief description of the columns included in the catalogues can be found in Appendix~\ref{App:mock_columns}.
%The 32 Uchuu-SDSS lightcone catalogues described in this section, along with the cubic box catalogues used for its construction, and the companion SDSS data set, are made publicly available at \url{http://www.skiesanduniverses.org}. A list and brief description of the columns included in the catalogues can be found in Appendix~\ref{App:mock_columns}.

\section{Properties of the Uchuu-SDSS galaxies}
\label{sec:properties}

\subsection{Galaxy properties}
\label{sec:gal_properties}

\begin{figure*}
    \centering
    \includegraphics[width=0.45\linewidth]{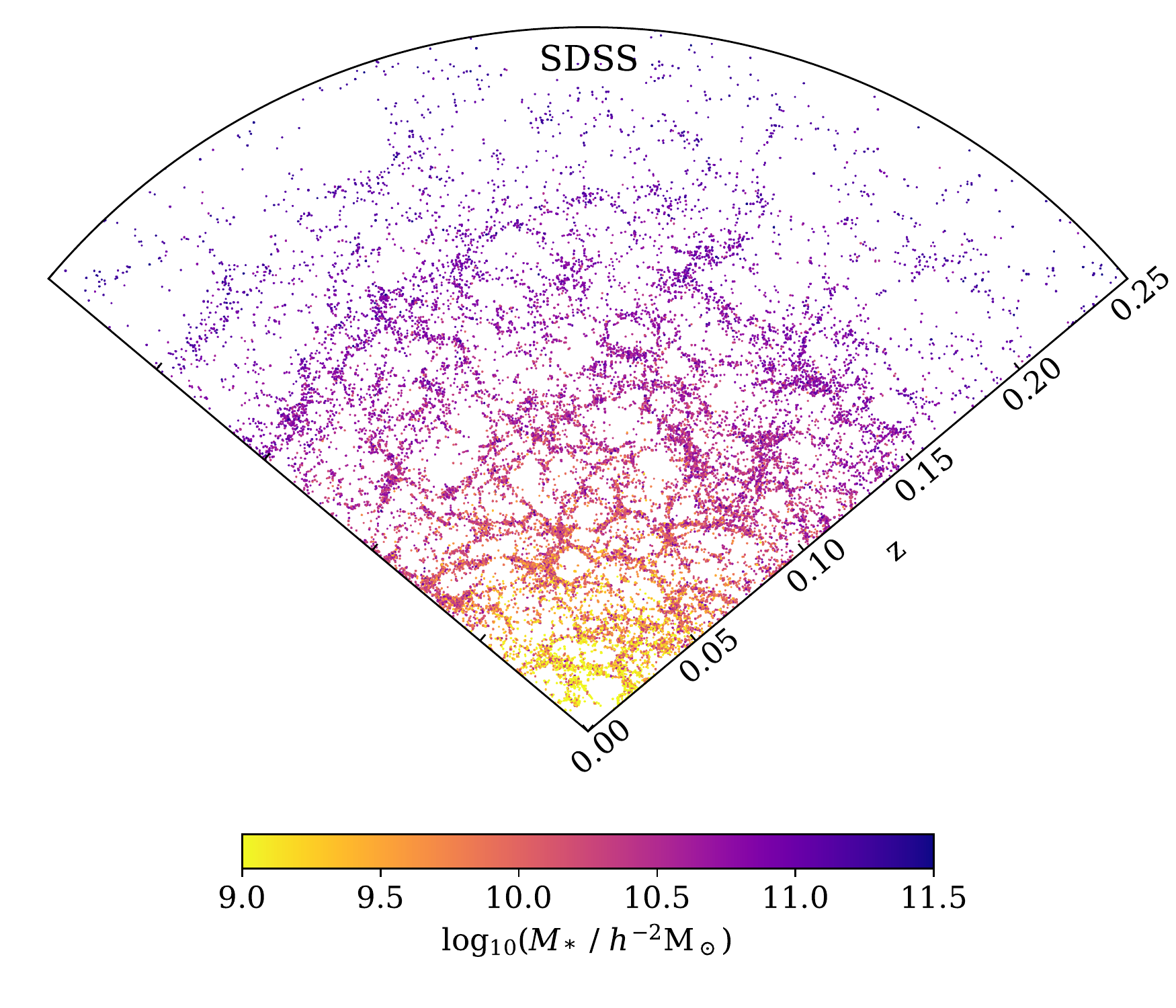}
    \includegraphics[width=0.45\linewidth]{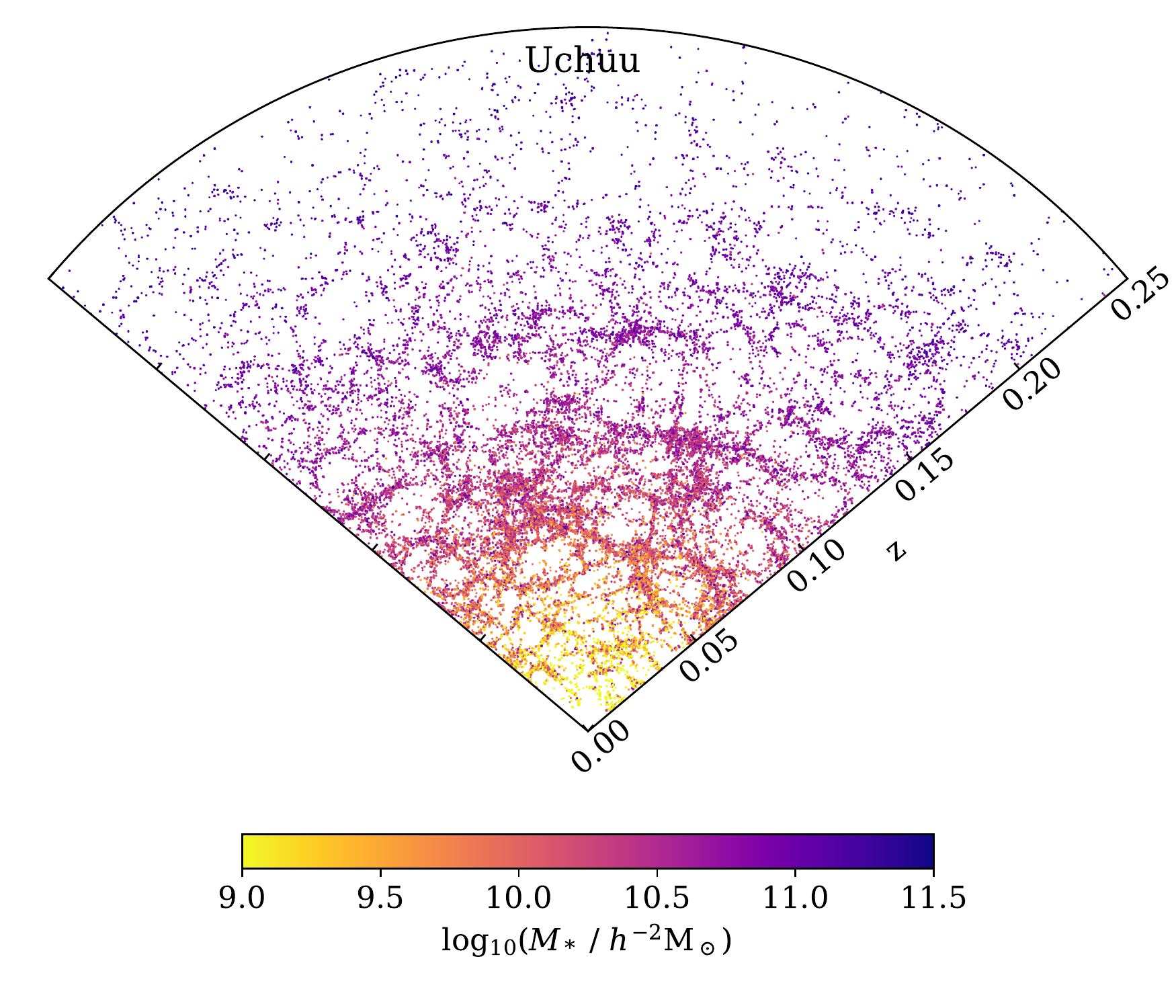}
    \caption{Slice through the SDSS galaxy catalogue (left), and Uchuu-SDSS lightcone (right), where galaxies are coloured 
    by their stellar mass. Galaxies are shown between $35 < \mathrm{dec} < 40~\deg$, and 
    $140 < \mathrm{RA} < 240~\deg$.
    }
    \label{fig:pie_plot}
\end{figure*}

\begin{figure}
    \centering
    \includegraphics[width=\columnwidth]{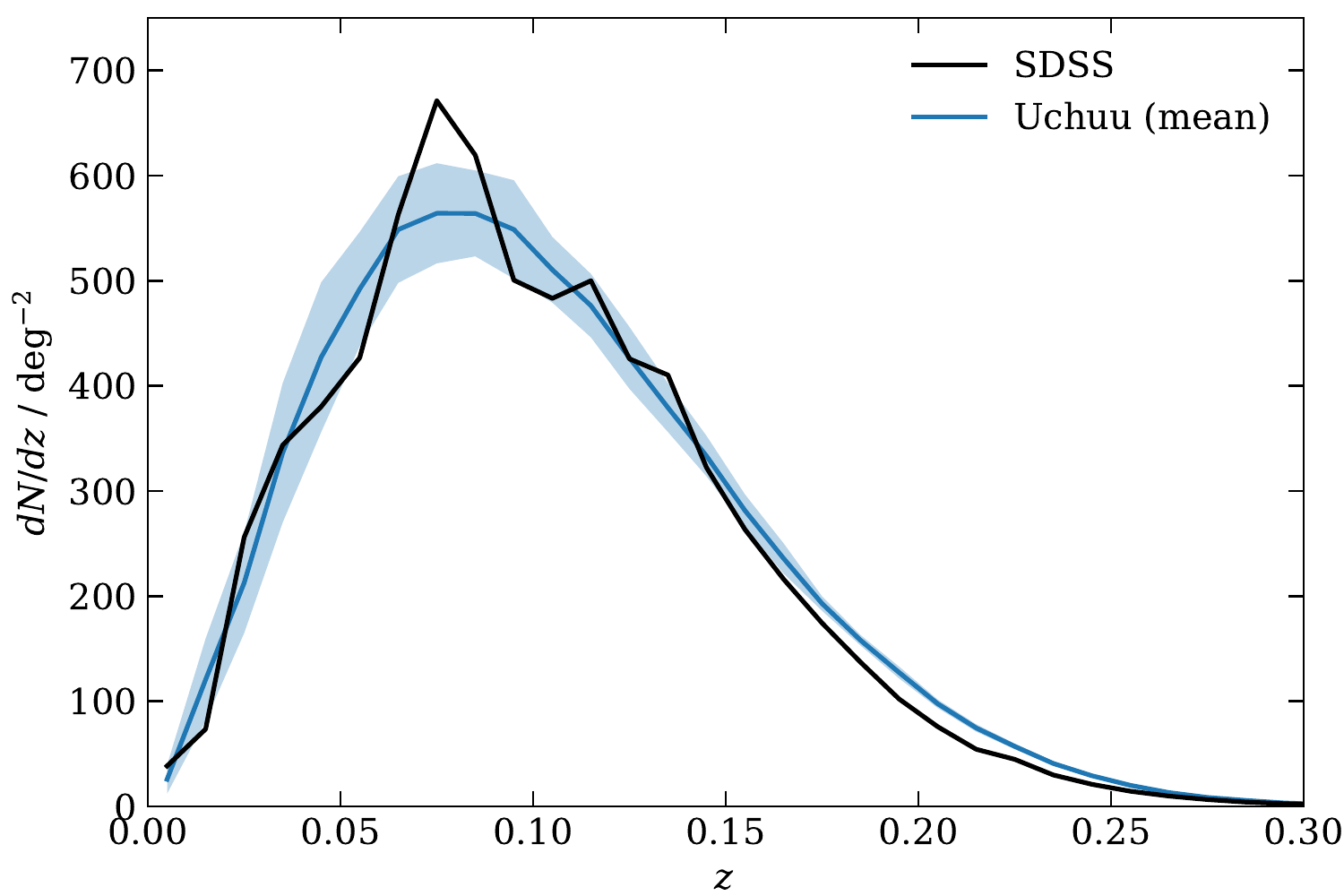}
    \caption{Redshift distribution of galaxies in the the Uchuu-SDSS lightcones, compared to the SDSS data, for galaxies with $r<17.6$. The blue curve is the mean of the 32 simulated lightcones, with the shaded region indicating the $1\sigma$ scatter. The black curve shows the redshift distribution of SDSS. Overall, we find good agreement between Uchuu and SDSS, with a small over-prediction of the number of galaxies in Uchuu at $z\sim 0.2$ due to differences in the luminosity functions.
    }
    \label{fig:dNdz}
\end{figure}

\begin{figure*}
    \centering
    \includegraphics[width=0.9\linewidth]{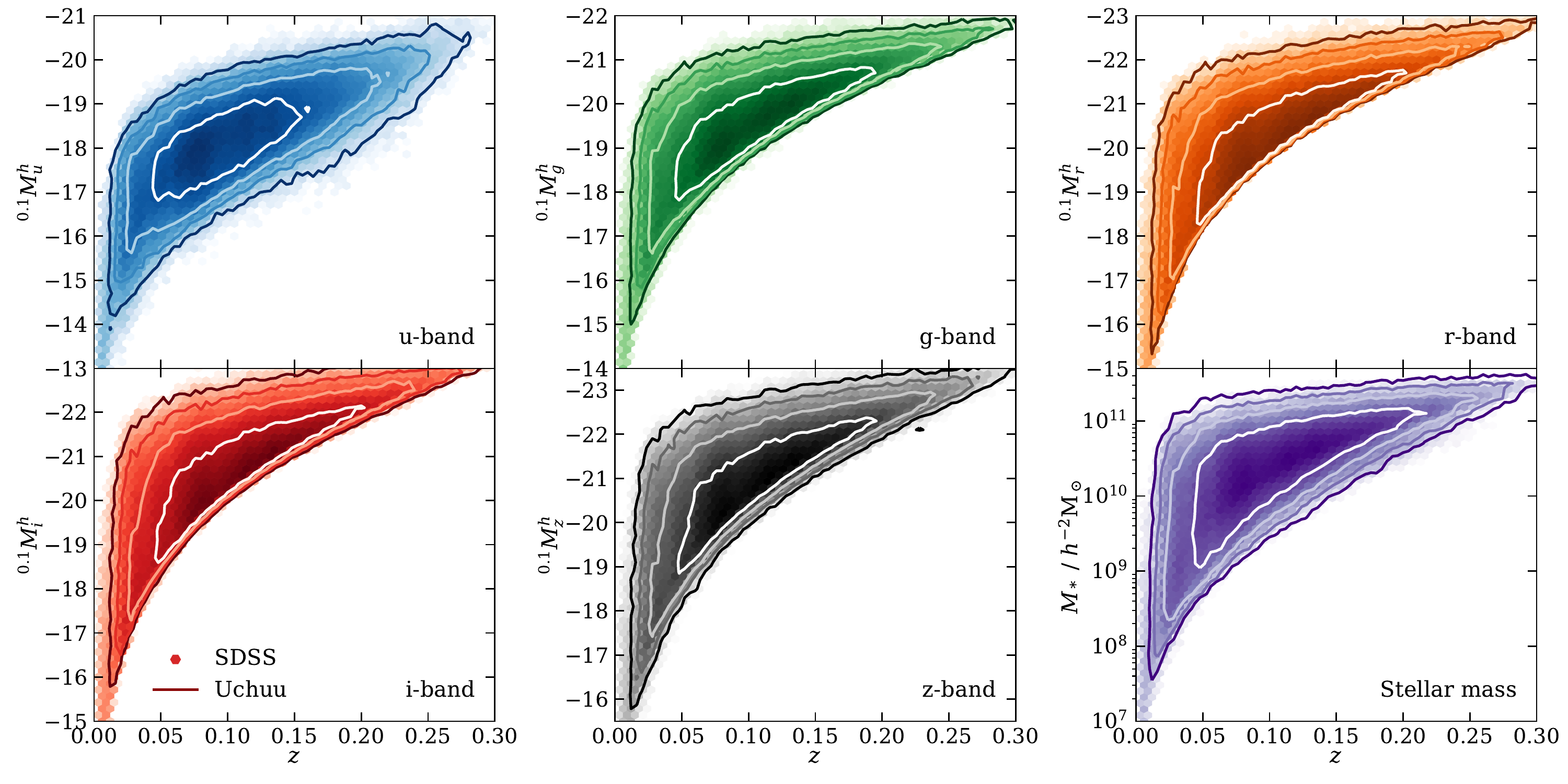}
    \caption{Absolute magnitudes and stellar masses of galaxies in one of the Uchuu-SDSS lightcones (contours), 
    as a function of redshift, compared to the data from SDSS (hexagons), for galaxies with 
    apparent magnitude $r<17.6$.
    The different panels show the magnitudes in the $ugriz$ bands, as indicated in the legend. 
    The lower right panel shows the MPA stellar masses.}
    \label{fig:lightcone_absolute_magnitude}
\end{figure*}

\begin{figure}
    \centering
    \includegraphics[width=0.95\linewidth]{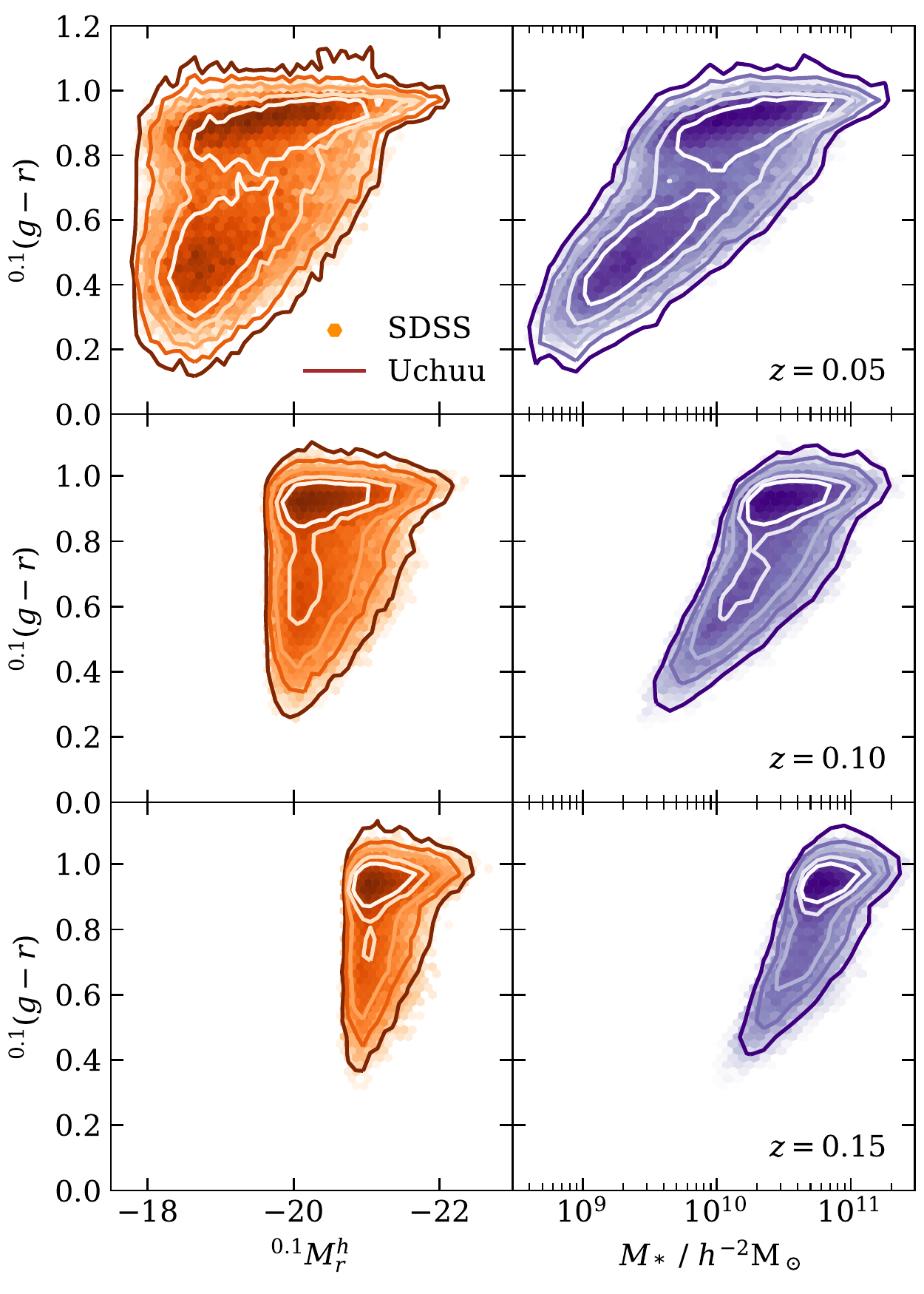}
    \caption{Distribution of ${}^{0.1}(g-r)$ colours in the Uchuu-SDSS lightcone (contours), compared with
    the colour distributions of SDSS (hexagons). The left-hand column shows the colour distribution
    with respect to the $r$-band absolute magnitude, at three different redshifts (from top to bottom, 
    $z=0.05$, $z=0.1$ and $z=0.15$). The right-hand column is the same, but with respect to the MPA
    stellar mass.}
    \label{fig:lightcone_g_r_colour}
\end{figure}

\begin{figure}
    \centering
    \includegraphics[width=0.95\linewidth]{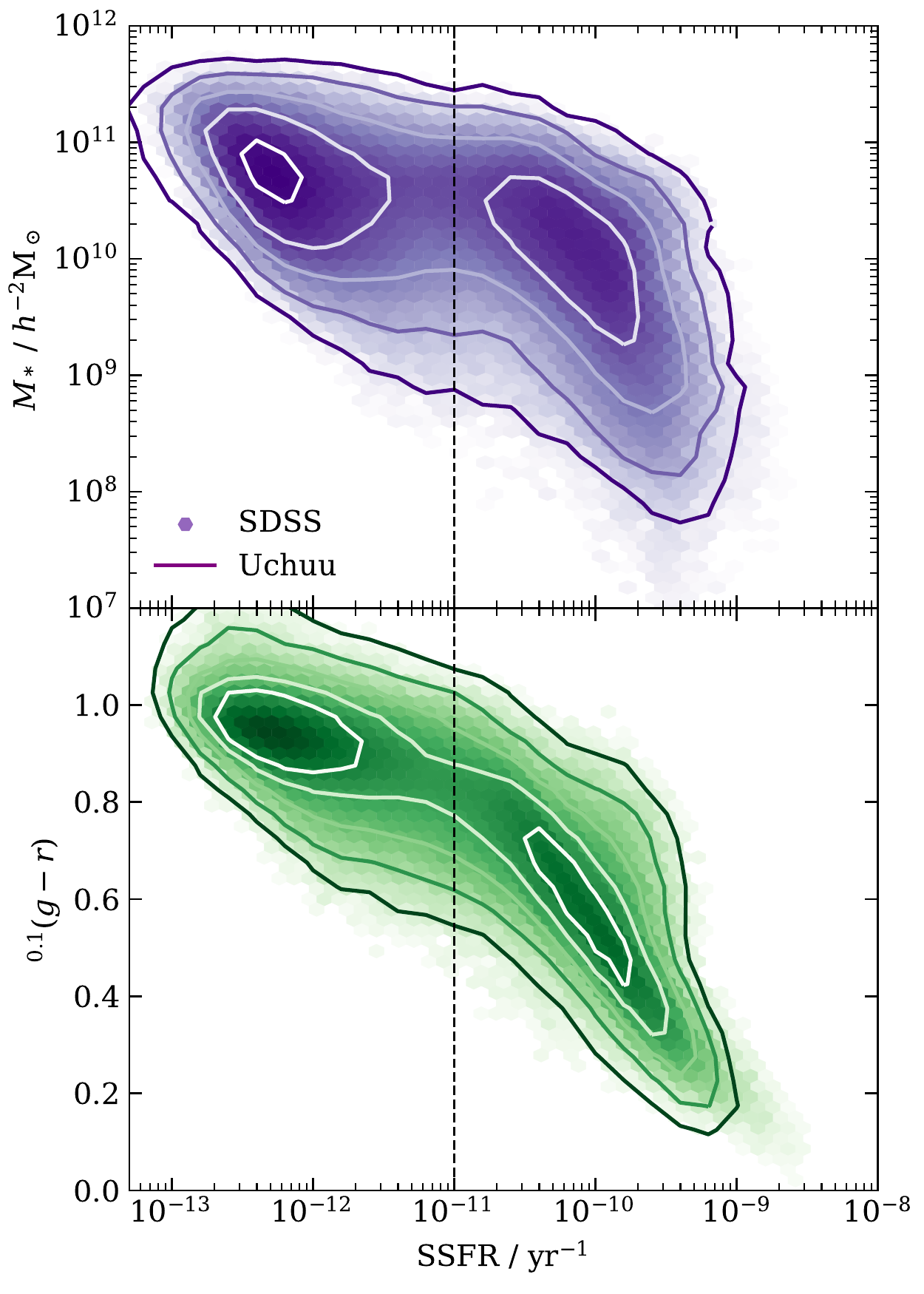}
    \caption{\textit{Top panel}: stellar mass vs sSFR for galaxies in SDSS (hexagons) and
    in the Uchuu-SDSS lightcone (contours). \textit{Bottom panel}: the ${}^{0.1}(g-r)$ colour vs sSFR. The vertical dashed line indicates the sSFR cut of $10^{-11}~\mathrm{yr}^{-1}$, which we use to split the bimodal distributions into samples of quiescent and star-forming galaxies.}
    \label{fig:lightcone_ssfr}
\end{figure}

In this section, we illustrate the various galaxy properties stored in the Uchuu-SDSS lightcones, and compare with the galaxy properties in the SDSS data.

Fig.~\ref{fig:pie_plot} shows galaxies in a thin slice of the SDSS catalogue and one of the
Uchuu-SDSS lightcones, where each galaxy has been coloured based on its stellar mass,
illustrating the similarities between the mock and data.
The density of galaxies falls with increasing redshift, since faint galaxies with low stellar masses
can only be observed close to $z=0$, while the brightest galaxies can be observed over
the full redshift range. The density of galaxies appears to be in good agreement between
the Uchuu-SDSS and SDSS data.

This can been seen quantitatively in Fig.~\ref{fig:dNdz}, which compares the redshift distribution 
of the 32 Uchuu-SDSS lightcones with SDSS. The Uchuu-SDSS lightcones are in good agreement with SDSS, peaking at $z \sim 0.08$. 
There is scatter between the 32 lighcones, due to cosmic variance, but the measurement
from SDSS is consistent with this scatter. 
%\Julia{From here...}
At higher redshifts ($z \sim 0.2$), there is a slight excess of galaxies in the Uchuu-SDSS lightcones compared to the data, due to differences in the luminosity function. While the target luminosity function used to construct the simulated lightcones is in good agreement with the SDSS data at low redshifts, we transition to the luminosity function measured in GAMA at higher redshifts. Using the GAMA luminosity function results in a higher number density of galaxies compared to the SDSS measurements, but the SDSS luminosity function is poorly constrained at this redshift.
%\Julia{... to here, I guess the intention is to explain the origin of the excess of galaxies in the n(z). The reason of the excess in the luminosity function is explained, but as it is written, it does not quite connect with the n(z)... Maybe rewrite this part?} \Alex{is this better?} \Julia{Yes! :)}

The absolute magnitudes of galaxies from one of the Uchuu-SDSS lightcones is shown in 
Fig.~\ref{fig:lightcone_absolute_magnitude}, compared to SDSS. In the $r$-band, magnitudes were assigned to
each galaxy to match an evolving target luminosity function from SDSS and GAMA measurements, and we find that this is able to reproduce well the distribution of absolute $r$-band magnitudes in the 
SDSS data. The magnitudes in the other bands were assigned by matching each simulated galaxy to a 
galaxy in the data, based on $r$-band magnitude, $g-r$ colour and redshift. By construction, the 
distribution of these magnitudes is also in good agreement with the SDSS data. 

The ${}^{0.1}(g-r)$ colour distributions in the Uchuu-SDSS lightcones are shown in the left-hand column of
Fig.~\ref{fig:lightcone_g_r_colour}, as a function of $\magr$,
in three narrow redshift bins at $z=0.05$, $z=0.1$ and $z=0.15$. 
In each redshift bin, the colour distribution is bimodal, with a red sequence of galaxies
and cloud of blue galaxies. The brightest galaxies are red, while fainter galaxies have a higher blue fraction.
The colour distributions show good agreement with the SDSS data, reproducing the same colour
evolution with redshift. There is a small discrepancy at low redshifts, where
the red sequence is more sloped in Uchuu-SDSS compared to in SDSS, since the colours in the Uchuu-SDSS lightcone were tuned to GAMA measurements. The right-hand
column shows the same colour distributions but as a function of the stellar mass.
Here we see that the blue galaxies tend to have lower stellar masses than the red galaxies.
Again, there is good agreement between the simulated and real data, with a slight discrepancy
in the red sequence at low redshifts. 

The distribution of star formation rates of galaxies in the lightcone is shown in Fig.~\ref{fig:lightcone_ssfr}, in comparison to the SDSS measurements.
The upper panel shows stellar mass against sSFR, while ${}^{0.1}(g-r)$ is plotted against 
sSFR in the lower panel. There is a clear bimodal distribution of quiescent galaxies with low star formation rates on the left, and star-forming galaxies on the right. The 
quiescent galaxies tend to have higher stellar masses than the star-forming galaxies. 
Applying a cut of $10^{11}~\mathrm{yr}^{-1}$ in sSFR is able to cleanly cut the galaxy
catalogue into these two samples, while we can see in the lower panel that a cut 
in ${}^{0.1}(g-r)$ would not work as well.

\subsection{Luminosity, mass and colour dependence of clustering}
In this section, we compare the clustering in our Uchuu-SDSS lightcones with the observational results from the volume-limited SDSS samples. We measure the first non-zero Legendre multipoles ($\ell = 0, 2, 4$) of the redshift-space two-point correlation function (TPCF), defined as,
    \begin{equation}
        \xi_\ell(s) = \left( 2\ell+1 \right)\int_{0}^{1}\xi^{s}\left(s,\mu\right)L_\ell(\mu)d\mu, \label{eq:mps}
    \end{equation}
where $s=|\mathbf{s}|$, $\mu=\pi/s$ is the cosine of the angle between the line-of-sight direction and the pair separation vector $\mathbf{s}$. $\xi^{s}\left(s,\mu\right)$ is the two point correlation function and $L_\ell$ is the $\ell^\mathrm{th}$-order Legendre polynomial. These quantities are computed using the publicly available code \texttt{FCFC} \citep[][]{FCFC}.

\begin{figure}
    \centering
    \includegraphics[width=\columnwidth]{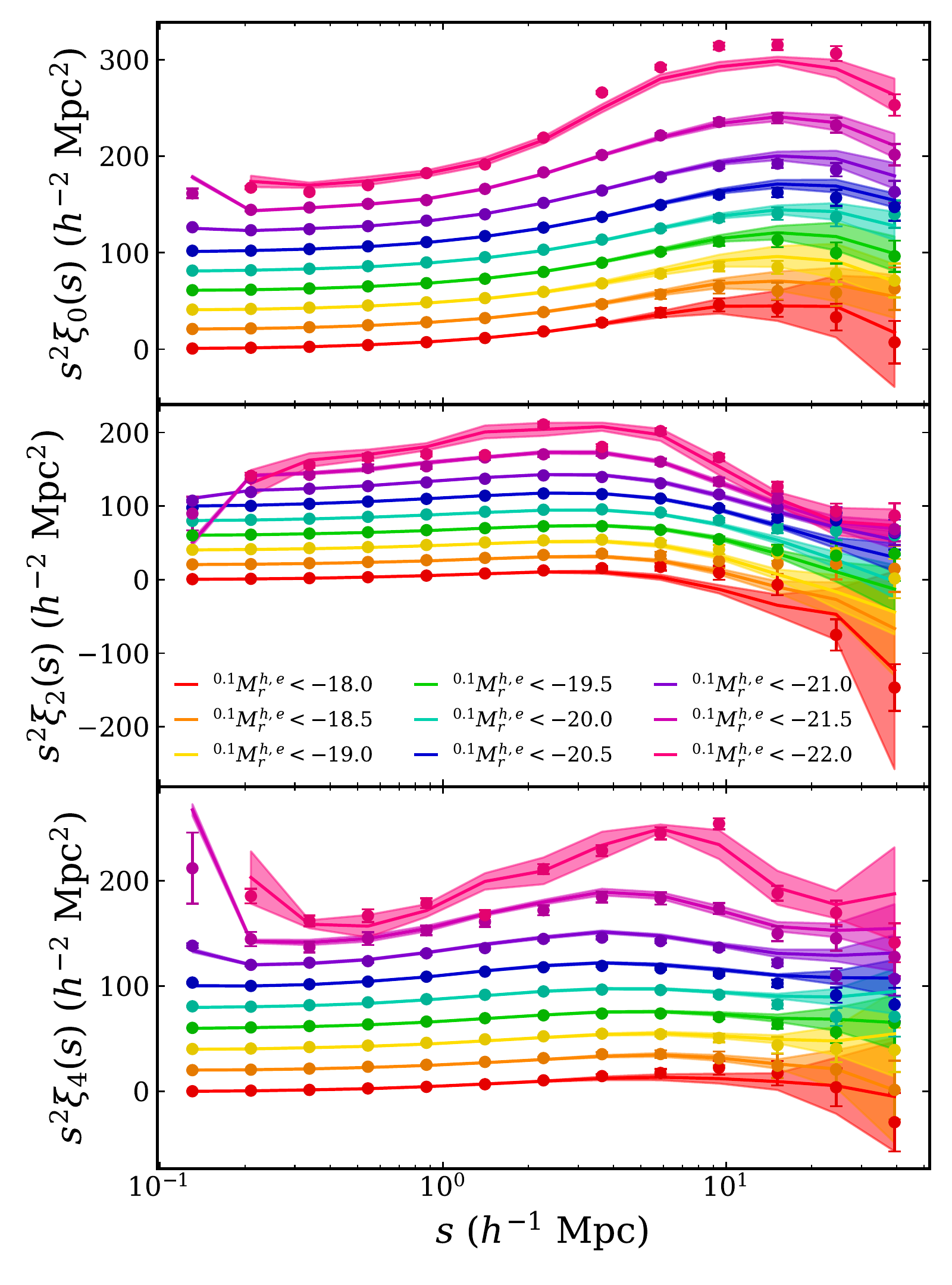}
    \caption{Monopole (top panel), quadrupole (middle panel) and hexadecapole (bottom panel) of the TPCF of Uchuu-SDSS and SDSS galaxies for several volume-limited samples corresponding to luminosity cuts (see Table~\ref{tab:sdss}). Solid lines indicate the average of the eight independent Uchuu-SDSS lightcones, while the shaded regions indicate the rms scatter. Data points indicate the clustering measured from the SDSS sample described in Section~\ref{sec:sdss}. For clarity, lines corresponding to different luminosity cuts have been offset by successive intervals $\SI{20}{\per\h\squared\mega\parsec\squared}$, starting from the lowest luminosity sample. Note that the lowest-$s$ point for the $\magr < -22$ sample has been removed due to low statistics.}
    \label{fig:xi_s}
\end{figure}

\begin{figure}
    \centering
    \includegraphics[width=\columnwidth]{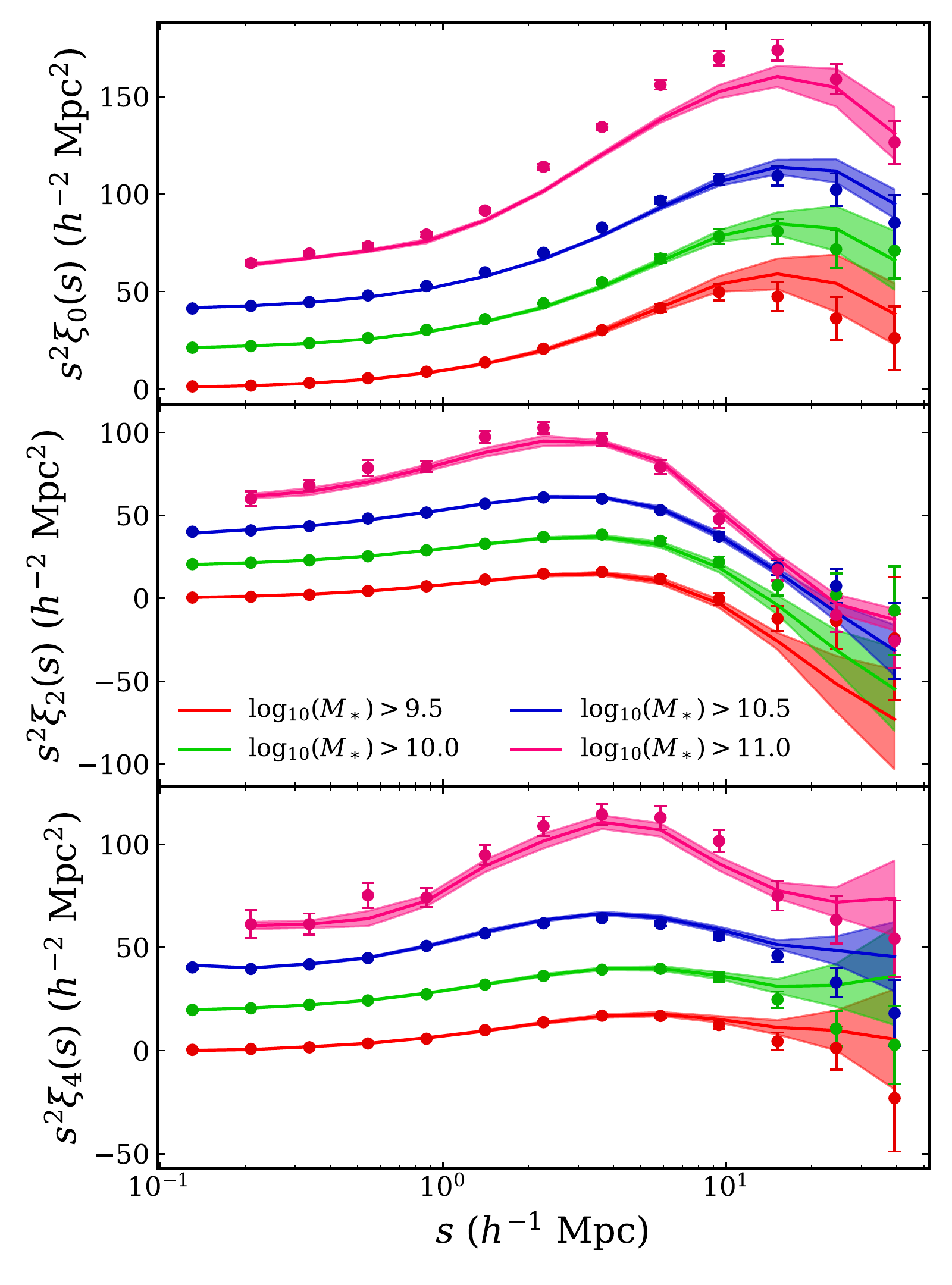}
    \caption{Similar to Fig.~\ref{fig:xi_s}. Here, the clustering for several $M_\ast$ cuts are depicted. For clarity, lines corresponding to different cuts have been offset by successive intervals $\SI{20}{\per\h\squared\mega\parsec\squared}$, starting from the lowest mass sample. Note that the lowest-$s$ point for the $\log_{10}(M_\ast) < 11.0$ sample has been removed due to low statistics.}
    \label{fig:xi_s_Mstar}
\end{figure}

Fig.~\ref{fig:xi_s} compares the monopole, $\xi_0(s)$, quadrupole, $\xi_2(s)$, and hexadecapole, $\xi_4(s)$, of the TPCF of 8 independent Uchuu-SDSS lightcone catalogues against the measurements from the SDSS dataset, for the set of volume-limited samples described in Table~\ref{tab:sdss}. The clustering of our Uhuu-SDSS galaxies is in good agreement with the data for all the volume-limited samples considered, despite the simplicity of our luminosity assignment model -- note that we neglect any potential dependence of our scatter parameter on redshift or luminosity, and consequently our luminosity assignment model has only one tunable parameter. The agreement is poorer for the $\magr<-22$ sample, with the Uchuu-SDSS lightcones underestimating the observed monopole. This suggests that the scatter between our halo mass proxy and galaxy luminosity is likely to decrease for bright galaxies $\magr \lesssim -22$, which is in agreement with previous findings \citep{Stiskalek21}.
%
% \begin{figure}
%     \centering
%     \includegraphics[width=\columnwidth]{xi_s_SDSS_colour.pdf}
%     \caption{The monopole of the redshift-space two-point correlation function of the box catalogue for several luminosity cuts. Data points indicate the SDSS measurements from~\citet{Guo15}. For clarity lines have been offset by successive intervals $\SI{20}{\per\h\squared\mega\parsec\squared}$, starting from the lowest luminosity sample.}
%     \label{fig:monopole_xi_SDSS}
% \end{figure}
% \begin{figure}
%     \centering
%     \includegraphics[width=\columnwidth]{wp_colour_SDSS_masaki_f1.0_-19_-20.png}
% \includegraphics[width=\columnwidth]{wp_colour_SDSS_masaki_f1.0_-20_-21.png}
%     \includegraphics[width=\columnwidth]{wp_colour_SDSS_masaki_f1.0_-21_-22.png}
%     \caption{The projected }
%     \label{fig:my_label}
% \end{figure}
Similarly, Fig.~\ref{fig:xi_s_Mstar} shows the two-point correlation function for several $M_\ast$ cuts. We find again good agreement between our simulated catalogues and the SDSS data. The poorer agreement for the highest mass cut at $\log_{10}(M_\ast) > 11$ suggests again a decrease in the scatter in the galaxy--halo connection at the high mass end, in agreement with previous studies \citep{Behroozi19}.

In order to study the colour-dependence of clustering, we split the sample into two populations of blue and red galaxies. We use a luminosity-dependent colour cut as introduced in \citet{Zehavi05}, \citep[eq.~13 in][]{Zehavi11}.
\begin{equation}
    {}^{0.1}(g-r)_\mathrm{cut} = 0.21 - 0.03~{}^{0.1}M_r^{h,e}.
    \label{eq:colour_cut}
\end{equation}
Galaxies above the ${}^{0.1}(g-r)$ cut belong to the red population, while galaxies below the cut belong to the blue population.

\begin{figure}
    \centering
    \includegraphics[width=\columnwidth]{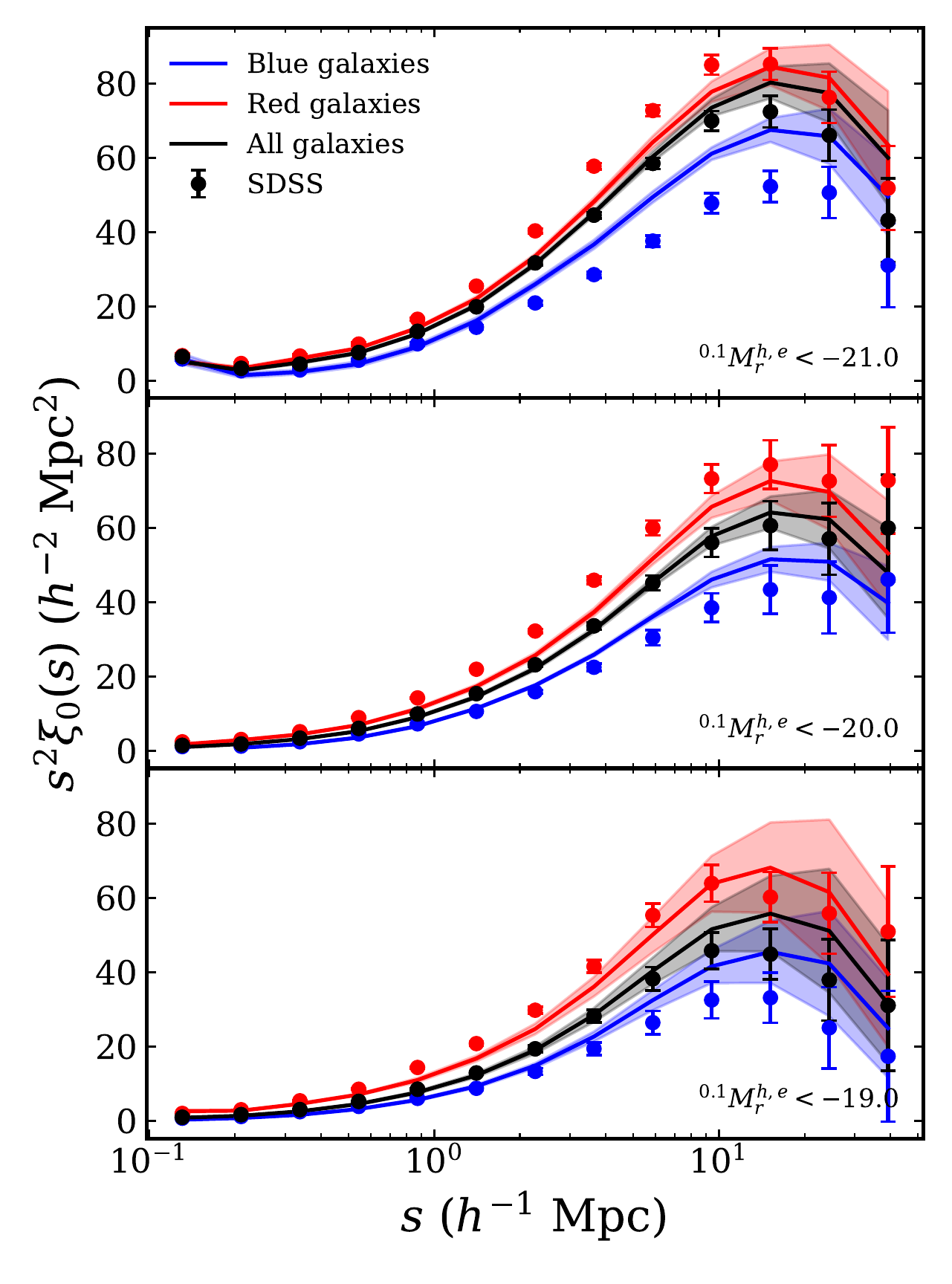}
    \caption{The monopole of the two-point correlation function for SDSS (points with error bars) and Uchuu-SDSS galaxies (solid lines), split into blue and red galaxies according to eq.~\ref{eq:colour_cut}. The dark solid lines indicate the average of the eight independent Uchuu-SDSS lightcones, while the shaded regions show the rms scatter. Each panel corresponds to a volume-limited sample, as indicated in the lower right corner.}
    \label{fig:xi_s_colour}
\end{figure}

Fig.~\ref{fig:xi_s_colour} compares the TPCF for blue, red and all galaxies of Uchuu-SDSS with that of the SDSS dataset, for a subset of the volume-limited samples in Table~\ref{tab:sdss}.
Uchuu-SDSS is in reasonable agreement with the SDSS data, although the agreement is visibly poorer than for the overall volume-limited samples, specially for the brightest samples. This points to limitations due to the simplicity of our colour-assignment algorithm. As described in Section~\ref{subsubsec:mag_colour_evolution}, colours in the lightcone are randomly drawn from our target colour distribution. %The clustering measurements do not improve significantly for the colour assignment method used in the cubic box catalogues (Section~\ref{subsubsec:colour_assignment}), where colours are assigned statistically for central galaxies, whereas for satellite galaxies some information about the galaxy's age is incorporated in a simple way. The free parameter $f$, which can control the the separation in the age of the blue and red population of galaxies, is set to $1$ for simplicity, and the full parameter range was not explored. In general, t
Our results could be improved, by using information about the halo age to assign colours, thus accounting for assembly bias. This would likely require to apply a model involving free parameters, which would need to be finely tuned in order to match the observed colour-dependent clustering.

% \begin{table}
% 	\centering
% 	\caption{This is an example table. Captions appear above each table.
% 	Remember to define the quantities, symbols and units used.}
% 	\label{tab:example_table}
% 	\begin{tabular}{cc} % four columns, alignment for each
% 		\hline
% 		Parameter name & Value \\
% 		\hline
% 		$\Omega_0$ & $0.3089$\\
% 		$\Omega_\mathrm{b}$ & $0.0486$ \\
% 		$\lambda_0$ & $0.6911$ \\
% 		$h$ & $0.6774$ \\
% 		$n_\mathrm{s}$ & $0.9667$ \\
% 		$\sigma_8$ & $0.8159$ \\
% 		$N$ & $12800$ \\
		
% 		\hline
% 	\end{tabular}
% \end{table}

\subsection{Stellar mass function}\label{sec:SMF}

In order to further validate our simulated galaxy catalogues, we calculate the stellar mass function (SMF) of the SDSS sample (see Section~\ref{sec:stellar_mass_sfr}) and compare it with that from the set of 8 independent Uchuu-SDSS lightcones (see Section~\ref{sec:assign_gal_properties_to_mock}). The SMF is estimated by the non-parametric $1/V_\mathrm{max}$ method widely used in deriving the galaxy luminosity function.
% \citep[see e.g.][]{Schmidt68,Johnston11}
% For a description of the method, we refer the reader to e.g. \citet{Schmidt68,Johnston11} \ca{I have removed method descriptions to save space, since it's standard material}. Note that in this section, $V_\mathrm{max}$ refers to volume and not to the halo maximum velocity defined in Section~\ref{subsubsec:luminosity_SHAM}.
To compute the SMF for SDSS and Uchuu samples, we select all galaxies in a redshift range of $0.02 \leq z \leq 0.2$ with stellar masses $M_\ast \geq 10^9~\hsqMsun$ and $r$-band apparent magnitudes $14.5 \leq r \leq  17.6$.

\begin{figure*}
    \centering
    \includegraphics[width=1.\linewidth]{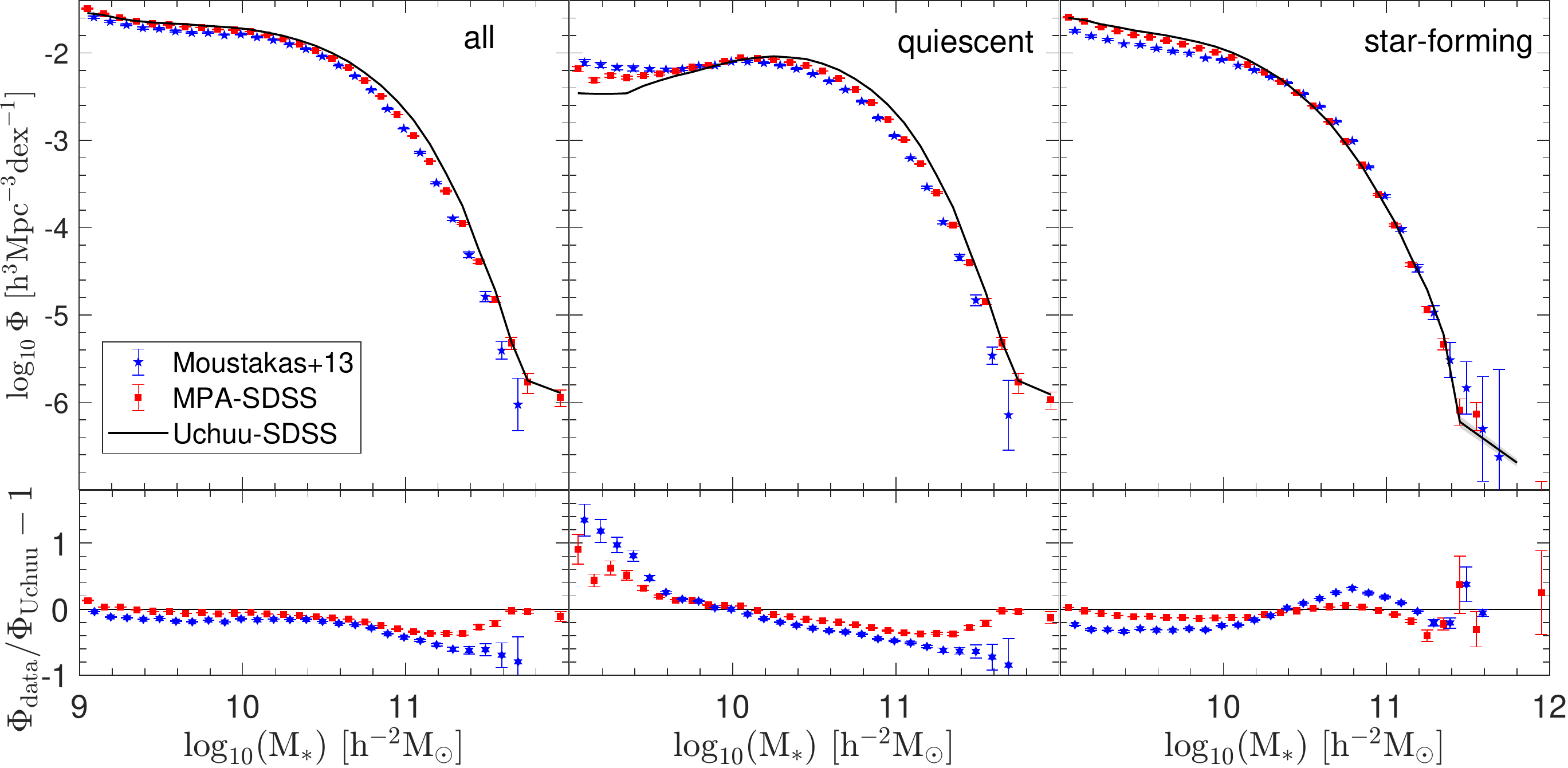} \vspace{-0.5 cm}
    \caption{Comparison of the SMF of the Uchuu-SDSS data (the mean of the 8 independent lightcones) with that obtained from the SDSS data with $0.02 \leq z \leq 0.20$, $\mathrm{M_\ast} \geq 10^9~h^{-2}M_\odot$, and $14.5 \leq r \leq 17.6$, in addition to the results from MPA and 
    \citet{Moustakas13}, as shown in the legends. Left, middle and right panels present the SMF for all, quiescent, and star-forming galaxies, respectively. 
    %We classify quiescent and star-forming galaxies according to the sSFR with $\mathrm{sSFR} \leq 10^{-11}~\mathrm{yr}^{-1}$ for quiescent and $\mathrm{sSFR} > 10^{-11}~\mathrm{yr}^{-1}$ for star-forming galaxies. Shaded black areas (very small) and bars represent $1\sigma$ Poisson error. 
    The lower panels show the relative fraction, $\Phi_\mathrm{data}/\Phi_\mathrm{Uchuu}-1$, of the SMF relative to that from Uchuu-SDSS. The plot shows a reasonable agreement among all three data sets, except for quiescent galaxies at the low- and high-mass ends.
%    \ca{do the Uchuu points not have errors, or are they too small? if the former is the case please add them. And add errors in quadrature for the residuals. Change Log by log in x-axis and $M_\mathrm{sun}$ by $M_\odot$}
    }
    \label{fig:SMF}
\end{figure*}

In Fig.~\ref{fig:SMF} (left panel), we present the SMF obtained from the mean of the 8 Uchuu-SDSS lightcones. Results are compared to SMF derived from the SDSS sample. We also compare our results with that obtained by 
\citet{Moustakas13}. In \citet{Moustakas13}, $M_\ast$ was determined utilizing iSEDfit, a suite of routines used to determine stellar masses, SFRs, and other physical properties of galaxies from the observed broadband SEDs and redshifts \citep[e.g.][]{Kauffmann03,Salim07}. 
The  Uchuu-SDSS is in reasonably good agreement with both MPA-SDSS and SMF obtained by \cite{Moustakas13}. 
%for $M_\ast \lesssim 10^{11.8}~\hsqMsun$. We believe that the %estimation of the MPA stellar mass of massive and luminous %galaxies is biased, which can be seen in the SMF at the high-%mass end with $M_\ast \gtrsim 10^{11.8}~\hsqMsun$. %\Alex{[comment] why is there a bias? }

%The figure shows that Uchuu and MPA-SDSS SMFs are consistent within $10\%$, and both are also in good agreement with the SDSS estimates from \citet{Bernardi13} and \cite{Moustakas13}, see Fig.~\ref{fig:SMF} (left panel). We also plot in the middle and right panels of Fig.~\ref{fig:SMF} the SMF of the quiescent and star-forming galaxies for each data set. We classify the SDSS galaxies into quiescent and star-forming according to the specific star formation rate (sSFR). We adopt $\mathrm{sSFR}=10^{-11}~\mathrm{yr}^{-1}$ as the threshold between quiescent and star-forming galaxies, where there is a minimum between the two peaks of the bimodal sSFR distribution, as seen in Fig.~\ref{fig:lightcone_ssfr}. 

The middle and right panels of Fig.~\ref{fig:SMF} show the SMF of quiescent and star-forming galaxies for each data set. %We classify the SDSS galaxies into quiescent and star-forming according to the specific star formation rate (sSFR).
We adopt $\mathrm{sSFR}=10^{-11}~\mathrm{yr}^{-1}$ as the threshold between quiescent and star-forming galaxies. This value corresponds to the minimum between the two peaks of the bimodal sSFR distribution shown in Fig.~\ref{fig:lightcone_ssfr}. The Uchuu-SDSS SMFs for quiescent and star-forming galaxies galaxies is consistent with those obtained from MPA-SDSS for all masses, although for quiescent galaxies the agreement is slightly poorer compared to the overall sample in the left panel. There is also a good agreement between the Uchuu SMFs and those obtained by \citet{Moustakas13}, with a notable offset at the low- and high-mass ends for quiescent galaxies. One possible reason for this difference is the different methods for the estimation of SFRs used by MPA/JHU and \citet{Moustakas13}, which result in different SFR values. %The SMF for the star-forming galaxies obtained from Uchuu is consistent with that derived from MPA-SDSS. %There is also a good agreement between the SMF of Uchuu and \cite{Moustakas13} with a maximum offset of $\sim 0.3$ at $\log{M_\ast} \sim  10.7~[\hsqMsun]$. 
%There is a notable offset between the SMF of Uchuu and that obtained from \citet{Bernardi13} for all stellar masses. One possible reason for this difference is that \cite{Bernardi13} classified galaxies into red and blue according to their color-magnitude diagram not according to sSFR.

% In summary, the SMF derived from the mocks is in very good agreement with the SMF of our SDSS sample. Also, the SMFs of all galaxies of our mock data are consistent with the results of the previous studies. We see some differences when the sample is split into quiescent and star-forming galaxies, but this is likely due to differences in how the galaxies were classified.
%\ca{I think we shouldn't mention Bernardi in the text if we don't show it in the figure}

\subsection{Halo occupation distribution}
\label{subsec:hod}

Galaxies are known to be biased tracers of the underlying dark matter density field. 
In order to better understand the connection between galaxies and haloes in  Uchuu-SDSS catalogues, we investigate the halo occupation distribution. We compute $\langle N_\mathrm{gal}(>L|M_\mathrm{halo})\rangle$ -- the mean number of galaxies brighter than a given $r$-band luminosity in a halo with virial mass $M_\mathrm{halo}$.
%In order to better understand the connection between galaxies and haloes in our mocks, we investigate halo occupation distribution (HOD), $\langle N(M_\mathrm{halo})\rangle$, which describes the mean number of galaxies in a halo with mass $M_\mathrm{halo}$ which meet some criterion on a galaxy property (in our case the number, of galaxies brighter than a given $r$-band luminosity threshold, $\magr$). Here, we use the virial mass of the halo as the relevant measure of the halo mass, $M_\mathrm{halo}$.

The luminosity-dependent HOD is modeled  with the functional form described in \citet{Zehavi11}. %\Alex{I guess instead of `method' you mean the functional form of the HOD? The method used to fit the HODs to the SDSS data in Z11 is different to what is done here.}
We write $\langle N_\mathrm{gal}(>L|M_\mathrm{halo})\rangle$ as a sum of the mean number of central and satellite galaxies.
%
%\begin{equation}
%    \langle N(>L|M_\mathrm{halo})\rangle = \langle N_\mathrm{cen}(>L|M_\mathrm{halo})\rangle + \langle N_\mathrm{sat}(>L|M_\mathrm{halo})\rangle.
%    \label{eq:hod_sum}
%\end{equation}
%
The mean occupation function of the central galaxies is modelled as a step-like function with a cutoff profile softened to account for the scatter between galaxy luminosity and halo mass, and
%\begin{equation}
%    \langle N_\mathrm{cen}(>L|M_\mathrm{halo})\rangle = \frac{1}{2}\left[1+\textrm{erf}\left(\frac{\log{M_\mathrm{halo}(L)}-\log{M_\mathrm{min}(L)}}{\sigma_{\log{M}}(L)}\right)\right],
%    \label{eq:hod_central}
%\end{equation}
%where %$\textrm{erf}(x)=\frac{2}{\sqrt{\pi}} \int_{0}^{x} e^{-t^2}dt$ 
%is the error function. 
the mean occupation of satellite galaxies is modelled as a power law modulated by a similar cutoff profile, see \citet{Zehavi11} for the details.
%\begin{equation}
%    \langle N_\mathrm{sat}(>L|M_\mathrm{halo})\rangle = \langle N_\mathrm{cen}(>L|M_\mathrm{halo})\rangle \times \left(\frac{M_\mathrm{halo}-M_{0}(L)}{M'_{1}(L)}\right)^{\alpha(L)}.
%    \label{eq:hod_satellite}
%\end{equation}
% By replacing eq.~\ref{eq:hod_central}~and~\ref{eq:hod_satellite} in eq.~\ref{eq:hod_sum} we can describe the distribution of the total number galaxies above a given luminosity threshold as a function of halo mass. 
This HOD model has five free parameters: the mass scale, $M_\mathrm{min}$, and width, $\sigma_{\log{M}}$, of the central galaxy mean occupation, and the cutoff mass scale, $M_{0}$, normalization, $M_{1}'$, and high-mass slope, $\alpha$, of the satellite galaxy mean occupation function. 

\begin{figure}
    \centering
    \includegraphics[width=\columnwidth]{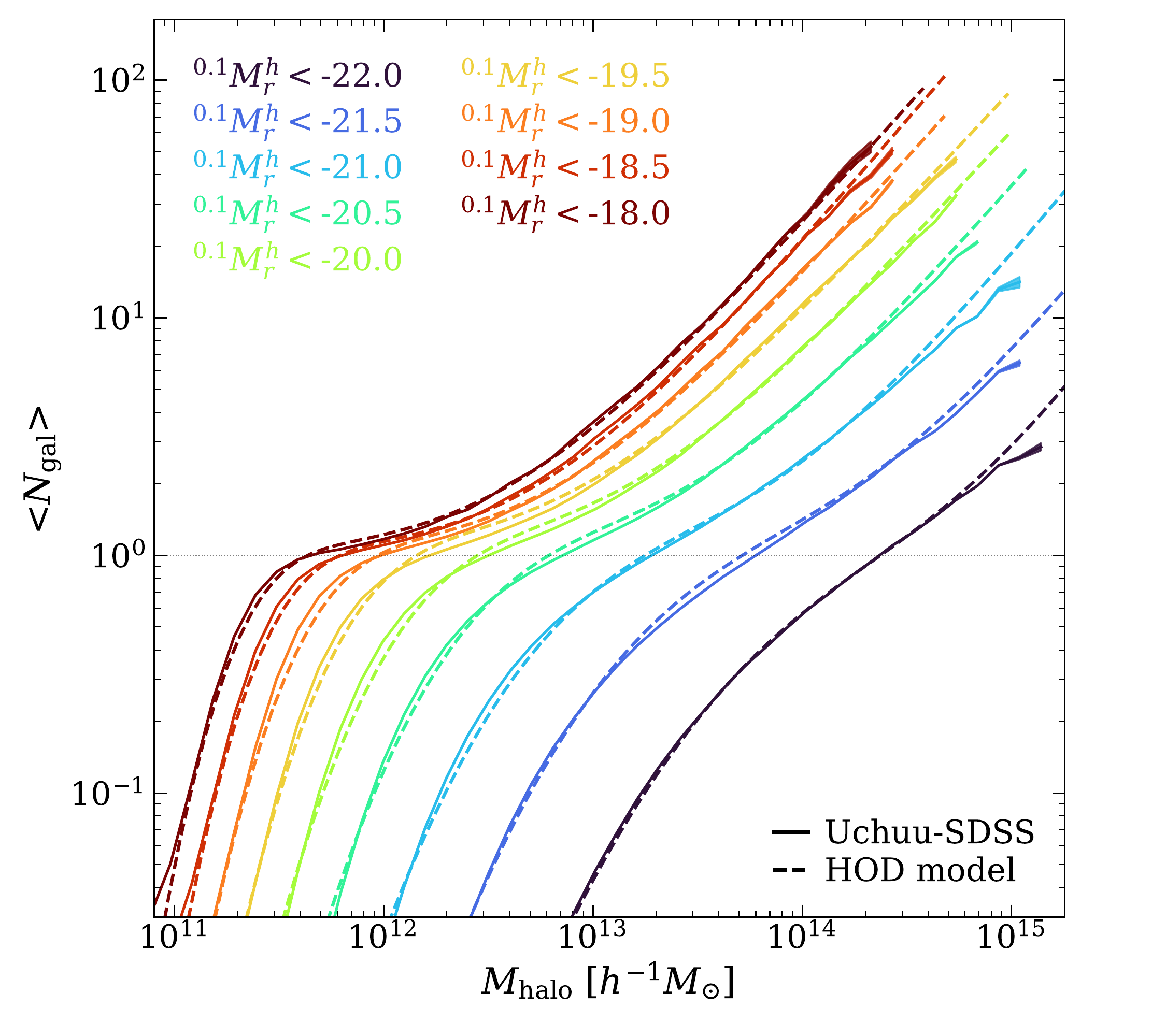}
    \caption{Measured mean halo occupation for the different luminosity threshold samples drawn from 8 independent Uchuu-SDSS lightcones as a function of halo mass.
    %\Paco{it is recommended to put the units in the x-axis of the plot. Btw. I'd suggest to keep the Mhalo units as they are in the Uchuu catalgos in Msun/h, since the luminosity thresholds are also given as a function of h. BUT this depends on the units adopted by Zehavi et al. for their halo masses. Please check.}). \Julia{Done! Zehavi also uses Msun/h}
    %The shaded area shows the standard deviation obtained from the average of the 8 independent Uchuu lightcones. 
    The dashed lines represents the best-fit models for the halo occupation functions. Best-fit HOD parameters are shown in Table~\ref{tab:hod_params} together with those obtained by \citet{Zehavi11}. %\ca{Maybe change the "this paper" legend (since all data is from this paper) to solid lines -> mocks, dashed lines -> best-fit model or something like that}}%\Alex{[comment] make the legend bigger. There are also other plots where the text is small, and font size could be increased } \Julia{Is this size ok?}\Alex{looks good!}
    The maximum host halo mass in each sample is related to their effective volume (see Table~\ref{tab:sdss}).
    %\ca{In the plot the values of $M_\ast$ are shown while in the label says $\log_{10}(M_\ast)$. Also, make sure to be consistent with Fig. \ref{fig:bao_hod} (plot in both either $\log_{10}(M_\ast)$ or $M_\ast$). If you choose $\log_{10}(M_\ast)$ show units in label (e.g. $\log_{10}(M_\ast/M_\odot)$) or mention in text}
    %\Alex{The fit to the -18 sample doesn't seem right. The slope for the satellites looks ok, but there is an offset}
    }
    
    \label{fig:all_hod}
\end{figure}

We fit this model to the average HOD obtained from  8 independent Uchuu-SDSS lightcones for the volume-limited samples corresponding to luminosity cuts described in Table~\ref{tab:sdss}. The best fitting HOD parameters are shown in Table~\ref{tab:hod_params}, along with the SDSS estimates from \citet{Zehavi11}. Fig.~\ref{fig:all_hod} shows the mean halo occupation of the Uchuu-SDSS galaxies and the best-fit HOD models. As seen in the figure, the HOD shifts towards more massive haloes as the luminosity threshold increases -- more luminous galaxies occupy more massive haloes. %\Julia{\textbf{REVIEW:} Note that at low halo masses there is a `turn up' of the HODs. When selecting the halo population to compute the HOD for each volume-limited sample, we are not taking a complete population of haloes (as we should be taking) due to the incompleteness of Uchuu halo mass function (see figures.~6,~7 from \citet{UchuuDR1}), which results in the HOD not decreasing to 0. In narrower redshift ranges (corresponding to fainter volume-limited samples), the incompleteness of the halo mass function is more significant, and so the `turn up' of the HOD is more noticeable. Note that, when fitting the model to the average Uchuu-SDSS HODs, we only use halo masses in the range where the mass function is complete.} %the data within the range of halo masses where the mass function is incomplete has been ignored.
%\ca{just checking if I understood correctly: do we use here the DESI halo catalogues or the SDSS ones?} \Julia{For the halo catalogue, I am using the BGS lightcones.}

Our results broadly agree with those of \citet{Zehavi11} over the wide range of brightness thresholds for all volume-limited samples. However, we observe a few discrepancies for some HOD parameters (see Table~\ref{tab:hod_params}). This is likely due to the difference in the %method of estimating the HOD.
methodologies. While we have computed the halo occupation directly from our values of $\magr$ and $M_\mathrm{halo}$ in the Uchuu-SDSS lightcones, \citet{Zehavi11} obtains it by %a parameterisation of 
fitting the projected correlation function of the observed SDSS data. All our best fit parameters, except $\alpha$, follow an ascending trend as a function of the luminosity threshold.

\setlength{\arrayrulewidth}{0.3pt}
\begin{table}
\centering
    \begin{tabular}{cccccc}
        \hline
        $^{0.1}M_r^\mathrm{max}$ & $\log{M_\mathrm{min}}$ & $\sigma_{\log{M}}$ & $\log{M_{0}}$ & $\log{M_{1}'}$ & $\alpha$ \\
        \hline
        \multirow{2}{*}{-18.0} & 11.34$\pm$0.09  & 0.29$\pm$0.20 & 11.07$\pm$0.08 & 12.60$\pm$0.03 & 0.99$\pm$0.03 \\
        & 11.18 & 0.19 & 9.81 & 12.42 & 1.04 \\  [0.75mm]
        \multirow{2}{*}{-18.5} & 11.48$\pm$0.08  & 0.31$\pm$0.16 & 11.08$\pm$0.07 & 12.73$\pm$0.04 & 1.03$\pm$0.03 \\
        & 11.33 & 0.26 & 8.99 & 12.50 & 1.02 \\  [0.75mm]
        \multirow{2}{*}{-19.0} & 11.66$\pm$0.06  & 0.35$\pm$0.12 & 11.16$\pm$0.07 & 12.83$\pm$0.02 & 0.99$\pm$0.02 \\
        & 11.45 & 0.19 & 9.77 & 12.63 & 1.02 \\ [0.75mm]
        \multirow{2}{*}{-19.5} & 11.85$\pm$0.10  & 0.38$\pm$0.16  & 11.37$\pm$0.09 & 12.95$\pm$0.04 & 0.95$\pm$0.03 \\
        & 11.57 & 0.17 & 12.23 & 12.75 & 0.99 \\ [0.75mm]
        \multirow{2}{*}{-20.0} & 12.12$\pm$0.08  & 0.45$\pm$0.11 & 11.48$\pm$0.09 & 13.17$\pm$0.04 & 0.97$\pm$0.04 \\
        & 11.83 & 0.25 & 12.35 & 12.98 & 1.00 \\ [0.75mm]
        \multirow{2}{*}{-20.5} & 12.43$\pm$0.06  & 0.52$\pm$0.06 & 11.49$\pm$0.11 & 13.47$\pm$0.03 & 1.00$\pm$0.04 \\
        & 12.14 & 0.17 & 11.62 & 13.43 & 1.15 \\ [0.75mm]
        \multirow{2}{*}{-21.0} & 12.86$\pm$0.12  & 0.62$\pm$0.09 & 11.69$\pm$0.12 & 13.83$\pm$0.09 & 1.07$\pm$0.14 \\
        & 12.78 & 0.68 & 12.71 & 13.76 & 1.15 \\ [0.75mm]
        \multirow{2}{*}{-21.5} & 13.32$\pm$0.14  & 0.68$\pm$0.09 & 11.79$\pm$0.14 & 14.25$\pm$0.09 & 1.08$\pm$0.16 \\
        & 13.38 & 0.69 & 13.35 & 14.20 & 1.09 \\ [0.75mm]
        \multirow{2}{*}{-22.0} & 13.98$\pm$0.30  & 0.81$\pm$0.22 & 11.86$\pm$0.22 & 14.76$\pm$0.19 & 1.26$\pm$0.30 \\
        & 14.06 & 0.71 & 13.72 & 14.80 & 1.35 \\ [0.75mm]
        \hline
    \end{tabular}
\caption{Best-fit HOD parameters for our volume-limited samples. For each luminosity cut the top row indicates the best-fit parameters for our Uchuu-SDSS lightcones, while the bottom row shows that for the SDSS data as reported in \citet{Zehavi11}. Error bars on our HOD parameters correspond to standard deviation errors on the parameters. Halo masses are in units of $h^{-1}M_{\odot}$. 
%\Paco{Please check. In the HOD figures it says Msun for the halo mass units}
}
\label{tab:hod_params}
\end{table}

\section{RSD and BAO measurements}
\label{sec:RSD_BAO}

In this section we study the BAO signal in the SDSS MGS. For this purpose we define a BAO sample (SDSSbao, see Section~\ref{subsec:sdss_samples}) in which this signal is enhanced similar to \citet{Ross15}. We also model the full shape of the TPCF to measure $f\sigma_8$ and obtain the anisotropic BAO distances. While we have the set of 32 Uchuu-SDSS lightcones to compare with the SDSSbao data, 
%. While these lightcones partially overlap, this number of mocks was created with the aim to provide the higher statistics required in order to study the BAO scales. %In section~\ref{subsec:light_mocks}, an extended set of partially overlapping 32 lightcone is introduced, which aim to provide the higher statistics required in order to study the BAO scales. 
in order to further improve our covariance errors on BAO scales, we generate an additional sample of 5100 light-cones, describe below, using a set of lower resolution $N$-body simulations run with the GLAM code \citep{Klypin18}.

%In the following, we describe the construction of our additional GLAM-SDSSbao mocks, and use them along with the original mocks in order to analyse RSD and BAO in the Uchuu suite.

In Section~\ref{subsec:GLAM_construction} we describe the method used to construct our 5100 GLAM-SDSSbao lightcones, while their clustering properties are explored in Section~\ref{subsec:GLAM_clustering}. Our RSD measurements are presented in Section~\ref{subsec:RSD_measurement}. Finally, we measure the isotropic BAO scale in Section~\ref{subsec:BAO_measurement}.
%\Alex{[comment] would be good to include an overview paragraph saying: In section ... we describe the GLAM mocks. Our RSD measurements are in Section ..., where our model is described in Section ... and results in Section... etc}

\subsection{Constructing the GLAM lightcones}
\label{subsec:GLAM_construction}

We generate 1275 GLAM simulations using the same cosmology and linear power spectrum as the Uchuu simulation. The GLAM simulations follow the evolution of $2000^3$ particles of mass $1.07\times 10^{10}\hMsun$ in a cubic box of size $1~\hGpc$ with $N_s=140$ timesteps, and mesh of $N_g=5800$. This numerical set-up yields a spatial resolution of $\Delta x=0.172~\hMpc$. The initial conditions are generated using the Zeldovich approximation starting at $z_\mathrm{ini}=105$.
The distinct haloes in GLAM are identified with the Bound Density Maximum halo finder \citep{Klypin97}. 

\begin{figure}
    \centering
    \includegraphics[width=\columnwidth]{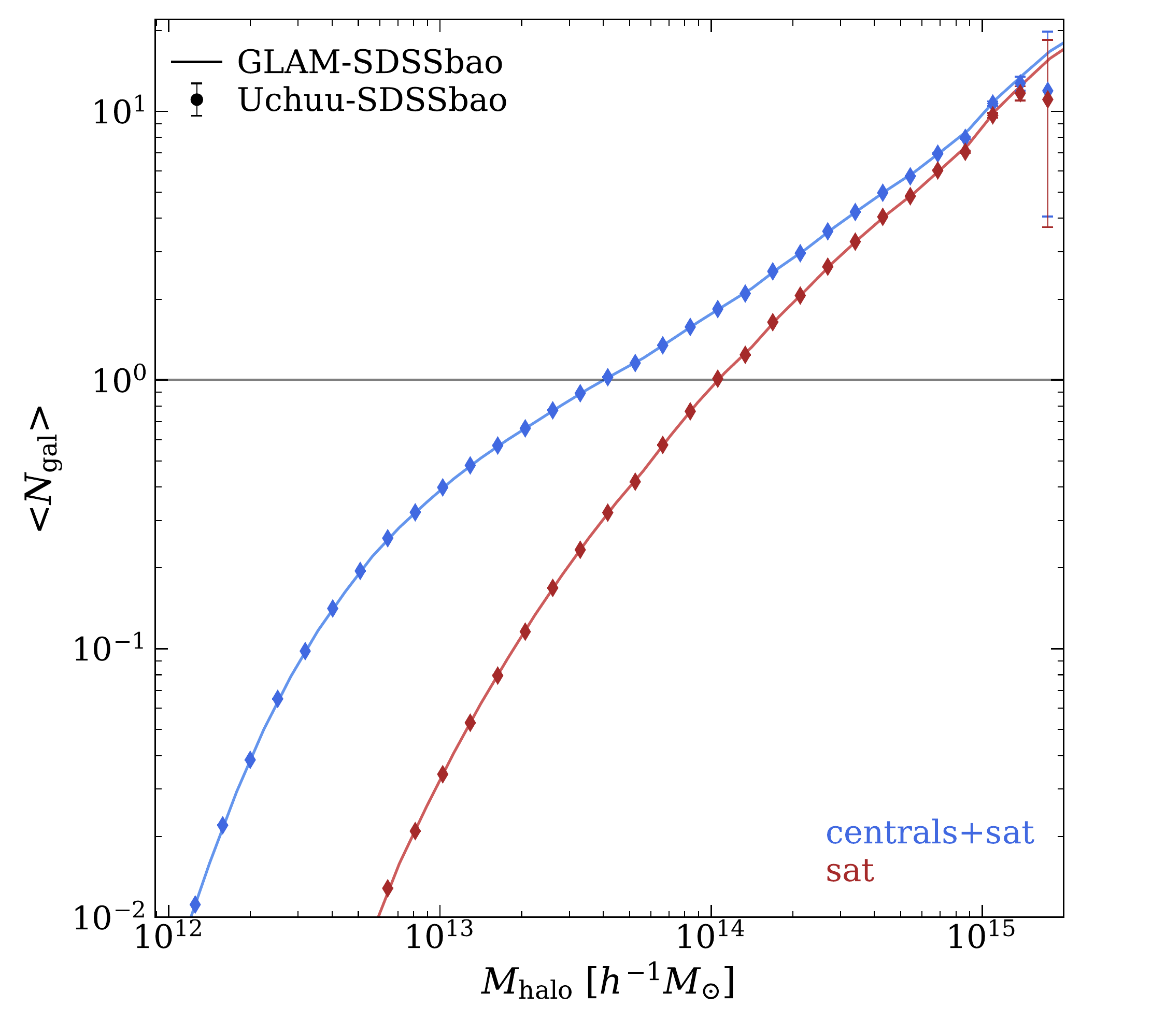}
    \caption{Mean halo occupation distribution from our Uchuu- and GLAM-SDSSbao catalogues, dots and solid line, for all galaxies and satellite galaxies only (in blue and red respectively). 
    %\Paco{Add units to Mhalo} 
    %The mean and the standard deviation of the Uchuu mocks is represented by dots and error bars while for the 1275 GLAM box galaxy mocks we have used a solid line and a shaded area. %\ca{do we need this figure? The HOD is reproduced by construction if I understood correctly} \Julia{Si, pero el HOD del Uchuu-SDSSbao no está en ninguna parte, hay que ponerlo porque es lo que se usa para GLAM}
    }
    \label{fig:bao_hod}
\end{figure}

Since the GLAM simulations are unable to resolve substructure inside distinct haloes, the SHAM method introduced in Section~\ref{sec:mock_construction} cannot be applied, and we thus resort to a statistical HOD method. First, we compute the HOD of the SDSSbao sample by applying the galaxy selection criteria in eq.~\ref{eq:hod_cuts} to our 8 independent Uchuu-SDSS lightcones (hereafter Uchuu-SDSSbao). 
%\ca{maybe describe explicitly how galaxies are assigned?} \Julia{Kind of explained in section 4.4?}. \Alex{[comment] say what galaxy sample the HOD is measured for}. 
We then use the 1275 GLAM halo catalogues available at the mean redshift of the BAO sample ($z \sim 0.1$) to generate a galaxy catalogue for each GLAM box by randomly drawing galaxies from the measured halo occupation statistics for each distinct halo. 
Fig.~\ref{fig:bao_hod} shows the mean HOD from Uchuu-SDSSbao used to populate with galaxies the GLAM simulations, along with the resulting mean HOD obtained from the 1275 GLAM galaxy catalogues. By construction, the HOD of Uchuu- and GLAM-SDSSbao galaxies are in agreement. 
It is important to note that the GLAM simulations are only able to resolve haloes larger than $10^{12}~\hMsun$. However, the HOD obtained from the high-fidelity Uchuu-SDSSbao lightcones does not extends to masses below this limit. % This results in our GLAM mocks having a deficit of central galaxies compared to the Uchuu mocks. However, this effect is unimportant at the BAO scales.
The resulting GLAM galaxy catalogues have an average density of $6\times 10^{-4}$ galaxies per unit volume (average $n(z)$ of the Uchuu-SDSSbao sample).

Once the GLAM halos for a given box are populated with galaxies, we adjust its number density to match that in the SDSSbao sample. For this step, we use the $n(z)$ presented in \citet{Ross15} as reference. Then we cut it to the northern contiguous region of the SDSS survey footprint. As with the Uchuu lightcones, by replicating the SDSS footprint across the full sky, we can generate a total of 4 independent SDSSbao lightcones from each GLAM box, which allows us to create a total of 5100 GLAM-SDSSbao lightcones. We decided not to apply the fibre collision correction to GLAM-SDSSbao since it effect is negligible on the scales we are going to make use of the lightcones (BAO scales).

\begin{figure}
    \centering
    \includegraphics[width=\columnwidth]{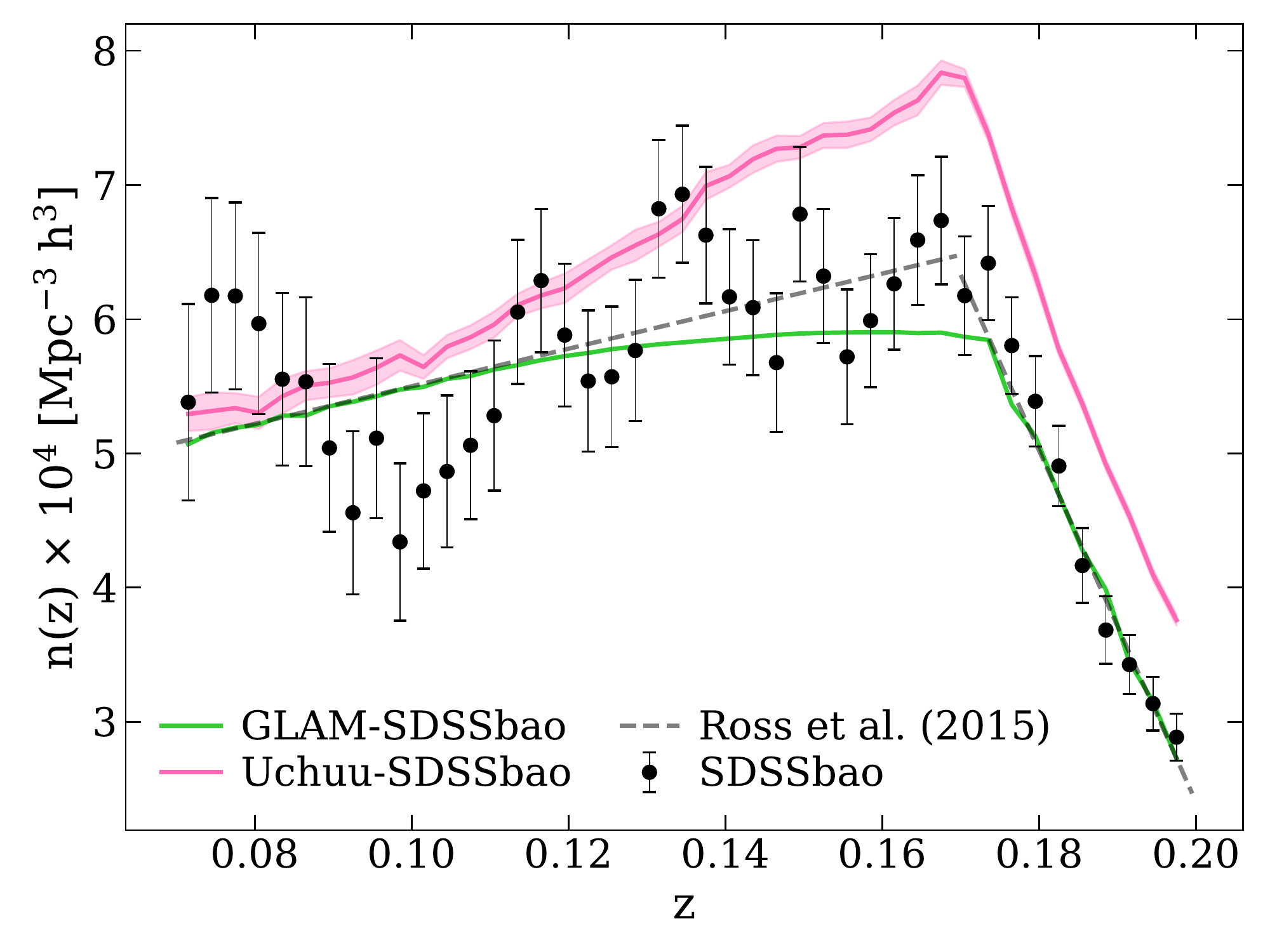}
    \caption{Number density of the SDSSbao data, Uchuu- and GLAM-SDSSbao lightcones. The green line represents the mean $n(z)$ of the 5100 GLAM-SDSSbao, while the magenta represents that from the 32 Uchuu-SDSSbao. The standard deviation for each of the samples is shown as a shaded area. The error bars that accompany the SDSSbao $n(z)$, represented with black dots, are taken from the standard deviation of the GLAM-SDSSbao catalogues.
    The deviation between GLAM and \citet{Ross15} around $z=0.16$ follows from the $n(z)$ of GLAM galaxy catalogues obtained from the HOD.}
    \label{fig:bao_nz}
\end{figure}

The GLAM-SDSSbao lightcones mean number density is shown in Fig.~\ref{fig:bao_nz} together with that from Uchuu-SDSSbao and the SDSSbao data. There is a 20\% slight excess of galaxies in Uchuu-SDSSbao as compared to the data and the GLAM-SDSSbao lightcones at high-redshift tail of the distribution. As explained in section \ref{sec:gal_properties}, this is because we use a target luminosity function to build the Uchuu galaxy catalogues that transitions from SDSS to GAMA, since the luminosity function from SDSS is poorly constrained at high redshifts. This difference does not seem to have a significant impact on the performance of our Uchuu-SBSSbao lightcones in reproducing within $1\sigma$ the SDSSbao clustering, however it may explain the small difference seen between Uchuu- and GLAM-SDSSbao (see Fig.~\ref{fig:xi_s_BAO}). %\Alex{[comment] add comment on differences in Uchuu number density at higher redshifts (uses GAMA LF, differences also seen in dN/dz in earlier figure).}
%{\bf AK: yes, the difference was mentioned before. This does not help much because we still do not know how important are results of the differences. We must say something here. The difference is substantial. My guess is we need to look at differences in correlation functions of uchuu and glam in fig.19. In spite of 20 per cent difference in abundance, clustering is very close.}

\subsection{Two-point correlation function and covariance matrix}
\label{subsec:GLAM_clustering}

In this section we explore the clustering on the BAO scales in the three data sets: SDSSbao, Uchuu- and GLAM-SDSSbao. 
%In all of them we study the BAO sample described by the cuts in eq.~\ref{eq:hod_cuts}. 

In order to optimise the BAO clustering signal-to-noise ratio, we weight both galaxies and randoms depending on the galaxy number density $n(z)$ using FKP weights \citep{Feldman94,Ross15}, i.e.
\begin{equation}  
    w_{\textrm{FKP}}=\frac{1}{1+P_{\textrm{FKP}} n(z)}.
    \label{eq:fkp_weigths}
\end{equation}
We set $P_{\textrm{FKP}} = 16\,000~h^{-3}\mathrm{Mpc}^{3}$
%\Paco{Is this vale the same adopted by Ross et al.?}\Julia{Yes!}
, which is close to the measured amplitude at $k = 0.1~\hMpc$. 

Further, from the 5100 GLAM-SDSSbao lightcones, we infer the covariance matrix, $C$, defined as 
%from 600 GLAM mocks

\begin{equation}
    C^{\ell_i \ell_j }(s_i,s_j)= \frac{1}{N-1}\sum^N_{n=1} [\xi^{(\ell_i)}(s_i) - \bar{\xi}^{(\ell_i)}(s_i)][\xi^{(\ell_j)}(s_j) - \bar{\xi}^{(\ell_j)}(s_j)],
    \label{eq:cov_mat}
\end{equation}
where $N$ is the number of lightcones.

\begin{figure}
    \centering
    \includegraphics[width=\columnwidth]{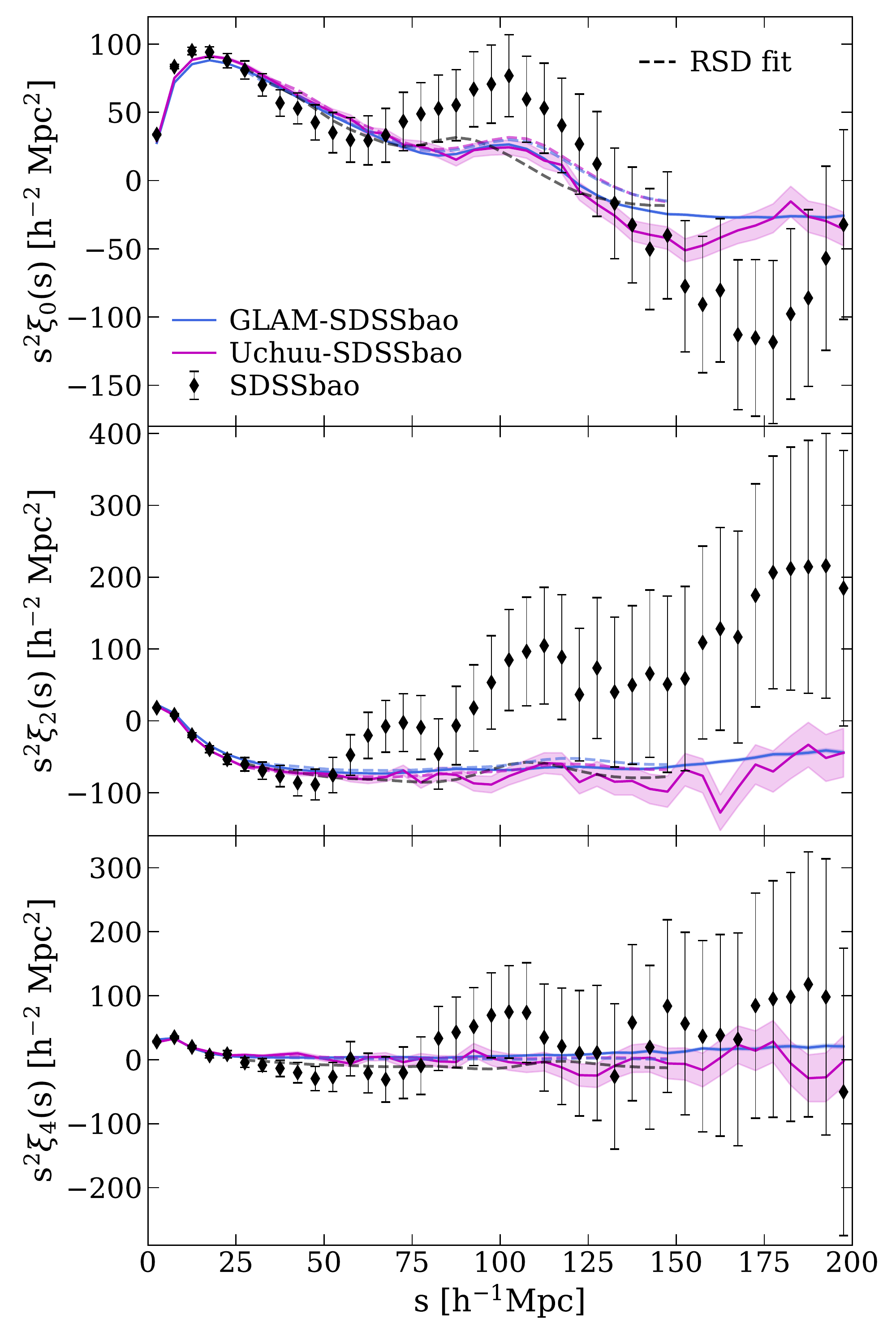}
    \caption{The monopole, quadrupole and hexadecapole of the two-point correlation function measured in our three SDSS BAO data set: SDSSbao (filled symbols), GLAM-SDSSbao (blue lines) and Uchuu-SDSSbao (pink lines). Errors have been estimated from the covariance matrix of the 5100 GLAM-SDSSbao lightcones. Observations and simulated data agree within 2$\sigma$ at all scales.  
    Following the same colour code, we also plot the best-fit RSD model (dashed lines) for each data, (see Section~\ref{subsec:RSD_measurement}).
    %{\bf AK: conclusions? I would remove the numbers of glam and uchuu mocks (technical and be found in the text) and how error-bars are estimated. Add summary of what we learn from the plot. EG SDSSbao and Uchuu-SDSSbao are within 2$\sigma$ at all scales. Note that considering that measurements are correlated, there is good agreement of uchuu, glam  with sdss.}
    %SDSSbao and Uchuu-SDSSbao are within 2$\sigma$ at all scales. Note that given that measurements are correlated, there is good agreement of Uchuu and GLAM with SDSS. 
    %\Julia{TBrD}
    %\ca{RSD analysis appears to underestimate the location of BAO peak for SDSS}
    }
    \label{fig:xi_s_BAO}
\end{figure}

Fig.~\ref{fig:xi_s_BAO} 
%\Paco{Could you please compare the error bars of the monopole in this figure with that shown in Fig. 18 for Uchuu, GLAM, and SDSSbao? Just to make sure that both are done with the same scaling given the number of mocks.} 
shows the monopole, quadrupole and hexadecapole of the two-point correlation function of the SDSSbao data, the mean of the 32 Uchuu-SDSSbao lightcones (described in Section~\ref{subsec:light_mocks}), and the mean of the 5100 GLAM-SDSSbao. We include all bins between 25 and $200~\hMpc$ in $5~\hMpc$ steps. The error bars in the SDSSbao TPCF are calculated from the diagonal elements of the GLAM-SDSSbao covariance matrix. 
For Uchuu and GLAM, we plot the error in the mean of the correlation functions as shaded regions. % is the same as that of the data, divided by the square root of the number of mocks used in computing the mean. 

The three data sets show a clear BAO peak at $s \sim \SI{100}{\per\h\mega\parsec}$. Uchuu- and GLAM-SDSSbao generally agree with the SDSSbao observational measurements within $2\sigma$. The GLAM measurements have better statistics than Uchuu, due to the huge number of lightcones in the GLAM suite. Moreover, Uchuu have a larger number density than GLAM, which translates into a lower clustering amplitude. Note that the GLAM lightcones, used to estimate uncertainties from its TPCF covariance matrix, match the SDSSbao number density (see Fig.~\ref{fig:bao_nz}).
%It is also worth mentioning that the two point correlation function of the GLAM mocks has been added to the plot for illustrative purposes as the errors used come from their covariance matrix. In no case the GLAM mocks (although they may have a $n(z)$ that matches the data more closely) replace the Uchuu ones, which are the mocks to be used in a comparison with the observed data. \ca{I'm not completely sure I understand these last two lines.}

%\begin{figure*}
%    \centering
%    \includegraphics[width=18.2cm]{figures/corr_matrix.pdf}
%    \caption{Correlation matrices measured from the 5100 GLAM mocks. The monopole is shown %in the left panel, with the quadupole in the middle, and hexadecapole on the right. %The colour indicates the level of correlation, where dark blue represents 100$\%$ %correlation, white indicates zero correlation, and light red means there is an %anti-correlation.
%    %Correlation matrices including the monopole (left plot), quadrupole (middle plot) and %hexadecapole (right plot) two-point correlation function from 5100 GLAM mocks. The %colour indicates the level of correlation, where dark blue represents 100$\%$ %correlation and light red means 50$\%$ anti-correlation.
%    }
%    \label{fig:corrm}
%\end{figure*}

\subsection{RSD measurements}
\label{subsec:RSD_measurement}

%==================================%
%     INTRODUCTION                 %
%==================================%

Galaxy clustering analyses that rely on two-point statistics are not sensitive to the growth rate of structure $f$ directly but instead to $f\sigma_8$, where $\sigma_8$ is the normalization of the linear power spectrum on scale of $8~\hMpc$.
%Thanks, Alex!

We measure the linear growth rate of structure, $f\sigma_8$, and the Alcock-Paczynski parameters, $\alpha_{\perp}$ and $\alpha_{\parallel}$, of the Uchuu-SDSS lightcones and SDSSbao sample described above using the two-point correlation function, $\xi(s,\mu)$.

The anisotropic Alcock-Paczynski parameters are defined as
\begin{equation}\label{eq:a_par}
    \alpha_{\parallel} = \frac{D_\mathrm{H}(z)r^{\text{fid}}_{d}}{D_\mathrm{H}^{\text{fid}}(z)r_{d}}
\end{equation}
and
\begin{equation}\label{eq:a_perp}
    \alpha_{\perp} = \frac{D_\mathrm{M}(z)r^{\text{fid}}_{d}}{D_\mathrm{M}^{\text{fid}}(z)r_{d}},
\end{equation}
%\ca{$D_\mathrm{A}$ here, while in the equations $D_\mathrm{M}$ is written}
where $H(z)$ is the Hubble parameter, $D_\mathrm{M}(z)$  is the angular diameter distance, $D_\mathrm{H}(z) = c/H(z)$ is the Hubble distance, and $r_d$ is the sound horizon at the drag epoch. Quantities with a `fid' superscript are calculated for the fiducial cosmology assumed during the analysis, while the quantities without a superscript exist in the true cosmology.

The analysis is performed using the Planck fiducial cosmology, as adopted for Uchuu, to convert the redshifts to comoving distances. 
If the assumed fiducial cosmology does not match the true cosmology, there is a scaling of the BAO peak position parallel and perpendicular to the line-of-sight (as given in eqs.~\ref{eq:a_par},~\ref{eq:a_perp}). 
%Thanks, Alex!
%As in the case of the isotropic BAO, it allows us to measure the difference between the observed cosmology and the fiducial cosmology. 
Thus, we should recover $\alpha_{\perp}=\alpha_{\parallel}=1$ from the Uchuu- and GLAM-SDSSbao lightcones. % as we are analysing the mocks with their fiducial cosmology.

From the covariance matrix $C$ computed in section~\ref{subsec:GLAM_clustering}, we can define the $\chi^2$ statistic as
\begin{equation}
    \chi^2 = \frac{N-p-2}{N-1} (\xi^{\text{Data}}-\xi^{\text{Model}})C^{-1}(\xi^{\text{Data}}-\xi^{\text{Model}})^T,
\end{equation}
where $p$ is the number of degrees-of-freedom being fitted and $N$ is the number of GLAM-SDSSbao lightcones, and we have included the Hartlap correction \citep{Hartlap_2006}. %The Hartlap correction \citep{Hartlap_2006} for the inverse of the covariance matrix is represented by the first term in the product.

%==================================%
%     THEORETICAL MODEL            %
%==================================%

 Our theoretical model for the TPCF, $\xi^{\text{Model}}$, is based on Lagrangian Perturbation Theory (LPT). 
 %It follows the displacements, $\boldsymbol{\Psi} (\boldsymbol{q},\tau)$, of fluid elements at starting position $q$ at some initial time, so that the observed position of the fluid elements is $\boldsymbol{x} = \boldsymbol{q} + \boldsymbol{\Psi}(\boldsymbol{q},\tau)$. 
%These displacements can be described by the usual Newtonian gravity in the expanding spacetime as:
%\begin{equation}
%    \ddot{\Psi}(q,\tau)  + \mathcal{H}\dot{\Psi}(q,\tau) = -\nabla_X \Phi
%\end{equation}
%where the dots refer to the conformal time derivatives, and $\Phi$ is the Newtonian gravitational potential, which is sourced by the overdensity of matter fluid elements $\delta_m$. 
%Using mass conservation, we get
%\begin{equation}
%    1+\delta_m(\boldsymbol{x},\tau)=\int d^3\boldsymbol{q} \delta_D(\boldsymbol{x}-\boldsymbol{q}-\boldsymbol{\Psi}(\boldsymbol{q},\tau)).
%\end{equation}
%One can solve then the displacements using the perturbation theory, where the first order solution is known as Zeldovich approximation.
To model the distribution of galaxies, one also needs to introduce a bias model that connects the matter density, $\delta_m$, and the galaxy density, $\delta_g$.  with two parameters $b_{1}$ and $b_{2}$ named the linear and quadratic Lagrangian bias, respectively. We can then obtain the model power spectrum, and include the effect of peculiar velocities to account for the RSD effect \citep{Seljak2011}. 

We also model the Fingers-of-God (FOG) effect in Fourier space, using the phenomenological Lorentz model \citep{Taruya_2010}:
%FOG equation
\begin{equation}
    P_\mathrm{FOG}(\boldsymbol{k}) = \frac{1}{1+(\boldsymbol{k}\cdot \hat{\boldsymbol{n}} \: \sigma_\mathrm{FOG})^2/2} P(\boldsymbol{k}).
\end{equation}
%Where we have followed \cite{Taruya_2010},
where $P(\boldsymbol{k})$ is the non-linear power spectrum without the FOG effect, $\sigma_\mathrm{FOG}$ is the one-dimensional velocity dispersion and $\hat{\boldsymbol{n}}$ is the normalised LOS direction vector.
%\Alex{[comment] make sure the parameters in these equations are defined}

The theoretical power spectrum $P_\mathrm{FOG}$ is obtained using the \texttt{MomentumExpansion} module of \texttt{velocileptors} package \citep[for more details, see][]{Chen_2020,Chen_2021}.
Finally, we obtain the correlation function by taking the Fourier transform of $P_\mathrm{FOG}$,
\begin{equation}
    \xi(\boldsymbol{x}) = \int d^3 \boldsymbol{k} e^{i\boldsymbol{k}\cdot \boldsymbol{x}} P_\mathrm{FOG}(\boldsymbol{k}).
\end{equation}
%We refer the reader to \cite{Chen_2020} for more details about \texttt{velocileptors}.
%Accounting for the Alcock-Paszynski effect happens by rescaling the distances relative to the "true" cosmology in directions parallel and perpendicular to the LOS. We have left out of consideration many details, such as resummation of long-wavelength displacements, loop corrections and so on, information on which can be found here \cite{Chen_2020}.

%==================================%
%     ANALYSIS                     %
%==================================%

We fit our RSD model to the correlation function multipoles from  three datasets SDSSbao, Uchuu-SDSSbao and GLAM-SDSSbao %obtained from the SDSSbao dataset
in the separation range $[25,145]~\hMpc$, with bins of width $5~\hMpc$. 
%The multipoles are shown in Fig.~\ref{fig:xi_s_BAO}.
%\Alex{[comment] I don't think this plot is needed. is it just the same as Fig 19 but with different binning? }
%with the mean of the 32 Uchuu mocks in blue, the SDSS data in orange, the mean of the 5092 GLAM mocks in green. The grey shaded region represents the 1$\sigma$ uncertainty region obtained from the diagonal terms of the 5092 GLAM mocks covariance matrix. 
In addition to the cosmological parameters $f\sigma_8$, $\alpha_{\parallel}$ and $\alpha_{\perp}$, we also estimate the Lagrangian biases $b_{1},b_{2}$ and the Fingers-of-God parameter $\sigma_\mathrm{FOG}$. Unphysical values of  parameters are avoided by setting the priors to $f\sigma_8>0$, $b_{1}>-1$ and $\sigma_\mathrm{FOG}>0$. The first Lagrangian bias is related to the Eulerian bias by $b_{1,\rm Eulerian}=1 + b_{1}$. We assume the effective redshift of the SDSSbao sample to be $z = 0.15$.

The correlation function multipoles corresponding to our best-fit models can be seen in Fig~\ref{fig:xi_s_BAO}. We observe good agreement with Uchuu-SDSSbao and GLAM-SDSSbao measurements. The RSD model predicts position of the BAO peak in the monopole that is too low as compared with the observed position of the peak. However, this disagreement is within the noise limits.
%\ca{comment on agreement with SDSS -- BAO peak seems to be displaced, why is this? Does this affect the reliability of our results?}.
%\Alex{Be consistent with notation for biases}

%\begin{figure}
   % \centering
   % \includegraphics[width=\columnwidth]{RSD_poles.pdf}
%    \includegraphics[scale=0.5]{RSD_poles.pdf}
   % \caption{First plot represents the monopoles $\ell=0$ of the correlation function plotted against the separation of the Uchuu mean mock, shown by the blue line, SDSS data, shown by the orange line, and GLAM mock mean from 5092 samples, shown by the green line. The second and the third plot show, accordingly, the quadrupole $\ell=2$ and the hexadecapole $\ell = 4$, with the same notation.}
 %   \label{fig:RSD_All_Plots}
%\end{figure}

The minima of $\chi^2$ is found using \texttt{iminuit} \citep{iminuit}, which achieves convergence near the minimum using the first and approximate second derivatives. Errors are estimated from the region of $\Delta \chi^2 =1$ of the marginalized $\chi^2$ distribution, and they are allowed to be asymmetric.
We also run Monte Carlo Markov chains (MCMC) with the \texttt{emcee} package \citep{emcee} in order to compute the likelihood surface of our set of fitted parameters. Their convergence is checked with the Gelman-Rubin convergence test \citep{Gelman1992, Brooks1998}.

%==================================%
%     RESULTS                      %
%==================================%

\begin{figure*}
    \centering
    \includegraphics[width=0.95\linewidth]{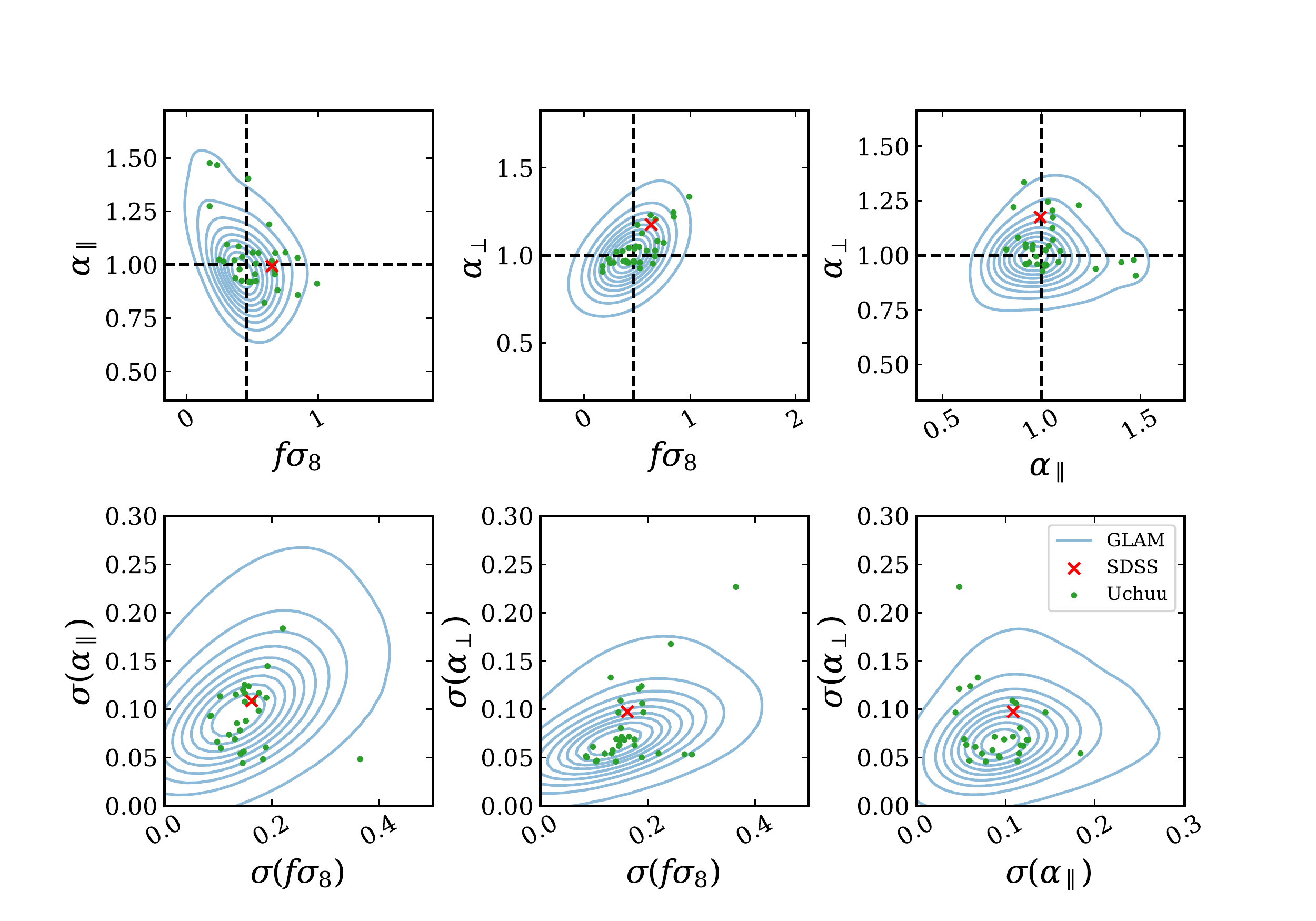}
    \caption{\textit{Upper row}: Parameter values of $f\sigma_8$, $\alpha_{\parallel}$ and $\alpha_{\perp}$ measured from the GLAM-SDSSbao (blue contours), Uchuu-SDSSbao (green points), and SDSSbao light-cones (red cross). Dashed black lines represent the expected values for the fiducial cosmology.
    \textit{Lower row}: The corresponding parameter errors. %for $f\sigma_8$, $\alpha_{\parallel}$, and $\alpha_{\perp}$. %The notation is the same as in the upper row. 
    %\Alex{The legend could be made a bit smaller so it doesn't overlap the contour. The + for sdss could be made a bit thinner (and maybe change to a x symbol instad of +) to make it stand out more from the Uchuu points - when zoomed out it's hard to see}
    %Slava: Made the SDSS labels larger, and rotated the x-labels for them not to overlap with the y-labels
    %\ca{I agree with Alex that a 'x' symbol would make it more visible. Also possibly choose another colour, since the contrast between orange and red is low.}
    %Slava: Done
    }
    \label{fig:RSD_Parameter_Spreads}
\end{figure*}

We first test our RSD pipeline on the GLAM-SDSSbao and Uchuu-SDSSbao light-cones. The best-fit results of the parameters $f\sigma_8$, $\alpha_{\parallel}$ and $\alpha_{\perp}$ are summarised in Fig.~\ref{fig:RSD_Parameter_Spreads}. We confirm that the theoretical model %using \texttt{velocileptors} 
recovers the fiducial Planck values within 1$\sigma$. %\Alex{[comment] the table referenced here is commented out}

%%
% \begin{table}
% \centering
% \setlength{\arrayrulewidth}{0.3pt}
% \begin{tabular}{cccc}
%   \hline
%   Parameter& $\chi^2$ minimization  & MCMC  & Expected \\
%   \hline
%   \\[-1em]
%   $f\sigma_8$ &$0.48^{+0.17}_{-0.17} $& $0.47^{+0.16}_{-0.16}$& 0.46\\
%   \\[-1em]
%   $\alpha_{\parallel}$ &$0.98^{+0.11}_{-0.08} $ &$0.98^{+0.12}_{-0.10}$  & 1\\
%   \\[-1em]
%   $\alpha_{\perp}$ & $0.94^{+0.10}_{-0.10}$& $0.96^{+0.09}_{-0.07}$ & 1 \\
%   \\[-1em]
%   \hline
% \end{tabular}
% \caption{RSD fit results of the Uchuu mocks, obtained by $\chi^2$ minimization in the first column, using Bayesian inference by running the Monte-Carlo chains in the second column, and the expected values in the third column}
% \label{tab:RSD_mocks_Results}
% \end{table}
%%
%----------------------
% Results on SDSS data
%----------------------

We then apply the pipeline to the SDSSbao sample. %, fitting it with the same configuration.
Our results are listed in Table~\ref{tab:RSD_Results}.
%We obtain the values of $f\sigma_8 =0.65^{+0.14}_{-0.14} $ from $\chi^2$ minimization and $f\sigma_8 =0.63^{+0.15}_{-0.14} $ from the MCMC sampling. 
%\ca{quoted values do not match the values in the table}
Both $\chi^2$ minimization and MCMC sampling methods provide consistent results, and the small difference in the errors and parameter values are attributed to a better treatment of the Fingers-of-God effect with MCMC chains. The parameter distributions for $\sigma_\mathrm{FOG}$ is non-Gaussian, as it is restricted to positive values, but the best-fit value is consistent with 0. Additionally, we compare our obtained $f \sigma_8$ and $b_1$ with \citet{Howlett_2015_b}, finding a good agreement, but %Since \citet{Howlett_2015_b} uses a different model for the parameter inference, not all parameters can been compared.
%Our results are consistent with \citet{Howlett_2015_b}. 
we obtain a $\gtrsim 30\%$ increase in precision on $f\sigma_8$ which it can be attributed to our better estimate of the covariance matrix (see Table~\ref{tab:RSD_Results}). 
%\Paco{[comment] The dashed black lines are missing in the figure}

%The SDSS multipoles, together with the residuals from the best-fit and uncertainties on them, represented by the shaded regions, are shown on Fig.~\ref{fig:RSD_SDSS_Fit_Poles}. 

%\begin{figure}
%    \centering
%    \includegraphics[width=0.75\columnwidth]{RSD_SDSS_Fit_Poles.png}
%    \caption{The upper plot presents correlation function multipoles with the best fit values and the mean of the SDSS mock correlation function multipoles, where red color stands for the monopole $\ell=0$, blue color is for the quadrupole $\ell-2$, and the green is for the hexadecupole $\ell=4$. The shaded regions stand for the uncertainties, inferred from 600 GLAM mocks. The dashed lines represent the best fit The lower plot present the differences between the fitted multipoles and their best fit. The notation is the same. }
%    \label{fig:RSD_SDSS_Fit_Poles}
%\end{figure}

\begin{table}
\centering
\begin{tabular}{cccc}
  \hline
  Parameter& $\chi^2$ minimization  & MCMC  & Reference \\ 
  \hline
  \\[-1em]
  $f\sigma_8$ & $0.65^{+0.17}_{-0.16}$ & $0.62^{+0.17}_{-0.17}$ & $0.63 ^{+0.24}_{-0.27}$  \\
  \\[-1em]
  $\alpha_{\parallel}$ & $0.99^{+0.11}_{-0.10}$ & $1.04^{+0.15}_{-0.10}$& N/A \\
  \\[-1em]
  $\alpha_{\perp}$ & $1.18^{+0.10}_{-0.08}$ & $1.17^{+0.07}_{-0.07}$& N/A  \\
  \\[-1em]
  
  $b_{1,\rm Eulerian}$ & $1.59^{+0.19}_{-0.19}$ & $1.62^{+0.22}_{-0.19}$& $1.36^{+0.29}_{-0.26}$  \\
  \\[-1em]
  $b_{2}$ & $-0.3^{+2.1}_{-1.6}$ & $-0.8^{+1.6}_{-1.4}$& N/A \\
  \\[-1em]
  $\sigma_\mathrm{FOG}[\hMpc]$ & $4.5^{+2.8}_{-4.5}$ &$3.6^{+2.5}_{-3.5}$& N/A \\
  \\[-1em]
  \hline
\end{tabular}
\caption{RSD fitted parameters from the SDSSbao data, obtained by $\chi^2$ minimization and using Bayesian MCMC inference. Only $f\sigma_8$ and $b_{1,\rm Eulerian}$ estimated values from \citet{Howlett_2015_b} are found in the literature.}
\label{tab:RSD_Results}
\end{table}

\begin{table}
\centering
\begin{tabular}{cccc}
  \hline
  Parameter& Uchuu  & GLAM  & Expected value \\ 
  \hline
  \\[-1em]
  $f\sigma_8$ & $0.48^{+0.15}_{-0.15}$ & $0.45^{+0.16}_{-0.16}$ & $0.46$  \\
  \\[-1em]
  $\alpha_{\parallel}$ & $0.94^{+0.11}_{-0.09}$ & $0.94^{+0.16}_{-0.10}$& 1.00 \\
  \\[-1em]
  $\alpha_{\perp}$ & $0.99^{+0.08}_{-0.07}$ & $1.00^{+0.09}_{-0.08}$& 1.00  \\
  \\[-1em]
    
  $b_{1,\rm Eulerian}$ & $1.42^{+0.17}_{-0.15}$ & $1.40^{+0.19}_{-0.17}$& N/A  \\
  \\[-1em]
  $b_{2}$ & $-0.4^{+1.6}_{-1.3}$ & $-0.6^{+1.6}_{-1.3}$& N/A \\
  \\[-1em]
  $\sigma_\mathrm{FOG}[\hMpc]$ & $0^{+9}_{-0}$ &$0^{+22}_{-0}$& N/A \\
  \\[-1em]
  \hline
\end{tabular}

\caption{RSD fitted cosmological parameters from the means of Uchuu and GLAM correlation functions, obtained using $\chi^2$ minimization in comparison with the values predicted by the fiducial cosmology.}
\end{table}

%----------------------
% Statistical tests
%----------------------

%\begin{figure*}
 %   \centering
  %  \includegraphics[width=0.95\linewidth]{RSD_hists.pdf}
   % \caption{\textit{Upper panels}: parameter distributions for $f\sigma_8$, $\alpha_{\parallel}$ and $\alpha_{\perp}$, where results from the GLAM mocks are shown by blue histograms, Uchuu mocks by orange histograms, and the red vertical line is from our fit to the SDSS measurements. The thick black lines show the expected values. \textit{Lower panels}: same as the upper panels, but showing the distribution of errors in $f\sigma_8$, $\alpha_{\parallel}$ and $\alpha_{\perp}$. \Alex{[comment] move the legend to the bottom right panel. Change 'Fits of mocks' to 'GLAM Mocks'} \ca{UCHUU -> Uchuu} \ca{also, it would be great if the full width figures could take a bit less of vertical space without sacrificing a lot of the better readability that they have now} \Javier{[comment] perhaps we could try to merge fig 23 and 24 and plot them in just two rows so they will be scaled down and use fewer vertical space}} 
 %   \label{fig:RSD_Parameter_Hists}
%\end{figure*}

%\begin{figure}
%    \centering
%    \includegraphics[width=\columnwidth]{RSD_mcmc.pdf}
%    %\includegraphics[width=\columnwidth]{RSD_mcmc.png}
%    \caption{Parameter posterior distributions obtained from the Monte-Carlo sampling with \texttt{emcee} from SDSS %data. The contours show the 68\% and 95\% confidence intervals.
%    \Javier{[comment] perhaps we could move fig 22 to appear on the right instead of below, or merge them}
%    \Alex{the labels giving the measured values of fsig8, etc, could be made bigger}}
%    \label{fig:RSD_MCMC}
%\end{figure}

The distribution of parameter values and their uncertainties can be seen in Fig. \ref{fig:RSD_Parameter_Spreads}, 
%where the results from the GLAM-SDSSbao mocks are shown in blue, the Uchuu-SDSSbao mocks in orange, our SDSSbao results in red, 
where the results from the SDSSbao sample together with those from GLAM and Uchuu are shown in blue, orange and red respectively,
and the black lines show the expected values of the parameters for the Planck fiducial cosmology. 
The results from GLAM- and Uchuu-SDSSbao are very consistent within each other and with the SDSSbao data, %and the comparison with the SDSS data also shows a very good consistency,
meaning that the mocks can be seen as a fair statistical representation of the data.
We confirm that the fits we have performed are valid and have an acceptable $\chi^2/\mathrm{dof}$, for GLAM mocks being $\chi^2/\mathrm{dof}= 1.00 \pm 0.18$, for Uchuu mocks being $ \chi^2/\mathrm{dof}= 1.06\pm 0.18 $ and for the SDSSbao sample being $\chi^2/\mathrm{dof} =0.99$. 
%\Paco{I guess the $\chi^2$ values given in Table 4 refer to $\chi^2/\mathrm{dof}$, isn't?} 

%Slava: This was a piece of text related to a now absent plot. Thank you for pointing it out!

%The separate measurements of $f\sigma_8$ from Uchuu mocks $\chi^2$ minimization with, together with the lines showing the expected values, fit of the mean of the correlation function measured, and the mean of the measured parameters with the dashed lines presenting the 1$\sigma$ deviation. \Julia{what?}
%\Paco{Sorry, I don't understand this.}

%\begin{figure}
%    \centering
%    \includegraphics[width=\columnwidth]{RSD_chi2s.pdf}
%    \caption{The normalised distribution of $\chi^2/\mathrm{dof}$ from the fits to the GLAM mocks (blue histogram) and %the Uchuu mocks (orange histogram). %fits is represented by the blue histogram, of the Uchuu mocks by the orange histogram, and 
%    The thick red line shows the $\chi^2/\mathrm{dof}$ for our fit to the SDSS data. %\Javier{thick black line -> thick red line} 
%    \Javier{uchuu mocks do not really look orange}}
%    \label{fig:RSD_Chi2s}
%\end{figure}

Finally we use our best-fit values of the BAO-inferred $\alpha_{\parallel}$ and $\alpha_{\perp}$ parameters to provide a measurement of the Hubble distance and the (comoving) angular diameter distance at the effective redshift of the SDSSbao sample, i.e. $D_\mathrm{H}(z=0.15) / r_d = 27.9^{+3.1}_{-2.7}$, $D_\mathrm{M}(z=0.15) / r_d = 5.1^{+0.4}_{-0.4}$.
%\ca{short description of $D_\mathrm{H}$, as it is not mentioned elsewhere in the text} 
% Slava: Changed the definition of alpha parallel and implemented DH definition there.
%\Paco{Please compute theses distances and their errors}. 
These distance results are valuable since at present only the spherically averaged distance $D_\mathrm{V}(z=0.15) / r_d$ has been reported from BAO isotropic measurements from a similar SDSSbao sample by \citet{Ross15}, see below for our own BAO isotropic analysis.

%\ca{do we also want to report the value of $\gamma$ we get, as mentioned in the introduction? I have removed mentions to $\gamma$ in the introduction as it’s not reported in the text. Feel free to restore if you feel it’s important / Let me know what you think}

%\Paco{Please check the $D_M$ definition as given in Alam et al. 2021, and update eq. 28. I guess $D_A$ and $D_M$ are both the comoving angular distance}

%\Pauline{Do we also want Fig.~\ref{fig:RSD_32_Uchuu}?}
%\begin{figure*}
    %\centering
    %\includegraphics[width=1.8\columnwidth]{RSD_32_UCHUU.png}
    %\caption{The first plot shows the $f\sigma_8$ measured for different Uchuu mocks as red triangles with with uncertainties, the orange solid %lines represent the fit of the mean multipoles of Uchuu mocks, with orange dashed lines showing the uncertainty of such fit, the green line %shows the mean value for the fits of the 32 mocks. The blue line represents the theoretical prediction from the given cosmology. The second and %the third plots show the same for the Alcock-Paczynski parameters  $\alpha_{\parallel}$ and $ \alpha_{\perp}$.}
 %   \label{fig:RSD_32_Uchuu}
%\end{figure*}
%

\subsection{Isotropic BAO measurements}
\label{subsec:BAO_measurement}
%Figure
\begin{figure*}
    \centering
    \includegraphics[width=0.65\columnwidth]{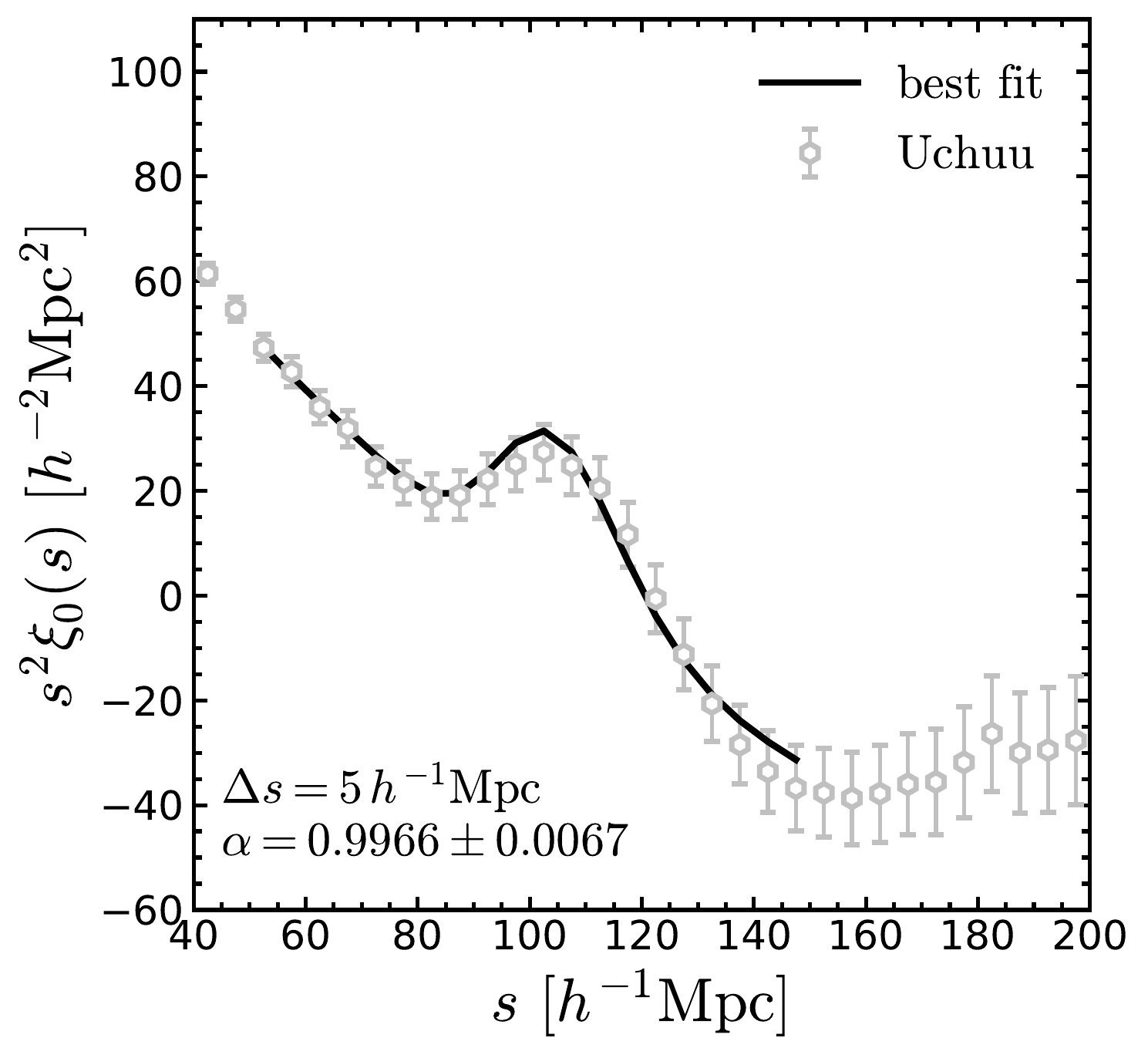}
    \includegraphics[width=0.65\columnwidth]{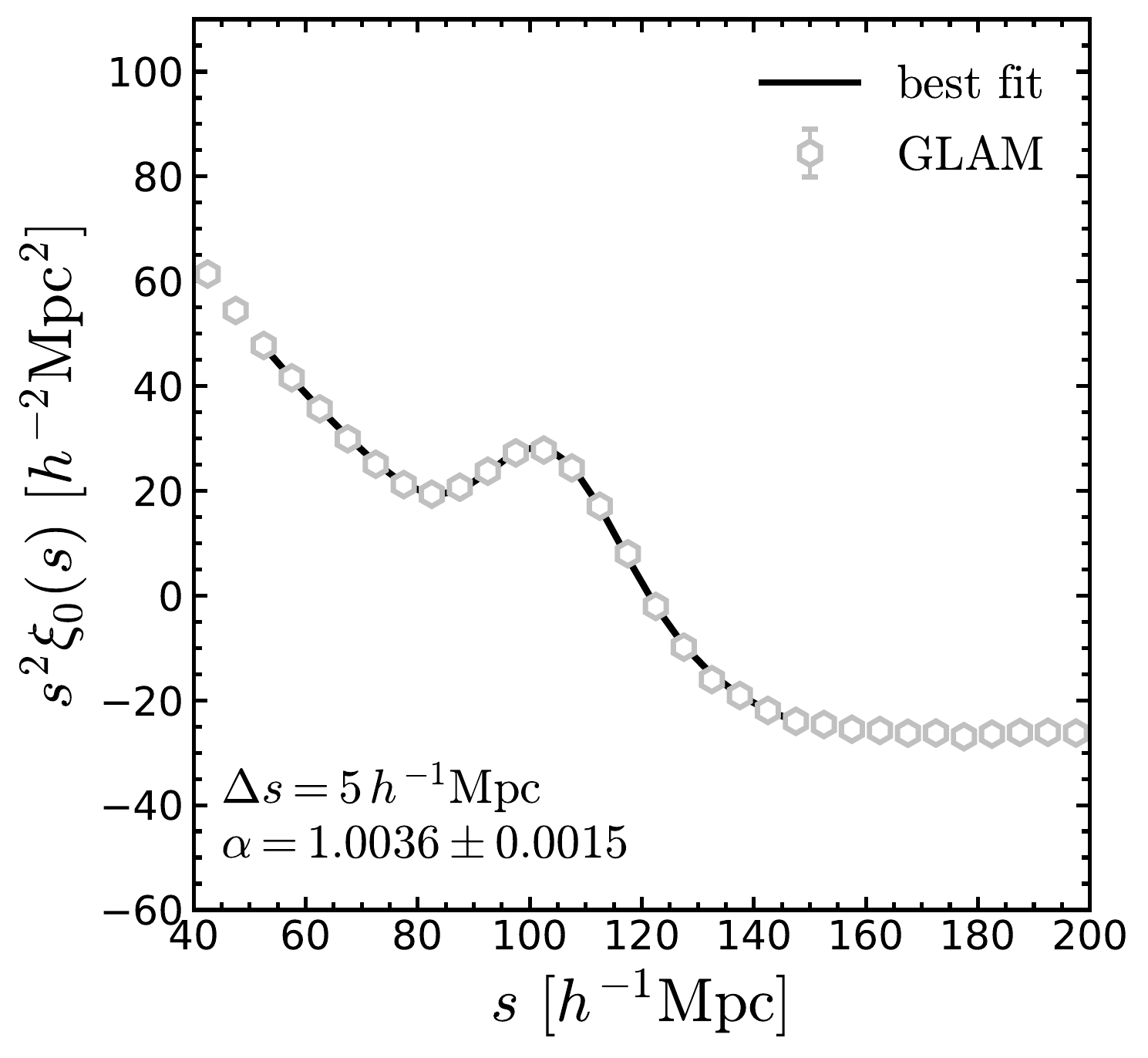}
    \includegraphics[width=0.65\columnwidth]{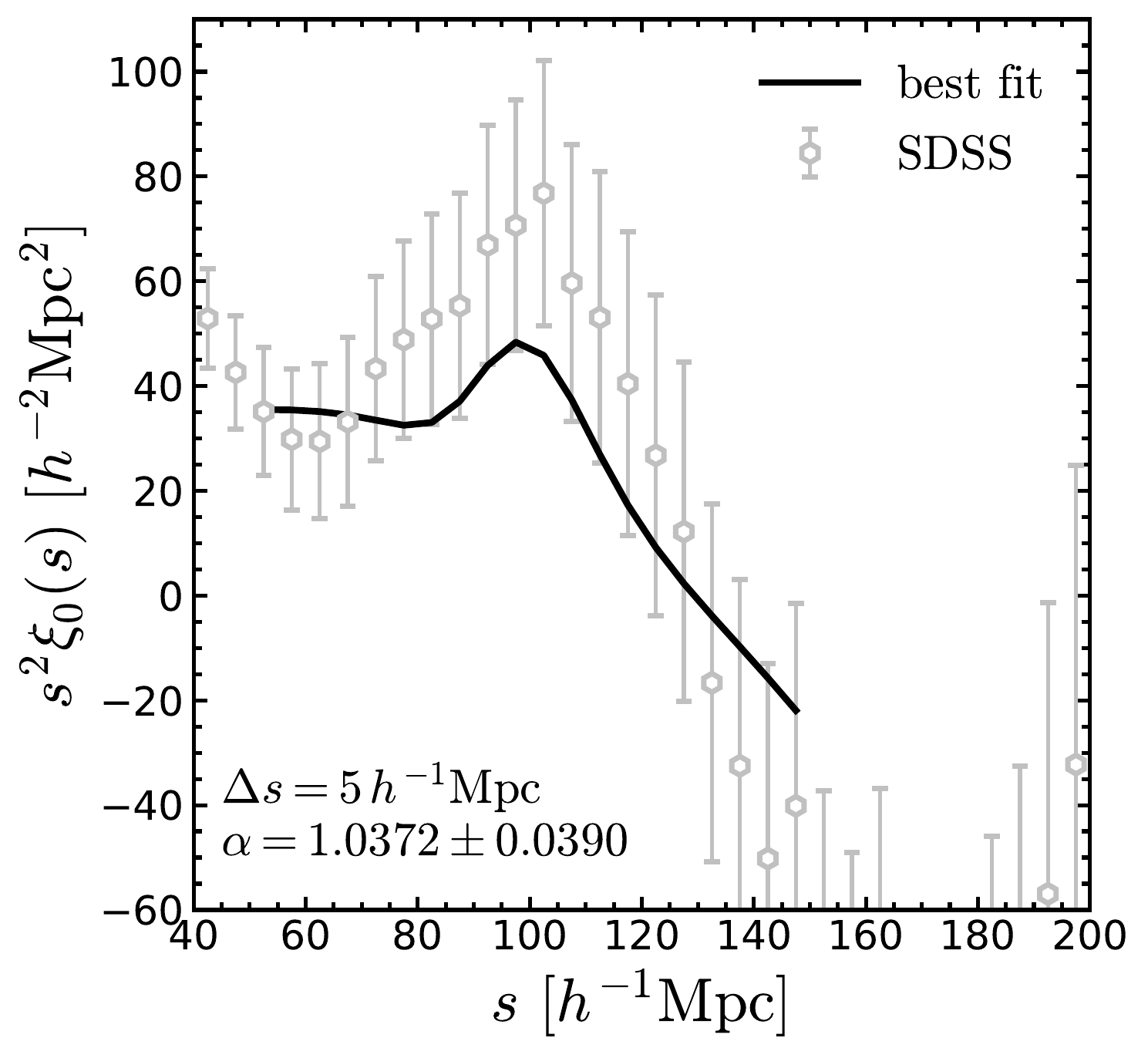}    
    \caption{Measured (grey dots) and best-fit (solid black lines) correlation function monopole from the Uchuu (left panel) and GLAM (middle panel) simulated light-cones, and SDSS observations (right panel) at $z=0.15$ using a bin width of $\Delta s = 5\,h^{-1}\rm{Mpc}$. The error bars for Uchuu and GLAM were rescaled by the square-root of the number of lightcones. %{\bf AK. Can barely see light grey symbols with errorbars. Can this be changed to, say blue?}
    %\Alex{The font size is too small}
    }
    \label{fig:bao_200-5}
\end{figure*}

%Figure
%\begin{figure*}
%    \centering
%    \includegraphics[width=0.65\columnwidth]{xi0_BAO_sdss-uchuu-new_200-8.pdf}
%    \includegraphics[width=0.65\columnwidth]{xi0_BAO_sdss-glam_200-8.pdf}
%    \includegraphics[width=0.65\columnwidth]{xi0_BAO_sdss-obs_200-8.pdf}    
%    \caption{The same as Fig.~\ref{fig:bao_200-5} but using a bin width of $\Delta s = 8\,h^{-1}\rm{Mpc}$.}
%    \label{fig:bao_200-8}
%\end{figure*}

We also measure the isotropic BAO scale from our Uchuu-SDSSbao and the SDSSbao data at $z=0.15$. The BAO signal can clearly be seen in the correlation function monopole measured from both data set (see Fig.~\ref{fig:xi_s_BAO}). The BAO scale can be extracted by fitting the monopole of the correlation function to a template that includes the dilation parameter, $\alpha$, 
\begin{equation}\label{eq:alpha}
    \alpha \equiv \frac{D_\mathrm{V}(z)r^\mathrm{fid}_\mathrm{d}}{D^\mathrm{fid}_\mathrm{V}(z)r_\mathrm{d}}\,,
\end{equation}
where
\begin{equation}\label{eq:DV}
    D_\mathrm{V}(z) = \left[cz(1+z)^2D^2_\mathrm{A}(z)H^{-1}(z)\right]^{1/3}\,,
\end{equation}
is the spherically average distance \citep{Eisenstein:2005su};
%$D_\mathrm{A}(z)$ is the angular-diametre distance, $r_\mathrm{d}$ is the sound horizon at the baryon drag epoch, $H(z)$ is the Hubble parameter and the superscript `fid' indicates the value of the distances in our fiducial cosmology, i.e., the Uchuu cosmology.  
%\Alex{[comment] some repetition from RSD section. Say more on how these equations relate to prev section}. 
$\alpha$ is related to the anisotropic Alcock-Paczynski parameters $\alpha_{\perp}$ and $\alpha_{\parallel}$ given in eqs.~\ref{eq:a_par} and \ref{eq:a_perp}, by
$\alpha = \alpha_{\parallel}^{1/3}\alpha_{\perp}^{2/3}$.

To obtain the best-fit $\alpha$ value, we use Bayesian statistics and maximise the likelihood by adopting the BAO model of the monopole of the redshift-space correlation function as presented in \citet{Ross15}. We perform the fit to the simulations and data monopoles on scales $50 < s < 150~\hMpc$, using separation bins of width $\Delta s = 5\hMpc$.
The measured and the best-fit model of the correlation function monopole from  the SDSSbao data and the simulated Uchuu and GLAM lightcones are shown in Fig.~\ref{fig:bao_200-5}.
We find $\alpha=0.9966\pm0.0067$, $\alpha=1.0036\pm0.0015$ and $\alpha=1.0372\pm0.0390$ from Uchuu, GLAM, and the SDSSbao observations, respectively. Our best-fit $\alpha$ value for the SDSS data and its error is consistent with that reported by \citet{Ross15}, i.e. $\alpha=1.057\pm0.037$. 

From our dilation parameter estimate we measure a spherically averaged distance $D_\mathrm{V}(z=0.15)/r_\mathrm{d} = 4.43 \pm 0.16$, in agreement with the value $D_\mathrm{V}(z=0.15)/r_\mathrm{d} = 4.47 \pm 0.17$ measured by \citet{Ross15}. The values of $D_\mathrm{V}$ from the Uchuu and GLAM lightcone measurements are  $D_\mathrm{V}(z=0.15)/r_\mathrm{d} = 4.257 \pm 0.028$ and $D_\mathrm{V}(z=0.15)/r_\mathrm{d} = 4.304 \pm 0.007$, respectively.
%$D_\mathrm{V}(z=0.15) = 664 \pm 25\,(r_{\rm d}/r^{\rm fid}_{\rm d})\,{\rm Mpc}$ measured by \citet{Ross15}. \Paco{Cesar, could you please compute $D_V$?}
%We additionally measure the BAO scale from the mean of the monopole of the correlation function of the mock catalogues and SDSS data but using bins of size $\Delta s = 8\hMpc$ (see Fig.~\ref{fig:bao_200-8}). We find consistent results with those from the measurements with $\Delta s = 5\hMpc$.

\section{Summary}
\label{sec:conclusions}
The cosmological interpretation of large galaxy surveys requires generation of high-fidelity simulated galaxy data. Here we describe and analyze publicly available Uchuu-SDSS lightcones: a set of simulated SDSS catalogues generated using the Uchuu $N$-body simulation. Uchuu is a large high-resolution cosmological simulation that follows the evolution of the dark matter across cosmic time in the Planck cosmology. The Uchuu-SDSS catalogues are tailored to reproduce the sky footprint and galaxy properties of the SDSS MGS observational sample. This facilitates the direct comparison between our simulated lightcones and the observational data to probe the Planck-$\Lambda$CDM cosmology model using the large-scale clustering signal.

The rest-frame $r$-band magnitudes, $\magr$, are assigned to the Uchuu haloes using the SHAM method, with a simple recipe for the scatter in the galaxy-halo connection. By construction, our scheme  reproduces  the SDSS and GAMA luminosity functions, and it reproduces to good accuracy the SDSS clustering (see Fig.~\ref{fig:tpcf_box_SDSS}), while using only one free parameter -- the scatter. The resulting galaxy catalogues computed at different redshifts are combined in spherical shells and cut to the SDSS sky footprint. 
By placing observers at different positions in the cubic box, we produce a set of 8 independent Uchuu-SDSS lightcones, with only a small mutual overlap in volume. Additionally, an extended set of 32 lightcones, which overlap for $z>0.175$, are generated to increase the statistics for our SDSS RSD and BAO measurements. Galaxy colours are also produced using a Monte Carlo method that randomly draws colours from the $\magr$ and redshift-dependent $g-r$ distribution obtained from the GAMA survey. Redshifts, $r$-band magnitudes and $g-r$ colours are used to match each simulated galaxy to a galaxy in the SDSS sample. This allows us to assign absolute magnitudes, apparent magnitudes and $k$-corrections in the $u$, $g$, $i$ and $z$-bands in addition to stellar masses and specific star formation rates.
%Absolute magnitudes, apparent magnitudes and $k$-corrections in the $u$, $g$, $i$ and $z$-bands, as well as stellar masses and specific star formations, are added by selecting the properties of the similar galaxy distribution in the SDSS sample. 
Finally, we implement the effect of fibre collisions by applying a nearest neighbour correction to a set of galaxies situated in close angular proximity.

Our Uchuu-SDSS lightcones are able to recover the galaxy redshift distribution (see Fig.~\ref{fig:dNdz}) and various galaxy properties from the SDSS survey to very high accuracy (see Figs.~\ref{fig:lightcone_absolute_magnitude},~\ref{fig:lightcone_g_r_colour}~and~\ref{fig:lightcone_ssfr}). We also compute the galaxy correlation functions  for several volume-limited samples corresponding to a wide range of luminosities (Fig.~\ref{fig:xi_s}) and stellar mass cuts (Fig.~\ref{fig:xi_s_Mstar}), finding a very good agreement with the SDSS clustering down to the smallest $100\, \hkpc$ scale.
%, despite the simplicity of our luminosity assignment procedure. %and scatter implementation.
The colour- and stellar-mass-dependent galaxy clustering (see Fig.~\ref{fig:xi_s_colour}) is  in general agreement with the SDSS results. Similarly, the simulated stellar mass function (Fig.~\ref{fig:SMF}) is in good agreement with that from SDSS data. We also provide the halo occupation distributions and parameters of the Uchuu-SDSS galaxies (Fig.~\ref{fig:all_hod}), which are in agreement with previous SDSS analyses.

We  explore the RSD and BAO signal in our Uchuu-SDSS lightcones and the SDSS data (see Fig.~\ref{fig:xi_s_BAO}). In order to obtain high-precision covariance matrices for the error estimates of the RSD and BAO measurements, we create a large number of GLAM simulations. We apply HOD method  to populate the GLAM halos with
galaxies, generating a total of 5100 GLAM-SDSSbao lightcones.
%In order to explore RSDs and the BAour lightcones (Fig.~\ref{fig:xi_s_BAO}) with the improved statistics required to produce covariance matrices, we employ a statistical HOD method informed by the HOD of our original Uchuu-SDSSbao sample mocks in order to populate with galaxies the GLAM simulations, obtaining a total of 5100 GLAM-SDSSbao mocks. 
We measure $f \sigma_8$ and the anisotropic BAO parameters, $\alpha_{\perp}$ and $\alpha_{\parallel}$, from a full-shape model fit of the TPCF, finding a very good agreement between the SDSSbao data and our GLAM- and high-fidelity Uchuu-SDSSbao lighcones (see Fig.~\ref{fig:RSD_Parameter_Spreads}). Our results are given in Table~\ref{tab:RSD_Results}. We obtain a $\gtrsim 30\%$ increase in precision on $f\sigma_8$, as compared to the previous measurement by \citep{Howlett_2015_b}, which  can be attributed to our better estimate of the covariance matrix. 

We use our best-fit values of the BAO-inferred $\alpha_{\parallel}$ and $\alpha_{\perp}$ parameters to provide a measurement of the Hubble distance and the (comoving) angular diameter distance at the effective redshift of the SDSSbao sample, i.e. $D_\mathrm{H}(z=0.15) / r_d = 27.9^{+3.1}_{-2.7}$, $D_\mathrm{M}(z=0.15) / r_d = 5.1^{+0.4}_{-0.4}$. We highlight that these distance results are valuable since at present only the spherically averaged distance $D_\mathrm{V}(z=0.15) / r_d$ has been reported by \citet{Ross15}. Finally, we measure the isotropic dilation scale, $\alpha$, of the BAO signal by fitting a model template of the BAO peak \citep{Ross15}, obtaining again a good agreement between the simulated and observational SDSS data (Fig.~\ref{fig:bao_200-5}).

Based on our results, we conclude  that the Planck \textLambda CDM cosmology nicely explains the observed statistics of the large-scale structure of the SDSS main galaxy survey.

%In summary, we have produced a suite of realistic lightcone mocks, which are constructed to reproduce the overall properties of SDSS, and are able to reproduce to very good accuracy a large range of its observational results. Our work illustrates the power of the Uchuu-SDSS catalogues as means to aid the interpretation of SDSS data and optimise the information that can be extracted from SDSS, as well as to constrain the connection between galaxies and their host dark matter haloes, and compare the observed LSS to the predictions from the Planck \textLambda CDM cosmology.

This work shows the great potential of the Uchuu simulation as a canvas for the creation of simulated galaxies (and quasars) for large surveys. The procedures presented in this paper can be readily applied for the creation of high-fidelity lightcones from Uchuu, and covariance errors from GLAM simulations, tailored for upcoming galaxy surveys such as DESI \citep{DESI16}, Euclid \citep{Laureijs11}, and LSST \citep{LSST2009}. This will aid the evaluation of analysis pipelines, the assessment of observational biases and systematic effects, and enable cosmological models to be probed.

\section*{Acknowledgements}

We thank Gary Mamon and Marko Shuntov for helpful discussions. We thank Instituto de Astrofisica de Andalucia (IAA-CSIC), Centro de Supercomputacion de Galicia (CESGA) and the Spanish academic and research network (RedIRIS) in Spain for hosting Uchuu DR1 in the Skies \& Universes site for cosmological simulations. The Uchuu simulations were carried out on Aterui II supercomputer at Center for Computational Astrophysics, CfCA, of National Astronomical Observatory of Japan, and the K computer at the RIKEN Advanced Institute for Computational Science. The Uchuu DR1 effort has made use of the \texttt{skun@IAA-RedIRIS} and \texttt{skun6@IAA} computer facilities managed by the IAA-CSIC in Spain (MICINN EU-Feder grant EQC2018-004366-P). 
%This work used the \texttt{DiRAC@Durham} facility managed by the Institute for Computational Cosmology on behalf of the STFC DiRAC HPC Facility (www.dirac.ac.uk). 
This work used the \texttt{DiRAC@Durham} facility managed by the Institute for Computational Cosmology on behalf of the STFC DiRAC HPC Facility (www.dirac.ac.uk). The equipment was funded by BEIS capital funding via STFC capital grants ST/K00042X/1, ST/P002293/1, ST/R002371/1 and ST/S002502/1, Durham University and STFC operations grant ST/R000832/1. DiRAC is part of the National e-Infrastructure.
CAD-P, AS, JE, FP, AK, JR thank the support of the Spanish Ministry of Science and Innovation funding grant PGC2018-101931-B-I00. CAD-P gratefully acknowledges generous funding from the John Simpson Greenwell Memorial Fund.
CH-A acknowledges support from the Excellence Cluster ORIGINS which is funded by the Deutsche Forschungsgemeinschaft (DFG, German Research Foundation) under Germany's Excellence Strategy - EXC-2094 - 390783311. 
TI has been supported by IAAR Research Support Program in Chiba University Japan, 
MEXT/JSPS KAKENHI (Grant Number JP19KK0344, JP21F51024, and JP21H01122), 
MEXT as ``Program for Promoting Researches on the Supercomputer Fugaku'' (JPMXP1020200109),
and JICFuS.

%%%%%%%%%%%%%%%%%%%%%%%%%%%%%%%%%%%%%%%%%%%%%%%%%%
\section*{Data Availability}

The 32 Uchuu-SDSS galaxy lightcones, the 6 Uchuu-box galaxy catalogues at redshifts ${z=\{0, 0.093, 0.19, 0.3, 0.43, 0.49\}}$, the 5100 Uchuu-SDSSbao galaxy lightcones, and the companion SDSS LSS catalogue used in this work are made available at \url{http://www.skiesanduniverses.org/Simulations/Uchuu/}, together With the information on how to read the data. For a list and brief description of the available catalogue columns, please see Appendix~\ref{App:mock_columns}.



\bibliographystyle{mnras}
\bibliography{references}

%\begin{thebibliography}{99}
%\end{thebibliography}

%%%%%%%%%%%%%%%%%%%%%%%%%%%%%%%%%%%%%%%%%%%%%%%%%%

%%%%%%%%%%%%%%%%% APPENDICES %%%%%%%%%%%%%%%%%%%%%

\appendix

\section{Content of the Uchuu-SDSS catalogues}
\label{App:mock_columns}
%Along with this paper, we make publicly available at \url{http://www.skiesanduniverses.org/Simulations/Uchuu/} our 32 Uchuu-SDSS lightcones, the 5100 GLAM-SDSSbao lightcones, the companion SDSS data set, and a subset of the Uchuu box mocks at redshifts ${z=\{0, 0.093, 0.19\}}$. 
Below is a list of the columns of each data set, along with a short description.

\subsection{Uchuu-SDSS galaxy lightcones}

%\Alex{[comment] be aware that MNRAS charge £50 for each extra page beyond 20. This information about the mocks is great, but could be moved to the online documentation if you want to reduce the extra charges.}
%\Tomo{Explaining columns with tables seems to be viewable, although the current format is ok.} \ca{this will all be moved to \url{http://www.skiesanduniverses.org/Simulations/Uchuu/}}

%Our Uchuu-SDSS mocks are cut to the footprint of our SDSS sample as described in \ref{subsec:light_mocks}. 
Each Uchuu-SDSS lightcone has $\sim590\,000$ galaxies in total (excluding the regions of low fibre-collision completeness), with the following columns:
\begin{itemize}
    \item \texttt{galaxy\_type}: indicates whether the galaxy is central or a satellite (0 for centrals, 1 for satellites).
    \item \texttt{ra}: right ascension (degrees).
    \item \texttt{dec}: declination (degrees).
    \item \texttt{z\_cos}: cosmological redshift.
    \item \texttt{z\_obs}: observed redshift (accounting for peculiar velocities).
    \item \texttt{z\_obs\_fib}: observed redshift including fibre collisions, i.e. fibre-collided galaxies get a nearest neighbour correction (see Section~\ref{subsec:fibre}).
    \item \texttt{k\_corr\_r}: $r$-band $k$-correction at a reference redshift $z=0.1$. There are similar columns for the other SDSS photometric bands i.e. \texttt{k\_corr\_u}, \texttt{k\_corr\_g}, \texttt{k\_corr\_i} and \texttt{k\_corr\_z}.
    \item \texttt{k\_corr\_r\_fib}: $r$-band $k$-correction, recomputed for fibre-collided galaxies using the nearest neighbour-corrected redshift, \texttt{z\_obs\_fib}.% \Paco{I guess there are similar columns for the other bands, isn't?}.\Alex{This is just for the r-band. I think Adam calculated this, using the GAMA colour-dependent k-corrections. Getting k-corrections for the other bands would be more complicated}
    \item \texttt{M\_r}: rest-frame $r$-band absolute magnitude $\magr$, $k$-corrected to $z=0.1$, with no E-correction. Similarly, the absolute magnitudes in the other bands are \texttt{M\_u}, \texttt{M\_g}, \texttt{M\_i} and \texttt{M\_z}.
    \item \texttt{M\_r\_fib}: $r$-band absolute magnitude accounting for fibre collisions, i.e. recomputed for fibre collided galaxies using the values of \texttt{kcorr\_r\_fib} and \texttt{z\_obs\_fib}. %\Paco{Are there similar columns for the other bands?} \ca{from the files I believe this only exists for the r-band}.
    \item \texttt{m\_r}: apparent $r$-band magnitude, $r$. The apparent magnitudes in the other bands are \texttt{m\_u}, \texttt{m\_g}, \texttt{m\_i} and \texttt{m\_z}.
    \item \texttt{g\_r}: rest-frame $g-r$ colour $k$-corrected to $z=0.1$, ${}^{0.1}(g-r)$. This is the colour from our colour-assignment algorithm.
    \item \texttt{g\_r\_obs}: observer-frame $g-r$ colour, from our colour-assignment algorithm.
    \item \texttt{stellar\_mass\_MPA}: MPA stellar mass ($\hsqMsun$).
    \item \texttt{stellar\_mass\_granada\_best}: Granada best-fitting stellar mass ($\hsqMsun$).
    \item \texttt{stellar\_mass\_granada\_median}: Granada median stellar mass ($\hsqMsun$).
    \item \texttt{ssfr\_MPA}: MPA specific star formation rate ($\log_{10}(1/\mathrm{Gyr})$).
    \item \texttt{ssfr\_granada\_best}: Granada best-fitting stellar mass ($\log_{10}(1/\mathrm{Gyr})$).
    \item \texttt{ssfr\_granada\_median}: Granada median stellar mass ($\log_{10}(1/\mathrm{Gyr})$).
    \item \texttt{idnn}: ID of the nearest neighbour in the original catalogue (-1 if no galaxies are within $55''$; -2 if the galaxy is collided but keeps its true spectroscopic redshift).
    \item \texttt{is\_collided}: indicates whether a galaxy is fibre-collided (is \texttt{True} if \texttt{idnn} is not either -1 or -2).
    \item \texttt{completeness}: fibre-collision completeness in the healpix pixel containing the galaxy.
    \item \texttt{snap}: snapshot number from the original Uchuu-box catalogue.
\end{itemize}
In addition, the following columns describe the DM host halo properties of the Uchuu-SDSS galaxies, taken from the original values in the Uchuu simulation halo catalogues.
\begin{itemize}
    \item \texttt{First\_Acc\_Scale}: scale factor at which current and former satellites first passed through a larger halo.
    \item \texttt{Macc}: halo virial mass at accretion, excluding unbound particles ($\hMsun$).
    \item \texttt{Mvir\_all}: mass enclosed within the virial overdensity, including unbound particles ($\hMsun$).% {\bf is this unit correct?}
    \item \texttt{Rvir}: virial radius ($\hkpc$).
    \item \texttt{Vpeak}: peak value of $V_\mathrm{max}$ over the halo history (physical $\si{\kilo\meter\per\second}$).
    \item \texttt{halo\_id}: halo ID.
    \item \texttt{halo\_mass}: mass enclosed within an overdensity of $200\rho_\mathrm{crit}$, where $\rho_\mathrm{crit}$ is the critical density at the snapshot redshift ($\hMsun$).
    \item \texttt{is\_cen}: indicates whether the halo is central (\texttt{True} for centrals, \texttt{False} for satellite).
    \item \texttt{pid}: ID of the parent central halo for satellite haloes, -1 for central haloes.
    \item \texttt{pos}: 3D position vector. The coordinate system of the Uchuu cubic box has been shifted so that the observer is at the origin, and there are periodic replications of the box (comoving $\hMpc$).
    \item \texttt{rs}: scale radius of a fitted NFW profile (comoving $\hkpc$)
    \item \texttt{vel}: 3D velocity (physical $\kms$).
    \item \texttt{vrms}: velocity dispersion (physical $\kms$).
\end{itemize}

Additionally, a set of 10 random catalogues, each containing $10$ times the number of galaxies of an individual catalogue, are provided in order to facilitate the analysis of the simulated catalogues. The randoms are obtained by sampling galaxies randomly from the Uchuu-SDSS catalogues, but reassigning them a random sky position chosen with uniform probability within the footprint of the lightcone.

%More information on how to read the catalogue is provided in the Skies \& Universe website. 

\subsection{Uchuu-box galaxy catalogues}
For the $(2~\hGpc)^3$ Uchuu-box galaxy catalogues (at redshifts ${z=\{0, 0.093, 0.19\}}$), there are two galaxy properties: \texttt{M\_r} and \texttt{g\_r}, defined as in the previous subsection.

The halo columns are the same as in the Uchuu-SDSS lightcones, with a few exceptions: \texttt{halo\_id} is renamed as \texttt{id}, and \texttt{halo\_mass} is renamed as \texttt{M200c}. The columns \texttt{pos}, \texttt{vel}, and \texttt{is\_cen} are removed, and the following are added:
\begin{itemize}
    \item \texttt{x}: x-position of the halo/galaxy (comoving $\si{\mega\parsec\per\h}$). Similarly for \texttt{y} and \texttt{z}.
    \item \texttt{vx}: x-velocity of the halo/galaxy (physical $\si{\kilo\meter\per\second}$). Similarly for \texttt{vy} and \texttt{vz}.
\end{itemize}
%
%Additionally, there are two galaxy properties \texttt{M\_r} and \texttt{g\_r}, defined as in the previous subsection.

\subsection{GLAM-SDSSbao galaxy lightcones}

Our 5100 GLAM-SDSSbao lightcones, described in \ref{subsec:GLAM_construction}, have the following columns:
\begin{itemize}
    \item \texttt{galaxy\_id}: indicates whether the galaxy is central or a satellite (1 for centrals, 2 for satellites).
    \item \texttt{ra}: right ascension (degrees).
    \item \texttt{dec}: declination (degrees).
    \item \texttt{z\_cos}: cosmological redshift.
    \item \texttt{z\_obs}: observed redshift (accounting for peculiar velocities).
\end{itemize}
In addition, the following columns describe the DM host halo properties of the galaxies, taken from the original values found in the GLAM simulation halo catalogues,
\begin{itemize}
    \item \texttt{Mtotal}: halo virial mass ($\hMsun$).
    \item \texttt{Rvir}: halo virial radius ($\hkpc$).
    \item \texttt{rs}: scale radius of a fitted NFW profile (comoving $\hkpc$).
    \item \texttt{pos}: 3D position vector. The coordinate system of the GLAM cubic box has been shifted so that the observer is at the origin, and there are periodic replications of the box (comoving $\hMpc$).
    \item \texttt{vel}: 3D velocity (physical $\kms$).
    \item \texttt{vlos}: velocity vector projected along line-of-sight, with the observer positioned at the origin ($\kms$).
\end{itemize}

\subsection{SDSS data sample}
We also make available the observed SDSS galaxy sample, which consists of $497\,536$ (excluding regions of low fibre-collision completeness).% \Paco{This statistics should be consistent with that given in A1. In A1 was about 600,000 galaxies including regions of low fiber-collision completness. I'd suggest to chose one criteria in both.}.
The columns, which are defined the same way as in the Uchuu-SDSS lightcones (see A1), with the same units, are as follows: 
\begin{itemize}
\item \texttt{indx}: ID of the galaxy. %\Alex{should this be `index'?} \ca{just checked, it is indeed index in the files}
    \item \texttt{ra}%: right ascension (degrees).
    \item \texttt{dec}%: declination (degrees).
    \item \texttt{z}: measured redshift, including fibre collisions.
    \item \texttt{k\_corr\_r}, \texttt{k\_corr\_u}, \texttt{k\_corr\_g}, \texttt{k\_corr\_i}, \texttt{k\_corr\_z}%: $r$-band $k$-correction at reference redshift $z=0.1$. Similarly for the other photometric bands i.e. \texttt{k\_corr\_u} for the $u$-band, \texttt{k\_corr\_g} for the $g$-band, \texttt{k\_corr\_i} for the $i$-band and \texttt{k\_corr\_z} for the $z$-band.
    \item \texttt{M\_r}, \texttt{M\_u}, \texttt{M\_g}, \texttt{M\_i}, \texttt{M\_z}%: rest-frame $r$-band absolute magnitude $\magr$, $k$-corrected to $z=0.1$, without the $e$-correction. Similarly for \texttt{M\_u}, \texttt{M\_g}, \texttt{M\_i} and \texttt{M\_z}.
    \item \texttt{m\_r}, \texttt{m\_u}, \texttt{m\_g}, \texttt{m\_i}, \texttt{m\_z}%: apparent $r$-band magnitude $m_r$. Similarly for \texttt{m\_u}, \texttt{m\_g}, \texttt{m\_i} and \texttt{m\_z}.
    \item \texttt{g\_r}%: rest-frame $g-r$ colour $k$-corrected to $z=0.1$, ${}^{0.1}(g-r)$.
    \item \texttt{g\_r\_obs}%: observer-frame $g-r$ colour.
    \item \texttt{mass\_MPA}%: MPA stellar mass ($\log(\si{\Msun\per\h\squared})$).
    \item \texttt{mass\_granada\_best}%: granada best-fit stellar mass ($\log(\si{\Msun\per\h\squared})$).
    \item \texttt{mass\_granada\_median}%: granada median stellar mass ($\log(\si{\Msun\per\h\squared})$).
    \item \texttt{ssfr\_MPA}%: MPA specific star formation rate ($\log_{10}(1/\mathrm{Gyr})$).
    \item \texttt{ssfr\_granada\_best}%: granada best-fit stellar mass ($\log_{10}(1/\mathrm{Gyr})$).
    \item \texttt{ssfr\_granada\_median}%: granada median stellar mass ($\log_{10}(1/\mathrm{Gyr})$).
    \item \texttt{idnn}%: for a fibre-collided galaxy, ID of the nearest neighbour, used to assign its redshift. -1 if the galaxy has its own spectroscopic redshift.
    \item \texttt{fgotten}: fibre-collision completeness in the region containing the galaxy.
\end{itemize}

%For more information on how to read the catalogue, please see the Readme file at \texttt{Uchuu-SDSS/data/README}. 
Additionally, a random file with $67$ times the number of galaxies of each GLAM-SDSSbao catalogue is made available to facilitate the analysis.

%\Alex{Are the GLAM catalogues also being made available?}

\section{The effect of scatter in SHAM}
\label{App:scatter}
%As discussed earlier in the paper 
As discussed in Section~\ref{subsubsec:luminosity_SHAM}, our SHAM model has only one free parameter $\sigma$, which roughly corresponds to the scatter in $\magr$ at a fixed value of our halo mass proxy, $V\mathrm{peak}$. In practice, the value of $\sigma$ affects mainly the amplitude of the TPCF for a given volume-limited sample, results are shown in Fig.~\ref{fig:tpcf_box_SDSS} for different values of the scatter. Increasing the value of $\sigma$ introduces a larger number of low $V_\mathrm{peak}$ galaxies into the volume-limited samples, which are weakly clustered (and also removes high $V_\mathrm{peak}$ galaxies, which are strongly clustered). This has the effect of decreasing the amplitude of the TPCF. The effect of scatter is consequently larger for strongly clustered volume-limited samples, becoming negligible for samples with magnitude threshold $\magr>-20$. In order to tune the scatter parameter, we optimise the goodness-of-fit of the TPCF monopole for the Uchuu snapshot at $z=0.093$ to the SDSS measurements, finding an optimal value of $\sigma=0.5$. We choose this snapshot since it is the closest to the median SDSS redshift $z=0.1$.
%
% \begin{figure}
%     \centering
%     \includegraphics[width=\columnwidth]{figures/TPCF_callibration.pdf}
%     \caption{Like Fig.~\ref{fig:tpcf_box_SDSS}, showing the monopole of the two-point correlation function for the snapshot closest to the median redshift of SDSS ($z=0.093$). The clustering is shown for different values of the scatter parameter $\sigma$, namely $\sigma = \{0.4, 0.5, 0.6\}$, by the dotted, sold and dashed curves, respectively.}
%     \label{fig:tpcf_box_SDSS_calibration}
% \end{figure}
%
This value is in agreement with previous observational estimations of the scatter using SHAM. For example, \citet{Trujillo-Gomez11} estimate the the scatter of $\magr$ at fixed halo circular velocity, by combining an estimate of the intrinsic scatter of the Tully-Fisher relation \citep{Verheijen01} with the scatter resulting from the distribution of dust extinction corrections in SDSS, arriving precisely at a value of $0.5$.

%%%%%%%%%%%%%%%%%%%%%%%%%%%%%%%%%%%%%%%%%%%%%%%%%%

% Don't change these lines
\bsp	% typesetting comment
\label{lastpage}
\end{document}